\documentclass[aps,prd,preprint,tightenlines,eqsecnum,showpacs,footinbib,amsmath,amssymb,amsfonts,nofootinbib]{revtex4}

\usepackage[usenames]{color}
\usepackage{dsfont}
\usepackage[unicode=true,pdfusetitle,
bookmarks=true,bookmarksnumbered=false,bookmarksopen=false,
breaklinks=false,pdfborder={0 0 0},backref=false,colorlinks=false]
{hyperref}
\usepackage{hyperref}
\usepackage{longtable}
\usepackage{bm}
\usepackage{graphicx}
\usepackage{subeqnarray}

\newcommand\edth{\;\raise1.0pt\hbox{/}\hskip-6pt\partial}
\newcommand\edthb{\;\overline{\raise1.0pt\hbox{/}\hskip-6pt\partial}}

\newcommand{\be}{\begin{equation}}
\newcommand{\ee}{\end{equation}}
\newcommand{\bea}{\begin{subeqnarray}}
\newcommand{\eea}{\end{subeqnarray}}
\newcommand{\beanosub}{\begin{eqnarray}}
\newcommand{\eeanosub}{\end{eqnarray}}
\newcommand{\gr}[1]{\mathbf{#1}}

\newcommand{\dbi}{\partial_{i}}
\newcommand{\dbj}{\partial_{j}}
\newcommand{\dhi}{\partial^{i}}

\newcommand{\lp}{\left(}
\newcommand{\lcr}{\left[}
\newcommand{\rp}{\right)}
\newcommand{\rcr}{\right]}
\newcommand{\ii}{\mathrm{i}}

\newcommand{\uline}[1]{\underline{#1}}

\def\dd{{\rm d}}
\def\HH{\mathcal{H}}

\newcommand{\MyB}{{\cal B}}
\newcommand{\myaleph}{{\varpi}}
\newcommand{\thetrad}{{\vartheta}}
\newcommand{\Thetrad}{{\Theta}}
\newcommand{\thetradUD}[2]{\thetrad^{#1}_{#2}}
\newcommand{\ThetradUD}[2]{\Thetrad^{#1}_{#2}}
\newcommand{\thetradDU}[2]{\thetrad_{#1}^{#2}}
\newcommand{\ThetradDU}[2]{\Thetrad_{#1}^{#2}}
\newcommand{\snumb}{{\cal N}}
\newcommand{\angleBianchiModel}{\varphi}
\newcommand{\obs}{{\rm o}}
\newcommand{\Rthreesymbol}{{\mathbb{R}^3}}
\newcommand{\ppuB}{{C}}

\newcommand{\Ti}{{\uline{i}}}
\newcommand{\Tj}{{\uline{j}}}

\newcommand{\tz}{{\uline{0}}}
\newcommand{\ti}{{\uline{i}}}
\newcommand{\tj}{{\uline{j}}}
\newcommand{\tk}{{\uline{k}}}
\newcommand{\tp}{{\uline{p}}}

\newcommand{\tq}{{\uline{q}}}
\newcommand{\ta}{{\uline{a}}}
\newcommand{\tb}{{\uline{b}}}
\newcommand{\tc}{{\uline{c}}}
\newcommand{\td}{{\uline{d}}}
\newcommand{\zz}{{\{0,0\}}}
\newcommand{\oz}{{\{1,0\}}}
\newcommand{\zo}{{\{0,1\}}}
\newcommand{\oo}{{\{1,1\}}}

\def\sva{\sigma_{_{\rm V}a}}
\def\svap{(\sigma')_{_{\rm V}a}}

\def\svb{\sigma_{_{\rm V}b}}
\def\svbp{(\sigma')_{_{\rm V}b}}

\def\stl{\sigma_{_{\rm T}\lambda}}
\def\stoneminusl{\sigma_{_{\rm T}(1-\lambda)}}
\def\stlp{(\sigma')_{_{\rm T}\lambda}}

\def\spar{\sigma_{_\parallel}}
\def\sparp{(\sigma')_{_\parallel}}
\def\stplus{\sigma_{_{\rm T}+}}
\def\stcross{\sigma_{_{\rm T}\times}}
\def\pva{\Pi_{_{\rm V}a}}
\def\pvb{\Pi_{_{\rm V}b}}
\def\ptl{\Pi_{_{\rm T}\lambda}}

\def\ppar{\Pi_{_\parallel}}

\def\kiu{\hat{k}^i}
\def\kid{\hat{k}_i}
\def\kju{\hat{k}^j}
\def\kjd{\hat{k}_j}
\def\klu{\hat{k}^l}
\def\kld{\hat{k}_l}

\newcommand{\troisj}[6]{\left(\begin{array}{ccc}
      #1 & #2 & #3 \\
      #4 & #5 & #6\end{array}\right)}

\begin{document}
\title{Weak-lensing by the large scale structure in a spatially anisotropic universe: theory and predictions}

\author{Cyril Pitrou}
\email{pitrou@iap.fr}
\affiliation{
\mbox{Institut d'Astrophysique de Paris, Universit\'e  Pierre~\&~Marie Curie - Paris VI}, 
\mbox{CNRS-UMR 7095, 98 bis, Bd Arago, 75014 Paris, France} \\ 
\mbox{Sorbonne Universit\'es, Institut Lagrange de Paris, 98 bis Bd Arago, 75014 Paris, France}
}

\author{Thiago S. Pereira}
\email{tspereira@uel.com}
\affiliation{
\mbox{Departamento de F\'isica, Universidade Estadual de Londrina,}\\
\mbox{Rod. Celso Garcia Cid, Km 380, 86057-970, Londrina, Paran\'a, Brazil.}
}

\author{Jean-Philippe Uzan}
\email{uzan@iap.fr}
\affiliation{
\mbox{Institut d'Astrophysique de Paris, Universit\'e  Pierre~\&~Marie Curie - Paris VI}, 
\mbox{CNRS-UMR 7095, 98 bis, Bd Arago, 75014 Paris, France} \\ 
\mbox{Sorbonne Universit\'es, Institut Lagrange de Paris, 98 bis Bd Arago, 75014 Paris, France,}\\
\mbox{Institut Henri Poincar\'e, 11, rue Pierre-et-Marie-Curie, 75005 Paris, France.}
\vskip1cm}


\date{\today}
\vskip1cm

\begin{abstract}
This article details the computation of the two-point correlators of the convergence, $E$- and $B$-modes of the cosmic shear induced by the weak-lensing by large scale structure assuming that the background spacetime is spatially homogeneous and anisotropic. After detailing the perturbation equations and the general theory of weak-lensing in an anisotropic universe, it develops a weak shear approximation scheme in which one can compute analytically the evolution of the Jacobi matrix. It allows one to compute the angular power spectrum of the $E$- and $B$-modes. In the linear regime, the existence of $B$-modes is a direct tracer of a late time anisotropy and their angular power spectrum scales as the square of the shear. It is then demonstrated that there must also exist off-diagonal correlations between the $E$-modes, $B$-modes and convergence that are linear in the geometrical shear and allow one to reconstruct the eigendirections of expansion. These spectra can be measured in future large scale surveys, such as Euclid and SKA, and offer a new tool to test the isotropy of the expansion of the universe at low redshift.
\end{abstract}

\maketitle
\tableofcontents
\pagebreak

\section{Introduction}\label{sec1}

\subsection{Motivations}\label{subsec1.1}

The standard model of cosmology describes our universe with a very simple  solution of general relativity describing a spatially homogeneous and isotropic spacetime, known as the Friedmann-Lema\^{i}tre solution. It is assumed to describe the geometry of our Universe smoothed on large scales. Besides, the use of the perturbation theory allows one to understand the properties of the large scale structure, as well as its growth from initial conditions set by inflation and constrained by the observations of the cosmic microwave background (CMB). It is a very successful model and allows one to deal with all existing observations in a consistent way with only 6 free parameters~\cite{planck16} from primordial nucleosynthesis (BBN) to today, involving mostly general relativity, electromagnetism and nuclear physics, that is physics below 100 MeV and well under control experimentally (see, e.g., Refs.~\cite{Uzan-Peter-anglais,slavabook,georgebook} for standard textbooks).

The construction of the cosmological model depends on our knowledge of micro-physics but also on a priori hypothesis on the geometry of the spacetime describing our universe. It relies on 4 main hypothesis (see Ref.~\cite{jpu2010} for a detailed description): (H1) a theory of gravity, (H2) a description of the matter and the non-gravitational interactions, (H3) symmetry hypothesis, and (H4) an hypothesis on the global structure, i.e. the topology, of the universe. The hypothesis H1 and H2, that refer to the physical theories, are not sufficient to solve the field equations and we need an assumption on the symmetries (H3) of the solutions describing our universe on large scales.

Among the generic conclusions of this standard model is the need  of a dark sector, including dark matter and dark energy, which emphasizes the need for extra degrees of freedom, either physical (new fundamental fields or interactions) or geometrical (e.g., a cosmological solution with lower symmetry). This has driven a lot of activity to test the hypotheses of the cosmological model. In that debate, weak-lensing is a key observation to test general relativity on cosmological scales~\cite{ub01} and to constrain the scale on which the fluid limit holds~\cite{dupfleu}. It complements tests of the other hypothesis such as the equivalence principle~\cite{testPE} and the Copernican principle~\cite{testCP}. Our {\em first motivation} is thus to provide a new test on the isotropy of the expansion at late time, hence providing a new test of the standard cosmological assumption. Any detection of a violation of a symmetry of the background spacetime would have important implications in terms of model building and on  the understanding of the dark sector.

While in the standard $\Lambda$CDM model the cosmological constant $\Lambda$ is the source of the acceleration of the universe, many models have been proposed to explain the acceleration of the cosmic expansion. The property of the dark sector is often modelled as a fluid with an equation of state, $P_{\rm de}=w\rho_{\rm de}$, relating its pressure to its energy density. Such a phenomenological parameterization allows to characterize the ability of different surveys to actually demonstrate that $w=-1$, as expected for a cosmological constant. Among the plethora of dark energy models, many enjoy an anisotropic pressure $\Pi_j^i$ and thus may trigger a phase of anisotropic expansion at late time when dark energy starts influencing the dynamics of the universe. This is for instance the case of magnetised dark energy~\cite{sharif10,Barrow:1997as}, solid dark matter~\cite{solidDM,solidDM2} induced by a network of frustrated topological defects, bi-gravity models~\cite{bigrav}, anisotropic dark energy~\cite{mota08,Urban:2009ke} and in models in which the backreaction~\cite{backreac} of the large scale structure on the background evolution is the source of the acceleration. This has led to the development of a phenomenological parameterisation of the anisotropic pressure in terms of an anisotropic equation of state as $\Pi_j^i=\Delta w^i_j\rho_{\rm de}$~\cite{Mota:2007sz,Koivisto:2007bp,linder,Pradhan:2012zt}. Our {\em second motivation} is thus to propose new observational tests on the anisotropic pressure of the dark energy sector, hence constraining another phenomenological deviation from a pure cosmological constant.\\

When concerned by anisotropic expansion, we can distinguish between 2
classes of models, that allows one to divide the different methods to
constrain anisotropy. Remind that any perturbed quantity, $X$ say,
such as the gravitational potential, the density contrast, etc. can
be split, in Fourier space, as the product of an initial
configuration and a transfer function as $X(t,{\bm k})= T_X(t,{\bm
  k}) X_i({\bm k})$. First, early anisotropic models (such as anisotropic inflation) have anisotropic initial conditions (in the sense that the correlation functions of the initial perturbed quantities is such that $\langle X_i({\bm k})X_i^*({\bm k}') \rangle\not=P_X(k)\delta({\bm k}-{\bm k}')$) while the transfer functions are independent of direction (i.e. $T_X(t,{\bm k})=T_X(t,k)$) because the geometry has isotropized at later times. Second, late time anisotropic models have been isotropic during most of the history of the universe (hence enjoying isotropic correlation functions, e.g. $\langle X_i({\bm k})X_i^*({\bm k}') \rangle=P_X(k)\delta({\bm k}-{\bm k}')$) while their transfer function at late time is anisotropic, i.e. $T_X(t,{\bm k})\not=T_X(t,k)$. These two types of models have a huge difference in the way one attacks observational constraints. In particular the propagation of light is only affected in the second class of models.

Without any source during inflation, any primordial anisotropy is washed out~\cite{ppu1,Anninos:1991ma} by the expansion. It was however demonstrated that it affects the construction of the Bunch-Davies state~\cite{ppu1} so that it lets very specific signatures on the primordial power spectrum~\cite{ppu2,Gumrukcuoglu:2007bx,dulaney10} and affects the onset of inflation~\cite{Kofman:2011tr}. Such deviation from isotropy can be constrained by CMB observations~\cite{Kogut:1997az,cmbaniso,Pullen:2007tu,Pullen:2010zy,Gumrukcuoglu:2010yc,Dulaney:2010sq,Chen:2014eua,ArmendarizPicon:2008yr}. An early, post-inflationary, anisotropy also affects the synthesis of light elements during primordial nucleosynthesis~\cite{BBN} (mostly because it affects the expansion rate).

Tests of a late time anisotropy have mostly focused on the Hubble
diagram from type~Ia
supernovae~\cite{sn1a,sn1b,sn1c,Ronggen11,linder,abm,Jimenez:2014jma,Schucker:2014wca,Kolatt:2000yg,colin2014,sn37}. An
anisotropic expansion will influence the transfer function so that it
can also be constrained by the study of the large scale
structure~\cite{dulaney08,dima08,Koivisto:2008ig,Cai:2013lja,Schwarz:2007wf,Antoniou:2010gw,Longo:2014wva,lss2,lss3}
and of the
CMB~\cite{battye09,Axelsson:2011gt,McEwen:2013cka,Sung:2010ek}. It was
argued that supernovae data leads to $\Delta w<2.1\times10^{-4}$
\cite{abm}, and that next generation galaxy surveys are capable of constraining anisotropies at the 5\% level~\cite{linder} in terms of the anisotropic equation of state.

In this article, we follow our former analysis~\cite{pup} on the imprint of a late time anisotropy on weak-lensing. According to the standard lore~\cite{slore}, in a homogeneous and isotropic background spacetime, weak-lensing by the large scale structure of the universe induces a shear field which, to leading order, only contains $E$-modes. It was demonstrated in Ref.~\cite{pup} that, even in the linear regime, anisotropic expansion will reflect itself in the existence of non-vanishing $B$-modes. The level of $B$-modes is used as an important sanity check during the data processing. On small scales, $B$-modes arise from non-linear effects~\cite{Bonvin2010} intrinsic alignments~\cite{intrinsic}, Born correction, lens-lens coupling~\cite{lenslens}, and gravitational lensing due to the redshift clustering of source galaxies~\cite{galcluster}. On large angular scales in which the linear regime holds, it was demonstrated~\cite{pup} that non-vanishing $B$-modes would be a signature of a deviation from the isotropy of the expansion, these modes being generated by the coupling of the background Weyl tensor to the $E$-modes. \\

While light propagation in strictly homogeneous Bianchi universes has
been widely investigated~\cite{oldSN,caceres09}, the analytic
computation of the Jacobi matrix was only determined
recently~\cite{Fleury2014} (see also
Ref.~\cite{Fanizza:2013doa}). This article focuses on the computation
of the Jacobi matrix taking into account cosmological perturbations at
linear order in a spatially homogeneous anisotropic Euclidean
spacetime of the Bianchi~$I$ family. We provide all the technical
tools (perturbation theory, light propagation, expression of the
observables). The application of our formalism is exposed in our
companion paper~\cite{ppulett15} in which we compute the expected
signals for the Euclid~\cite{euclid} and SKA~\cite{ska} observations.

Among our main results, we emphasize that, as soon as local isotropy does not hold at the background level, there exist a series of weak-lensing observables that allow one to fully reconstruct the background shear and thus test spatial isotropy.  More precisely, as a consequence of the non-vanishing of the $B$-modes, it can be demonstrated that
\begin{enumerate}
\item the angular correlation function of the $B$-modes, $C_\ell^{BB}$, is non vanishing~\cite{pup}, and scales as the square of the ratio of the geometric shear to the Hubble expansion rate, $\sigma^2/{\cal H}^2$;
\item the $B$-modes correlate with both the $E$-modes and the convergence $\kappa$ leading to the off-diagonal cross-correlations $\langle B_{\ell m}E^{\star}_{\ell\pm1 \,m-M}\rangle$ and  $\langle B_{\ell m}\kappa^{\star}_{\ell\pm1 \,m-M}\rangle$ in which $E_{\ell m}$ and $B_{\ell m}$ are the components of the decomposition of the $E$- and $B$-modes of the cosmic shear in (spin-2) spherical harmonics and $\kappa_{\ell m}$ the components of the decomposition of the convergence in spherical harmonics. These two correlators scale as $\sigma/{\cal H}$;
\item the deviation from isotropy also generates off-diagonals correlations among $\kappa$ and $E$ modes, $\langle E_{\ell m} {E}^{\star}_{\ell\pm2 \,m-M}\rangle$,  $\langle{\kappa}_{\ell m} {\kappa}^{\star}_{\ell\pm2 \,m-M}\rangle$, and $\langle{E}_{\ell m} {\kappa}^{\star}_{\ell\pm2 \,m-M}\rangle$. These three correlators scale as $\sigma/{\cal H}$;
\item for each type of correlator, there are five values of $M$ so that in principle they can be used to reconstruct the five components of the geometric shear $\sigma_{ij}$.
\end{enumerate}
This last point is very important since it exhibits a rigidity between independent observables that can be used to control systematic effects.

\subsection{Structure of the article}\label{subsec1.2}

Section~\ref{sec2} summarizes the description of the spacetime at the background level (\S~\ref{subsec2.1}) and for linear perturbations (\S~\ref{subsec2.2}). For the sake of clarity, the theory of gauge invariant perturbations is detailed in Appendix~\ref{secA}. It also introduces the parameterisation of an anisotropic dark energy sector. The main variables required to describe the evolution of the background spacetime are summarized in Table~\ref{table1}.

\begin{table}\label{table1}
\begin{longtable}{|c|lc|}
\hline
\hline 
Symbol & Meaning & Appears at Eq.:\tabularnewline
\hline 
\hline
\endhead
${\mu,\nu,\dots}$ & Formal space-time indices &  -- \tabularnewline
$i,j,\dots$ & Cartesian spatial indices &  -- \tabularnewline
$\Ti,\Tj,\dots$  & Spatial tetrad indices  & (\ref{spatial_triad}) \tabularnewline
$\tz$  & Time tetrad index & (\ref{Time_tetrad}) \tabularnewline
$a$ & Average scale factor &(\ref{e:metric2})\tabularnewline
$H$ & Cosmic time Hubble expansion rate &(\ref{g00})\tabularnewline
$\beta_i$ &  Log of directional scale factors & (\ref{e:def_gammaij})\tabularnewline
$\hat\sigma_{ij}$ & Geometrical (cosmic time) shear &(\ref{e:decbeta})\tabularnewline
$\sigma_{ij}$ & Geometrical (conformal time) shear & (\ref{ConfTimeShear})\tabularnewline
$\thetrad_{\tz}^{\,\,\nu}$ & Timelike vector of background tetrad & (\ref{e.213}) \tabularnewline
$\thetrad_{\ti}^{\,\,\nu}$ & Spacelike vector of background tetrad & (\ref{spatial_triad})  \tabularnewline
$\Thetrad_{\tz}^{\,\,\nu}$ & Timelike vector of perturbed tetrad & (\ref{DefTetrad}) \tabularnewline
$\Thetrad_{\ti}^{\,\,\nu}$ & Spacelike vector of perturbed tetrad &  (\ref{DefTetrad})\tabularnewline 
$\Delta w_i^j$& Equation of state of dark energy anisotropic stress &(\ref{DefEOSPi}) \tabularnewline
$\beta_{ij}$& Homogeneous perturbation of the Euclidean metric. & (\ref{DefMatrixBeta}) \tabularnewline
\hline
\caption{Table of most used quantities describing the background spacetime.}
\end{longtable}
\end{table}

Section~\ref{sec3bix} describes the propagation of a light bundle (\S~\ref{subsec3.1}) and presents in \S~\ref{subsec3.2}  the central equation for our analysis, namely the Sachs equation
$$
   \frac{\dd^2}{\dd v^2}{\cal D}^a_b={\cal R}^a_c{\cal D}^c_b\,,
$$
for the $2\times2$ Jacobi matrix ${\cal D}_{ab}$, the decomposition of which is presented in \S~\ref{subsec3.3}. It concludes by specifying these general results to the case of a Bianchi~$I$ spacetime (\S~\ref{subsec3.4}), focusing on the technical but useful use of a conformal transformation. The main variables required to describe the evolution of a geodesic bundle are summarized in Table~\ref{table2}.

\begin{table}\label{table2}
\begin{longtable}{|c|lc|}
\hline
\hline 
Symbol & Meaning & Appears at Eq.:\tabularnewline
\hline 
\hline
\endhead
$k^\mu$ & Null geodesic tangent vector & (\ref{e.geo3}) \tabularnewline
$\tilde k^\mu$ & Conformally null geodesic tangent vector & (\ref{e.defk}) \tabularnewline
$z$ & Redshift & (\ref{def_z}) \tabularnewline
${\bm n}_\obs$ &  Initial observed direction &  (\ref{Defn0})  \tabularnewline
${\bm n}_a$ &  Sachs basis &  (\ref{basis2d})  \tabularnewline
${\bm n}_\pm$ & Helicity basis  &  (\ref{e:helicity}) \tabularnewline
$\eta_a$ &  Components of the connecting vector in the Sachs basis& (\ref{e.gde})  \tabularnewline
${\cal R}_{ab}$ &  Optical tidal matrix &  (\ref{e.gde}) \tabularnewline
${\cal D}_{ab}$ &  Jacobi matrix & (\ref{e:defDab})  \tabularnewline
$\tilde{\cal D}_{ab}$ &  Conformal Jacobi matrix & (\ref{e.defDab})  \tabularnewline
$\bar D_A$& Background angular diameter distance & (\ref{DecompositionDab})  \tabularnewline
$D_A$ &  Angular diameter distance &  (\ref{eq:decomposition_Jacobi}) \tabularnewline
$\kappa$ & Convergence & (\ref{DecompositionDab})  \tabularnewline
$\gamma_{ab}$ & Cosmic shear &  (\ref{DecompositionDab}) \tabularnewline
$V$ &  Rotation & (\ref{DecompositionDab})  \tabularnewline
$\kappa_{\ell m}$ & Multipolar coefficients of the convergence &  (\ref{e:322a}) \tabularnewline
$V_{\ell m}$ &Multipolar coefficients of the rotation  &  (\ref{e:322b}) \tabularnewline
$E_{\ell m}$ & Multipolar coefficients of the cosmic shear $E$-modes & (\ref{dec_gpm})  \tabularnewline
$B_{\ell m}$ & Multipolar coefficients of the cosmic shear $B$-modes &  (\ref{dec_gpm}) \tabularnewline
\hline
\caption{Table of most used quantities describing the propagation of a geodesic bundle.}
\end{longtable}
\end{table}

\begin{figure}[ht!]
\centering
\includegraphics[width=16cm]{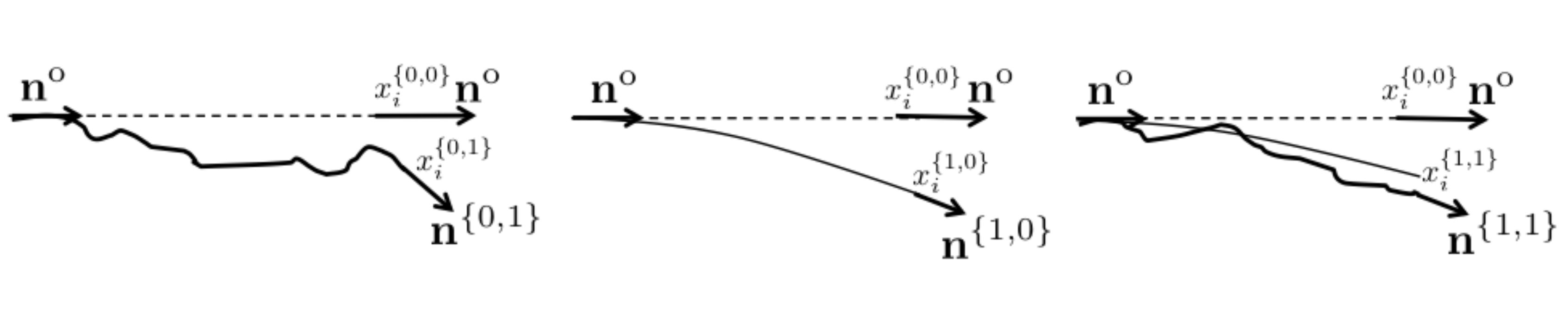}
\centering
\caption{Comparison of the geodesic in the approximation at order $\{n,p\}$. In order to adopt an observer based point of view, we need to relate the local direction of propagation ${\bm n}^{\{n,p\}}$ to ${\bm n}^\obs$. The transports for the 3 orders of perturbations are respectively detailed in Eqs.~(\ref{defxi10}) and (\ref{e.x01}) that determine $x^{i{\{n,p\}}}(\chi,{\bm n}^\obs)$ that can be further split in a radial component $\delta r^{{\{n,p\}}}(\chi,{\bm n}^\obs)$ and an orthoradial contribution that defines the deflection angle $\alpha^{a{\{n,p\}}}(\chi,{\bm n}^\obs)$.}
\label{Fig1}
\end{figure}

\begin{figure}[ht!]
\centering
\includegraphics[width=10cm]{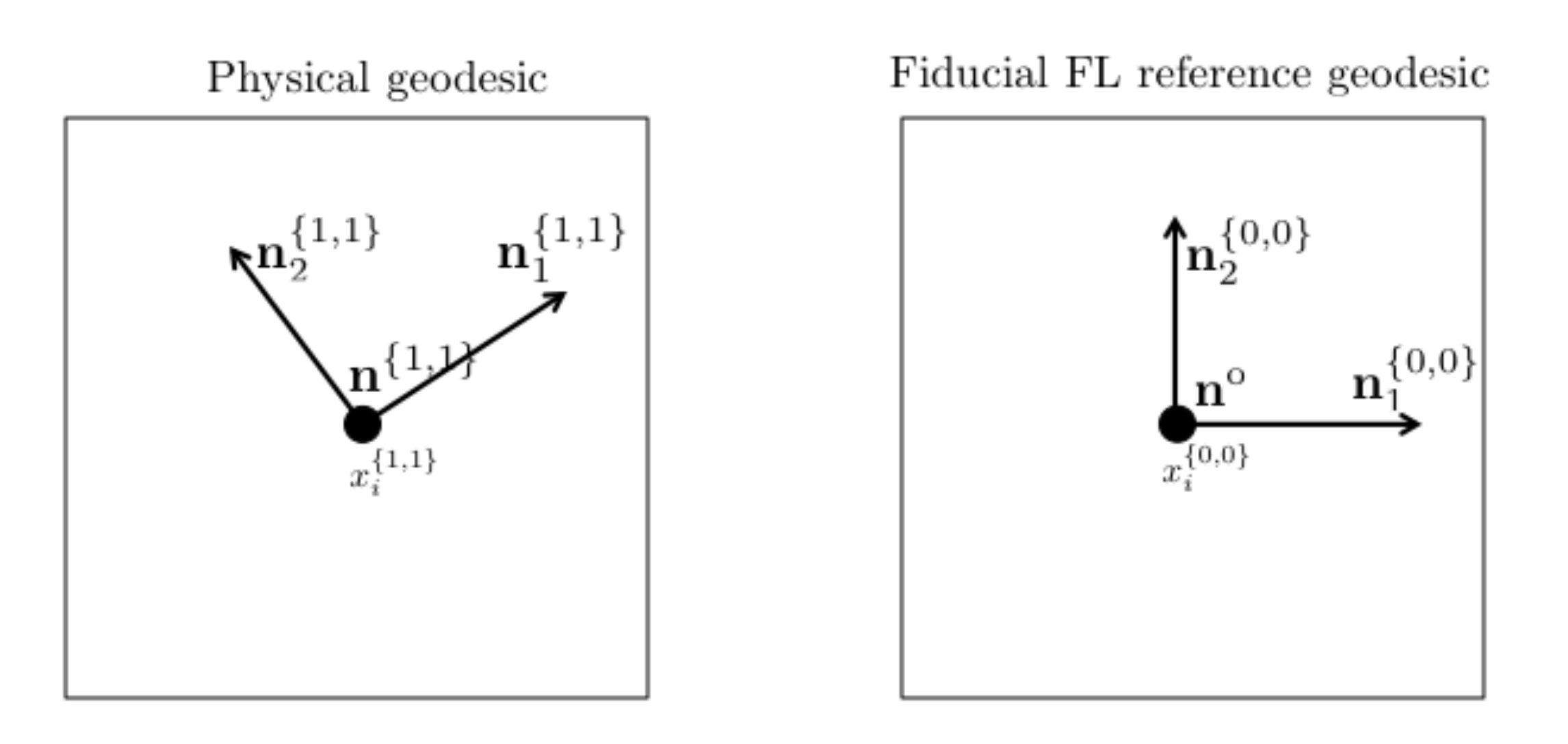}
\centering
\caption{The local Sachs basis at a point $P_{\{n,p\}}$ on the geodesic has to be transported to the point $P_\zz$ on the fiducial null geodesic of the background Friedmann-Lema\^{\i}tre spacetime. This implies to perform a transport on the tangent space at the same time that the point of observation is transported in real space. The transports for the 3 orders of perturbations are respectively detailed in Eqs.~(\ref{def_varpi00}), (\ref{varn01}) and (\ref{Projrules}) for ${\bm n}$ and in Eqs.~(\ref{def_varpi}), (\ref{varnsachs01}) and (\ref{Projrules}) for the Sachs basis.}
\label{Fig2}
\end{figure}

Since the geometric shear is obviously small, we develop in Section~\ref{sec5} an approximation scheme referred to as {\em small shear limit} in which ${\sigma/\HH}$ is considered as a small parameter. We then use a  two-parameter  expansion scheme in which both  ${\sigma/\HH}$ and the perturbations of the metric, say $\Phi$, are small. Thus, a given order $\{n,p\}$ corresponds to term of order $({\sigma/\HH})^n\Phi^p$.  In this approximation, the structure of our computation is the following. We start from the fact that the Sachs equation can be rewritten as (see Eq.~(\ref{MasterBonvin}))
$$
\frac{\dd^{2}{\cal D}_{ab}}{\dd\chi^{2}}+\frac{1}{k^{0}}\frac{\dd k^{0}}{\dd\chi}\frac{\dd{\cal D}_{ab}}{\dd\chi}=\frac{1}{\left(k^{0}\right)^{2}}{\cal R}_{ac}{\cal D}_{cb}\,,
$$
where $\chi$ is the coordinate along the lightcone in the background Friedmann-Lema\^{\i}tre spacetime (see \S~\ref{seczzzz}). At order $\zz$, ${\cal R}_{ab}^\zz=0$ and $k^{0\zz}=-1$ so that the equation takes the form
$$
\frac{\dd^{2}{\cal D}^\zz_{ab}}{\dd\chi^{2}}=0\,
$$
and can be integrated trivially (see \S~\ref{subsec6.1}). We then expand this equation order by order so that it formally takes the form (since $k^0_\zz=-1$)
$$
\frac{\dd^{2}{\cal D}^{\{n,p\}}_{ab}}{\dd\chi^{2}}= S^{\{n,p\}}\,
$$
in which the source term contains contribution from ${\cal R}_{ab}$ and $k^0$ up to order ${\{n,p\}}$, and from ${\cal D}_{ab}$ at lower order. The effects to be taken into account are then
\begin{enumerate}
\item the tensor and vector contributions to ${\cal R}_{ab}$, which starts at order $\oo$ and the contribution of the scalar modes at the relevant order;
\item the evolution of all the perturbative modes, that is of the transfer functions, which is decomposed as
$$
 T_X({\bm k},t) = T^\zo_X(k,t)  + T^\oo_X({\bm k},t) 
$$
since the order $\zz$ and $\oz$ correspond to homogeneous solutions. This requires to solve the equations of Appendix~\ref{secA}.
\item In order to determine $k^0$, we also need to solve perturbatively the geodesic equation.
\item A source observed in direction ${\bm n}^\obs$ at distance
  $\chi$, is located at a spacetime point $P_{\{n,p\}}$ and its contribution depends on the local direction of the tangent vector to the geodesic in ${\bm n}_{\{n,p\}}$, which determines the local Sachs basis in $P_{\{n,p\}}$. We shall thus proceed with 2 operations:
\begin{enumerate}
\item transport $P_{\{n,p\}}$ to $P_\zz$ (see Fig.~\ref{Fig1}),
\item transport ${\bm n}_{\{n,p\}}$ and the local Sachs basis (see Fig.~\ref{Fig2}).
\end{enumerate}
This is what we call the {\em central geodesic approximation} and the
possibility to go beyond this approximation is sketched in
Appendix~\ref{SecFullLensing}. We however stick to this approximation,
which is sufficient in the small shear approximation. At lowest order,
it corresponds to the usual Born approximation but at higher order
there are post-Born corrections to include.
\end{enumerate}

Section~\ref{SecAngularCorrelations} describes the computation of the angular correlation. Our philosophy is to adopt an observer point of view, that is, to compute all quantities on the celestial sphere of the observer. Given the previous perturbative expansion scheme, any observable $X^s$ of spin $s$ can then be formally expressed as (see Eq.~(\ref{e:GeneralLOS}))
$$
 X^s(\chi_{_{\rm S}}, \bm{n}_\obs) \bm{m}^{s}_{\obs} =
\int_0^{\chi_{_{\rm S}}} S^{X_s}(\chi_{_{\rm S}}, \chi,\bm{n}_\obs) \bm{m}^{s}_\obs \dd\chi\,,
$$
for a source $S^{X^s}$ located  at $\chi_{_{\rm S}}$ and observed in direction ${\bm n}^\obs$. According to the spin $s$ of the quantity we can expand in the proper spherical harmonics with respect to ${\bm n}^\obs$. This allows one to define the expansion of all the quantities in term of spherical harmonics. As a byproduct, we demonstrate in \S~\ref{sec4.3} that  the 5 off-diagonal correlators $\langle B_{\ell m}E^{\star}_{\ell\pm1 \,m-M}\rangle$ ,  $\langle B_{\ell m}\kappa^{\star}_{\ell\pm1 \,m-M}\rangle$ , $\langle E_{\ell m} {E}^{\star}_{\ell\pm2 \,m-M}\rangle$,  $\langle{\kappa}_{\ell m} {\kappa}^{\star}_{\ell\pm2 \,m-M}\rangle$, and $\langle{E}_{\ell m} {\kappa}^{\star}_{\ell\pm2 \,m-M}\rangle$ are non-vanishing.\\

Equiped with all these tools, we compute these correlators and the angular power spectra of the $E$- and $B$-modes in Section~\ref{sec6} order by order. Order~$\oz$ recovers the non-perturbative analysis of Ref.~\cite{Fleury2014} while order $\zo$ recovers the standard case of lensing by large scale structure in the linear regime; \S~\ref{subsec6.3} gives all the details of the computation at order $\oo$. This allows us to discuss the dominant contribution in Section~\ref{sec7}. In particular, we argue that the dominant term for the cosmic shear is given by
$$
 \gamma_{ab}=-\alpha^c D_cD_{\langle a}D_{b\rangle}\varphi
$$
where $\alpha^c$ is the deflection angle at order $\oz$, $\varphi$ the deflection potential, and $D_a$ the covariant derivative on the celestial sphere.\\

Many technicalities are gathered in the appendices: linear perturbation theory (\S~\ref{secA}), the expressions of the geometric quantities at first order in perturbation that are need to compute the source term of the Sachs equation (\S~\ref{secB}), details on the lensing method (\S~\ref{SecFullLensing}) and a catalog of useful mathematical identities (\S~\ref{secD}). Throughout this work we adopt units in which $c=1$.

\begin{table}\label{table3}
\begin{longtable}{|c|lc|}
\hline
\hline 
Symbol & Meaning & Appears at Eq.:\tabularnewline
\hline 
\hline
\endhead
$\Sigma$ & Scalar shear on the sphere  & (\ref{e.defSigma})\tabularnewline
 $\Sigma_a$&  Vector shear on the sphere& (\ref{e.defSigmaab})\tabularnewline
 $\Sigma_{ab}$& Tensor shear on the sphere  & (\ref{e.defSigmaab})\tabularnewline
 $D_a$ &  Covariant derivative on the sphere& (\ref{def_DaS2}) \tabularnewline
 $\edth, \edthb$&  spin-raising and lowering operators&  (\ref{def_spinder})\tabularnewline
 $\Sigma_{2m}$& Multipole of the scalar shear &  (\ref{def.Sig2m})\tabularnewline
$\Sigma_\pm$  & First derivative of the scalar shear in the helicity basis  &  (\ref{def_Sigpm})\tabularnewline
$\bm{m}_\obs^{s}$   &spin-$s$ polarization basis   &  (\ref{def_polar})\tabularnewline
  $S^{X_s}$  & Source of the field $X_s$  &  (\ref{e:GeneralLOS})\tabularnewline
 $T^{X_s}_{\ell m}$    & Anisotropic transfer function of the field $X^s$ &  (\ref{source_transfer})\tabularnewline
  $ {}^{X_s}\!T^{LM}_{\ell m}$    & Multipoles of the anisotropic transfer function &  (\ref{e.def_ccc})\tabularnewline
$E^{X_s}_{\ell m }$, $B^{X_s}_{\ell m }$      &  Multipoles of the $E$- and $B$-modes of the field $X^s$ &  (\ref{dec_EB})\tabularnewline
 $C_\ell^{EE}$     &  Angular power spectrum of the $E$-modes &  (\ref{def_CEE})\tabularnewline
   $C_\ell^{BB}$    & Angular power spectrum of the $E$-modes  &  (\ref{def_CBB})\tabularnewline
   $ \MyB$    & Scalar perturbation of the spatial metric &  (\ref{DefSigmaB})\tabularnewline
$\alpha^{a}$    & Deflection angle  &  (\ref{xi10})\tabularnewline
 $ \myaleph^a $         & Perturbation of propagating direction &  (\ref{def_varpi})\tabularnewline
 $\varphi$                   & Delflection potential &  (\ref{defphipot})\tabularnewline
  $\snumb(\chi)$   &  Source distribution&  (\ref{e.ceezo})\tabularnewline
 $P(k)$ & Primordial power spectrum   &  (\ref{iso_p_of_k})\tabularnewline
\hline
\caption{Table of most used quantities describing the propagation of a geodesic bundle.}
\end{longtable}
\end{table}

\section{Spacetime structures}\label{sec2}

\subsection{Background spacetime}\label{subsec2.1}

\subsubsection{Geometry}

At the background level, the universe is described by a spatially Euclidean, homogeneous, and locally anisotropic solution of the Einstein equation filled with a perfect fluid. Its metric takes the general form (see Refs.~\cite{Ellis:2006ba,Krasinski:2003zzb,Ellis:1998ct,Ryan:1975jw,Pontzen2007} for general references on Bianchi spacetimes)
\bea
    \dd s^2&=&-\dd t^2+\sum_{i=1}^3X_i^2(t)(\dd x^i)^2  \,,    \label{e:metric1}\\
                 &=&-\dd t^2+a^2(t)\gamma_{ij}(t)\dd x^i\dd x^j \,,   \label{e:metric2}
\eea
where $a(t)\equiv\sqrt{X_1(t)X_2(t)X_3(t)}$ is the volume averaged scale factor and $t$ the cosmic time. We define the tangent vector to the fundamental comoving observer by $u_\mu\dd x^\mu\equiv\dd t$. It is normalized such that $u_\mu u^\mu=-1$. The spatial metric $\gamma_{ij}$ and its inverse $\gamma^{ij}$ can be decomposed as
\begin{equation}\label{e:def_gammaij}
      \gamma_{ij}(t)=\exp[2\beta_{i}(t)]\delta_{ij}\,,\qquad \gamma^{ij}(t)=\exp[-2\beta_{i}(t)]\delta^{ij},
\end{equation}
with the constraint
\begin{equation}\label{e:sum_beta}
   \sum_{i=1}^3\beta_i=0\,
\end{equation}
that ensures that the comoving volume remains constant (i.e, $\dot\gamma=\gamma^{ij}\dot\gamma_{ij}=0$). Note that, as a consequence of Eq.~(\ref{e:def_gammaij}), some spatial directions should contract while others grow~\cite{ppu2}.  Note also that that there is no sum on $i$ in the definition of $\gamma_{ij}$ and Latin indices $\{i,j,k\dots\}$ are raised with $\gamma^{ij}$ and lowered with $\gamma_{ij}$. 

The geometrical shear is defined as 
\begin{equation}\label{e:decbeta}
     \hat\sigma_{ij}\equiv\frac{1}{2}\dot\gamma_{ij}\,
\end{equation}
where a dot refers to a derivative with respect to cosmic time. We shall also use the conformal time $\eta$ defined by $a(\eta) \dd \eta = \dd t$, and denote derivatives with respect to it by a prime. Thus, the conformal shear is  defined as 
\be\label{ConfTimeShear}
      \sigma_{ij} \equiv \frac{1}{2} \gamma'_{ij}= a \hat\sigma_{ij}\,.
\ee
The amplitude of the shear is defined by
\begin{equation}\label{e:def_sigma2}
 \hat \sigma^2 \equiv \hat\sigma_{ij}\hat\sigma^{ij} =
 \sum_{i=1}^3\dot\beta_i^2\qquad {\rm and} \qquad 
 \sigma^2 \equiv \sigma_{ij}\sigma^{ij} =
 a^2 \sum_{i=1}^3\dot\beta_i^2 = \sum_{i=1}^3{\beta_i'}^2.
\end{equation}

\subsubsection{Decomposition of the geometric shear}

The shear, being a symmetric and traceless spatial tensor ($\sigma^i_i=0$), has 5 degrees of freedom, three of which correspond to the Euler angles necessary to express the shear in a general basis. By choosing the Cartesian basis~(\ref{e:def_gammaij}), we have set these three angles to zero so that we are left with only 2 degrees of freedom, namely the three $\beta_i$ with the constraint~(\ref{e:sum_beta}). The components of the shear can thus be expressed as
\begin{equation}\label{e:expr_sigma_ij}
   \hat\sigma_{ij}(t)=\dot \beta_i \exp[2\beta_{i}(t)]\delta_{ij}\,,\qquad
   \hat\sigma^{ij}(t)=\dot \beta_i \exp[-2\beta_{i}(t)]\delta^{ij}\,,\qquad
   \hat\sigma^{i}_{\,j}(t)=\dot \beta_i \delta^{i}_j\,.
\end{equation}
These two independent degrees of freedom can also be decomposed as a magnitude and an angle $\angleBianchiModel$. The first is related to the scalar shear while the choice of the angle defines which of the spatial directions are initially expanding. These variables can be obtained by performing the decomposition
\begin{equation}\label{e:dec_beta}
   \beta_i(t) \equiv \ppuB_i W(t)\,,
\end{equation}
with the 3 constants $\ppuB_i$ given by
\begin{equation}\label{e:def_Bi}
 \ppuB_i=\sqrt{\frac{2}{3}}\mathcal{S}\sin\left(\angleBianchiModel+\frac{2\pi}{3}i\right),\qquad i\in\{1,2,3\}\,
\end{equation}
and where $\mathcal{S}$ is constant. This parameterization obviously satisfies the required constraints,
\begin{equation}
   \sum_i \ppuB_i=0\,,\quad\sum_i \ppuB_i^2=\mathcal{S}^2\,.
\end{equation}
Therefore, once the Cartesian basis is chosen, we can choose the two constants $(\angleBianchiModel,\mathcal{S})$ to describe the  two degrees of freedom of the shear since  $\hat\sigma^2=(\mathcal{S} \dot W)^2$.

\subsubsection{Spatial triad}

It is convenient to introduce a spatial triad -- a set of three orthonormal vectors and co-vectors; the normalization being defined from $\gamma_{ij}$ and $\gamma^{ij}$ -- related directly to the Cartesian coordinates $x^i$. Their components in the coordinates basis read
\be\label{spatial_triad}
   \thetradDU{\Ti}{\,\,j} = \exp[-\beta_{i}(t)] \delta_i^j \,,\qquad 
    \thetradUD{\Ti}{\,\,j} =  \exp[\beta_{i}(t)] \delta^i_j\,.
\ee
In such a triad basis, the shear components take the simple form
\be\label{e.213}
  \hat\sigma_{\Ti\Tj} =\hat\sigma^{\Ti\Tj}= \dot \beta_i \delta_{ij} \qquad  \sigma_{\Ti\Tj} = \sigma^{\Ti\Tj} = \beta'_i \delta_{ij}.
\ee
Thus, this triad can easily be extended to a tetrad by using the observer's 4-velocity as the normalized time-like vector
\be\label{Time_tetrad}
  \thetradDU{\tz}{\,\,\mu} = \delta_0^\mu = u^\mu, \qquad \thetradUD{\tz}{\,\,\mu} = \delta^0_\mu=-u_\mu\,.
\ee

\subsubsection{Description of matter and field equations}

Concerning the matter sector, we assume it is composed of a pressureless matter fluid and a dark energy component. The dark sector is then described by a fluid whose energy-momentum tensor enjoys a non-vanishing anisotropic stress,
\begin{equation}\label{e:Tmunu_de}
    T^{\mu}_{\,\,\nu}=(\rho+P)u^\mu u_\nu+P \delta^\mu_\nu+\Pi^\mu_{\,\,\nu}\,.
\end{equation}
The anisotropic stress tensor is symmetric ($\Pi_{\mu\nu}=\Pi_{\nu\mu}$), traceless ($\Pi^\mu_\mu=0$) and transverse ($u_\mu\Pi^\mu_\nu=0$) which means that  it has only 5 degress of freedom encoded in its spatial part $\Pi_{ij}$. Unless we define a microscopic model, we need to use an equation of state for $\Pi^i_j$. We decompose it as
\begin{equation}\label{e:dec_Pi}
    \Pi^i_j\equiv\rho_{\rm de}\Delta w^i_j\,,
\end{equation} 
so that the pressure tensor takes the general form
\begin{equation}\label{DefEOSPi}
   P_i^j = \rho_{\rm de}\left(w\delta_i^j + \Delta w_i^j\right)\,,
\end{equation}
where $w$ is the equation of state relating the isotropic pressure to the energy density and $\Delta w_i^j$ is an equation  of state for the anisotropic pressure. From a phenomenological point of view, this corresponds to an extension of the dark energy sector, similar to the ansatz~(1) of Ref.~\cite{abm}, which will allow us to address the question of the deviation from the standard cosmological constant reference ({\em i.e.} $w=-1$ and $\Delta w_i^j=0$).

Defining the Hubble expansion rate by
\begin{equation}\label{e:def_H}
   H=\dot a/a,
\end{equation}   
the background equations~\cite{ppu1} take the form
\bea\label{gall}
    3H^2&=&\kappa(\rho_{\rm m}+\rho_{\rm de})+\frac{1}{2}\hat\sigma^2\,,\slabel{g00} \\
    (\hat\sigma^i_j)\dot{} &=& -3H\hat\sigma^i_j + \kappa\Pi^i_j\,. \slabel{gTT}  \\
    \dot\rho_{\rm m} &=& -3H\rho_{\rm m}\,, \slabel{e:dT_mat} \\
     \dot\rho_{\rm de}&=&-3H(1+w)\rho_{\rm de}-\hat\sigma_{ij}\Pi^{ij} \slabel{e:dT_de} \,.
\eea
The first equation is the analogous of the Friedmann equation in presence of a spatial shear, the second is obtained from the traceless and transverse part of the Einstein equation and dictates the evolution of the shear. The last two equations are the continuity equations for the dark matter ($P=\Pi^i_j=0$) and dark energy sector. We have set\footnote{In order to easily check the homogeneity of the equations, we recall that \[[H]\sim M_P\,,\quad [\rho]\sim M_P^4\,,\quad [\kappa]\sim M_P^{-2}\,,
\quad [\sigma]\sim M_P\,,\quad [\Pi]\sim M_P^4\]} $\kappa =8\pi G \equiv M_P^{-2}$.

\subsubsection{Dynamics}

The set of equations~(\ref{gall}) can be formally integrated. As usual, the dark matter energy density scales as
\begin{equation}\label{e:rmat_sol}
 \rho_{\rm m}=\rho_{\rm m0}\left(\frac{a_0}{a}\right)^3\,.
\end{equation}
Equation~(\ref{gTT}) has a first integral given by
\begin{equation}\label{e:sigij_sol}
 \hat\sigma^i_j=\left(\frac{a_0}{a}\right)^3\left[\mathcal{C}^i_j
+\kappa\int\Pi^i_j\left(\frac{a}{a_0}\right)^2\frac{\dd (a/a_0)}{H}\right]
\end{equation}
where $\mathcal{C}^i_j$ is a constant tensor representing the decaying mode of the shear.
Note that if the term proportional to $\mathcal{C}^i_j$ is not
negligible then the shear is not proportional to the anisotropic
stress so that $\sigma_{ij}$ and $\Pi_{ij}$ cannot be diagonalized in the same basis.
Integrating Eq.~(\ref{e:dT_de}) leads to
\begin{equation}\label{e:rde_sol1}
\rho_{\rm de}=\left(\frac{a}{a_0}\right)^{-3(1+w)}\left[\rho_{\rm de0}-
\int\hat\sigma_i^j\Pi^i_j\left(\frac{a}{a_0}\right)^{2+3w}\frac{\dd(a/a_0)}{H}\right]\,,
\end{equation}
or, if one uses the decomposition~(\ref{e:dec_Pi}), as
\begin{equation}\label{e:rde_sol2}
  \rho_{\rm de}=\rho_{\rm de0} \left(\frac{a}{a_0}\right)^{-3(1+w)} \exp\left[-
   \int\hat\sigma^j_i\Delta w^i_j\frac{\dd a}{aH} \right].
\end{equation}
In the particular case where $w=-1$, this latter equation teaches us that the dark energy density does not remain constant.

\subsection{Linear perturbations}\label{subsec2.2}

Cosmological perturbation theory around a Bianchi~$I$ background spacetime, in the Bardeen formalism, was first investigated in Ref.~\cite{ppu1,ppu2}. The perturbed spacetime has a metric of the form
\be\label{eq:perturbedmetric2}
\dd s^2 = a^2[-(1+2A)\dd\eta^2+2B_i\dd x^i\dd\eta+(\gamma_{ij}+h_{ij})\dd x^i\dd x^j]
\ee
where $A$ is a free scalar function, $B_i\equiv \dbi B + \bar{B}_i$ and $h_{ij} \equiv  2C \left(\gamma_{ij}+\frac{\sigma_{ij}}{\HH}\right)+
2\dbi\dbj E + 2\partial_{(i}E_{j)}+2E_{ij}$ defined together with the usual transversality and trace-free conditions $\dbi \bar{B}^i=0=\dbi E^i, E^i_i=0=\dbi E^{ij}$.

As summarized in \S~\ref{secA.2}, one can define 2 scalar ($\Phi$ and $\Psi$), 2 vector ($\Phi_i$) and 2 tensor ($E_{ij}$) degrees of freedom, which are gauge invariant; see Eq.~(\ref{eq.A25}). Similarly, one can define gauge invariant variables for the matter sector, leading to 4 scalar variables ($\delta\hat\rho$, $\delta\hat P$, $\hat v$ and $\hat\pi^S$ respectively for the density, pressure, velocity and anisotropic stress), 4 vector variables ($\hat{\bar v}^i$ and $\hat\pi_i^V$ ) and 2 tensor variables ($\hat\pi_{ij}^T$) the expressions of which are gathered in Eqs.~(\ref{eq.A33}) and~(\ref{eq.A35}).

Appendix~\ref{secA} summarizes all the techniques and results needed to study the perturbations, including the definition of the Fourier transform (\S~\ref{sec:FourierMode}), and the construction of the gauge invariant variables (\S~\ref{secA.2}). It then derives the full set of Einstein equations (\S~\ref{secA.3}) and the conservation equations (\S~\ref{secA.4}).

Among the important features that differ from the standard perturbation theory around a Friedmann-Lema\^{\i}tre spacetime, let us mention 
\begin{itemize}
\item the fact that only the components $k_i$ of the wave-(co)vector are constant so that both $k^i$ and $k$ are time dependent - see e.g. Eq.~(\ref{kiprime});
\item the fact that the scalar-vector-tensor modes do not decouple;
\item the fact that, even at late time, the two Bardeen potentials are not equal because of the anisotropic stress.
\end{itemize}

\section{Weak-lensing in a general spacetime}\label{sec3bix}

This section provides the definitions and equations describing the propagation of a geodesic bundle (geodesic equation and Sachs equation) in a general spacetime and in the particular case of a Bianchi $I$ universe.

\subsection{Geodesic bundle}\label{subsec3.1}

Weak-lensing is concerned with the deformation of an infinitesimal bundle of light rays propagating in curved spacetimes. It is thus related to the geodesic deviation equation.

\subsubsection{Geodesic equation}

The central quantity in the geodesic equation, obtained as the eikonal limit of Maxwell's equations, is the wave-vector of an electromagnetic wave, 
$k^\mu(v)\equiv\dd{x}^\mu/\dd v$, where $v$ is an affine parameter of a given geodesic $x^{\mu}(v)$ and defined such that it is zero at the observer and increases towards the source. We shall be working in the eikonal approximation where $k^\mu$ is a null vector satisfying the geodesic equation
\begin{equation}\label{e.geo3}
  k^\nu\nabla_\nu k^\mu=0. 
\end{equation}  
If we parameterize the bundle of null geodesics by $x^\mu(v,s)$, where $s$ is a continuous parameter labeling each ray of the bundle, then each ray has a wave-vector given by $k^\mu(v,s)= \partial x^\mu/\partial v$, whereas the vector $\eta^\mu = \partial x^\mu/\partial s$ gives the infinitesimal separation between two neighbouring geodesics of the bundle.
The photon wave-vector can always be decomposed in components respectively parallel and orthogonal to $u^\mu$ as
\begin{equation}\label{e.34}
  \hat{k}^\mu\equiv U^{-1}k^\mu = -u^\mu + n^\mu\,,
\end{equation}
where $n^\mu$ are the components of the local directional vector $\bm{n}$, defined such that
\be
   u^\mu n_\mu=0\,,\quad n_\mu n^\mu =1\,.
\ee
Once the geodesic equation is solved, any comoving observer with four-velocity $u^\mu$, normalized such that $u^\mu u_\mu=-1$, defines the redshift of a source by
\begin{equation}\label{def_z}
  1+z(v,{\bm n}_\obs)\equiv \frac{(k_\mu u^\mu)_v}{(k_\mu u^\mu)_\obs}\,,
\end{equation}
where $v$ is the affine parameter that specifies the position of the source down the lightcone and 
\be\label{Defn0}
{\bm n}_\obs\equiv{\bm{n}}(v=0)
\ee
is the direction of observation. The energy of a photon at a given redshift is
\be
   U(v,{\bm n}_\obs)=U_\obs[1+z(v,{\bm n}_\obs)]\,,\qquad U_\obs=(k^\mu u_\mu)_\obs\,.
\ee

By definition, the local spacelike vector $\bm{n}$ is a function of the affine parameter $v$ and of the direction of observation observer ${\bm n}_\obs$, that is the spacelike vector pointing along the line of sight. 

\subsubsection{Geodesic deviation equation}

A (narrow) light beam is a collection of neighbouring light rays. The behaviour of any such geodesic, with respect to an arbitrary reference one, is described by the separation (or connecting) vector $\eta^\mu$. Asuming that all the rays converge at a given event $O$ (the location  of the observer), $\eta^\mu(0)=0$. The evolution of $\eta^\mu(v)$ along the beam is governed by the geodesic deviation equation
\be
   \frac{\dd^2\eta^\mu}{\dd v^2} = {R^\mu}_{\nu\alpha\beta} k^\nu k^\alpha\eta^\beta\,,
\ee
where ${R^\mu}_{\nu\alpha\beta}$ is the Riemann tensor.

\subsubsection{Sachs basis and screen space}

For any observer whose worldline intersects the light beam at an event different from $O$, the beam has a non-zero extension, since a priori $\eta^\mu \not= 0$. The observer can thus project it on a \emph{screen} to characterize its size and shape. This screen is by essence a 2-dimensional spacelike hypersurface, and chosen to be orthogonal to the local line of sight $n^\mu$.  Two such spatial vectors required to construct a basis for the tangent space, $\bm{n}_a$ with $a=\{1,2\}$, are defined by the requirement that
\be\label{basis2d}
   n_{a}^\mu n_{b\mu}=\delta_{ab}\,,\quad n_{a}^\mu u_{\mu}=n_{a}^\mu n_{\mu}=0\,.
\ee
With these definitions we can construct a tensor which projects any geometrical quantity on the two dimensional surface orthogonal to ${\bm n}$
\begin{equation}\label{e:defSmunu}
  S_{\mu\nu}\equiv g_{\mu\nu}+u_\mu u_\nu -n_\mu n_\nu\,.
\end{equation}
Then, with the help of the orthogonality relations~(\ref{basis2d}), this two-dimensional screen basis can be parallel propagated along null geodesics as~\cite{Lewis2006} 
\begin{equation}\label{e:prop_n}
  S_{\mu \sigma} k^\nu \nabla_\nu n_a^\sigma=0.
\end{equation}

A basis satisfying the condition~(\ref{basis2d}) and propagated according to Eq.~(\ref{e:prop_n}) is called a {\em Sachs basis}. It is important to note that the basis formed by the vectors ${\bm n}_a$ is defined up to an overall rotation around ${\bm{n}_\obs}$. We can fix this freedom by introducing a spherical basis at the observer (i.e., at $v=0$) by demanding that $\{\bm{n}^\obs,\bm{n}_1^\obs,\bm{n}_2^\obs\}=\{\bm{n}^\obs_r,\bm{n}^\obs_\theta,\bm{n}^\obs_\varphi\}$. With this choice, the integration of Eq.~(\ref{e:prop_n}) allows us to define a unique three-dimensional basis $\{\bm{n}_r,\bm{n}_\theta,\bm{n}_\varphi\}(\bm{n}^\obs,\hat v)$ at each point along the geodesics; see Ref.~\cite{pup}. Furthermore, it will be convenient to define a helicity basis as
\begin{equation}\label{e:helicity}
  \bm{n}_\pm\equiv \frac{1}{\sqrt{2}}\left(\bm{n}_\theta \mp \ii \bm{n}_\varphi\right)=\frac{1}{\sqrt{2}}\left(\bm{n}_1 \mp \ii \bm{n}_2\right),,
\end{equation}
whose components in the $\bm{n}_a$ basis read
\be
n_\pm^a=\bm{n}_\pm\cdot\bm{n}_a=\frac{1}{\sqrt{2}}(\delta_1^a \mp \ii \delta_2^a) 
\ee
and are, by construction, constant.

\subsection{Sachs equation}\label{subsec3.2}

The screen projection of the connecting vector, $\eta_a\equiv n_a^\mu\eta_\mu$, represents the relative position on the screen of the two light spots associated with two rays separated by $\eta^\mu$. Similarly, and if we set by convention $U_\obs=1$, $\theta_a \equiv  (\dd \eta_a/\dd v)_\obs$ represents the angular separation of those rays, as observed from $O$. 

The geodesic equation can be recast as~\cite{Uzan-Peter-anglais} an equation for $\eta_a$ as
\begin{equation}\label{e.gde}
\frac{\dd^2\eta_a}{\dd v^2} = {\cal R}_{ab}\eta^b\,,\qquad
  {\cal R}_{ab}\equiv{R}_{\mu\nu\alpha\beta}k^\nu k^\alpha n_a^\mu n_b^\beta
\end{equation} 
${\cal R}_{ab}$ is the screen projected Riemann tensor which can be split into its symmetric traceless part ${\cal R}_{\langle ab \rangle}$ and its trace part ${\cal R}\equiv{\cal R}_{ab}\delta^{ab}/2$. It is also referred to as the \emph{optical tidal matrix}. Furthermore, thanks to the linearity of Eq.~(\ref{e.gde}), one can decompose the connection vector on the geodesic to its initial derivative as
\be\label{e:defDab}
\eta^a(v) = {\cal D}^a_b(v) \left(\frac{{\dd\eta^b}}{{\dd v}}\right)_{v=0}\,.
\ee
This defines the {\it Jacobi map} ${\cal D}_{Ãab}$ that satisfies the Sachs equation~\cite{sachs,Uzan-Peter-anglais}
\begin{equation}\label{gde2}
   \frac{\dd^2}{\dd v^2}{\cal D}^a_{\,b}={\cal R}^a_{\,c}{\cal D}^c_{\,b}\,,
\end{equation}
subject to the following initial conditions:
\be\label{initialconditions}
{\cal D}^a_{\,b}(0) =0\,,\quad \frac{\dd{\cal D}^a_{\,b}}{\dd v}(0)=\delta^a_b\,.
\ee

\subsection{Decomposition of the Jacobi matrix and observables}\label{subsec3.3}

The Jacobi matrix entering the Sachs equation~(\ref{gde2}) encodes all the information about the deformation of a light beam when propagating through a curved spacetime. This $2\times2$ matrix can be decomposed in different ways.

The usual decomposition is described in terms of a convergence $\kappa$, a rotation $V$, and a shear $\gamma_{ab}$ as
\begin{equation}\label{DecompositionDab}
   {\cal D}_{ab} (v) \equiv \bar D_A(v) \left[(1+\kappa) I_{ab} + V \epsilon_{ab} + \gamma_{ab}\right] 
\end{equation}
with
\be
\epsilon_{ab}= 2\ii n^-_{[a}n^+_{b]}\,,\quad \gamma^a_{\,\,a}=0\,,
\ee
and where screen-basis indices $a$ and $b$ are manipulated with $I_{ab}\equiv S_{\mu\nu}n^{\mu}_a n^{\nu}_b=\delta_{ab}$, that is with a two-dimensional Euclidian metric. 

A canonical decomposition was introduced in Ref.~\cite{Fleury2014} as
\begin{equation}\label{eq:decomposition_Jacobi}
{\cal D}_{ab} (v) \equiv D_A(v)
\underbrace{\begin{bmatrix}
\cos\psi & \sin\psi \\
-\sin\psi & \cos\psi
\end{bmatrix}
}_\text{rotation}
\underbrace{
\exp
\begin{bmatrix}
-\Gamma_1 & \Gamma_2 \\
\Gamma_2 & \Gamma_1
\end{bmatrix}
}_\text{cosmic shear}.
\end{equation}
According to this decomposition, the real size and shape of the light source is obtained from the image by performing the following transformations: (i) an area-preserving shear $(\Gamma_1,\Gamma_2)$, (ii) a global rotation $\psi$, (iii) a global scaling. The latter defines the angular distance as
\be
D_A(v) \equiv \sqrt{{\rm det} {\mathcal D}_{ab}(v)}\,,
\ee
which does not assume any background spacetime and perturbative expansion. On the other hand, the definition~(\ref{DecompositionDab}) introduces the background angular distances $\bar D_A$. Both are related by
\be
   D_A(v) \simeq \bar D_A(v) [1+\kappa(v)]\,.
\ee
As for the deformation of the source shape, it is given by the reduced shear 
\be
\frac{{\mathcal D}_{\langle a b \rangle }}{\sqrt{{\rm det}{\mathcal D}_{ab}} }\simeq \frac{\gamma_{ab}}{(1-\kappa)}\,.
\ee

Each one of the above observables are defined on our past lightcone, and, as such, they are functions of $\bm{n}_\obs$ and $v$. The convergence and the rotation are scalar functions, and therefore can be expanded in terms of scalar spherical harmonics as
\bea
\kappa(\bm{n}^\obs,v)&=&\sum_{\ell,m}\kappa_{\ell m}(v)Y_{\ell m}(\bm{n}^\obs)\,,\label{e:322a}\\
V(\bm{n}^\obs,v)&=&\sum_{\ell,m}V_{\ell m}(v)
Y_{\ell m}(\bm{n}^\obs)\,.\label{e:322b}
\eea
The cosmic shear, on the other hand, being a spin two quantity, can be expanded in terms of the polarization basis as
\begin{equation} 
\label{e:gammapm}
  {\gamma}_{ab}(\bm{n}^\obs,v) \equiv \sum_{\lambda=\pm}{\gamma}^\lambda (\bm{n}^\obs,v) {n}^{\lambda}_a {n}^{\lambda}_b \,.
\end{equation}
The coefficients $\gamma^{\pm}$ can be further expanded in terms of $E$- and $B$-modes on a basis of spin-2 spherical harmonics as
\be\label{dec_gpm}
{\gamma}^\pm (\bm{n}^\obs,v) = \sum_{\ell,m} \left[{E}_{\ell m}(v)\pm \ii 
{B}_{\ell m}(v) \right]Y_{\ell m}^{\pm 2}(\bm{n}^\obs)\,.
\ee

It should be stressed that we adopt an observer-based point of view. This means that all quantities are expressed in terms of $(\bm{n}^\obs,\hat v)$. In general, $\bm{n}(\bm{n}^\obs,\hat v)\not=\bm{n}^\obs$, with the obvious exception of, e.g., Friedmann-Lema\^{\i}tre spacetimes and spacetimes with a local spherical symmetry for an observer located at the center of symmetry. Therefore, one of the difficulties in obtaining cosmological observables as a function of $v$, or equivalently as a function of the redshift $z$, lies in the determination of these coefficients.

\subsection{Particular case of a Bianchi~$I$ spacetime}\label{subsec3.4}

\subsubsection{Geodesic equation}\label{subsec3.4a}

The Bianchi~$I$ spacetime enjoys 3 Killing vectors, $\partial_i$, that allow one to construct 3 conserved quantities, $g(\partial_i,k)=k_i$, along
any geodesic. It implies that
\be\label{e41}
 k_i = \mathrm{cst}, 
\ee
on each geodesic so that
\bea\label{e41b}
  k^i = \frac{\gamma^{ij} k_j}{a^2}.
\eea
$k$ being a null vector, one concludes that $\omega^2\equiv (k^t)^2=g^{ij} k_i k_j$ with
\begin{equation}\label{Defomegatilde}
\omega \equiv \frac{1}{a}\sqrt{\sum_{i=1}^3 (\hbox{e}^{-\beta_i} k_i)^2}.
\end{equation}
It follows that the components of the direction of observation vector $n^\mu$ are given by
\begin{equation}\label{di}
n_i = k_i/\omega\,,\qquad n^i = k^i/\omega\,.
\end{equation}
The constants of motion $k_i$ are then directly related to the direction in which the observer in $O$ needs to look to detect the light signal, i.e. the direction of the source $n^\mu_\obs$. The redshift of a source is then given by
\begin{equation}\label{e.defz}
  1+z({\bm n}_\obs,t_{\rm S}) \equiv \frac{\omega_{\rm S}}{\omega_\obs} = \frac{a_\obs}{a(t_{\rm S})} \sqrt{ \frac{\sum_{i=1}^3{\left[\hbox{e}^{-\beta_i(t_{\rm S})} k_i\right]}^2}{\sum_{i=1}^3{\left[\hbox{e}^{-\beta_i(t_\obs)} k_i\right]}^2}}.
\end{equation}
It is always possible to choose the normalisation such that $a_\obs=1$
and $\beta_{i}(t_\obs)=0$, but we do not make that choice here.

\subsubsection{Jacobi matrix}\label{sec_conforme}

The study of the Sachs equation is simplified after performing a conformal transformation of the metric by a scale factor $a$, 
\be
{g}_{\mu\nu}=a^2 \tilde g_{\mu\nu}\,.
\ee
It can be checked that any null geodesic for $g_{\mu\nu}$, affinely parametrized by $v$, is also a null geodesic for $\tilde{g}_{\mu\nu}$, affinely parametrized by $\tilde{v}$ with $\dd v = a^2 \dd\tilde{v}$. The associated wave four-vectors then read $\tilde{k}^\mu = a^2 k^\mu$. Since the four-velocities of the comoving observers for both geometries are respectively $u=\partial_t$ and $\tilde{u}=\partial_\eta$, so that $\tilde{u}^\mu = a\,u^\mu$, we deduce that
\begin{equation}\label{e:cenergy}
\omega \equiv g_{\mu\nu} u^\mu k^\nu 
				= a^{-1} \tilde{g}_{\mu\nu} \tilde{u}^\mu \tilde{k}^\nu 
				\equiv a^{-1}\tilde\omega.
\end{equation}
The 3+1 decomposition of $\tilde{k}^\mu$ is therefore
\begin{equation}\label{e.defk}
\tilde{k}^\mu = \tilde\omega (-\tilde{u}^\mu+\tilde{n}^\mu)
\end{equation}
with $\tilde{n}^\mu\equiv a\,n^\mu$ implying $\tilde n_\mu = n_\mu/a$ and
\begin{equation}\label{ditilde}
\tilde n_i = \frac{\tilde k_i}{\tilde \omega}\,.
\end{equation}

The Sachs basis~$(\tilde{n}_a^\mu)$ for the conformal geometry is then related to the original one~(\ref{basis2d}) by
\begin{equation}
  \tilde{n}_a^\mu = a \, n_a^\mu\,,\qquad \tilde{n}^a_\mu = a^{-1} \, n^a_\mu.
\end{equation}
One can indeed check that the orthonormality~(\ref{basis2d}) and the parallel transport conditions~(\ref{e:prop_n}) are preserved by the conformal
transformation with the use of the projection matrix $\tilde
S_{\mu\nu}=a^{-2}{S}_{\mu\nu}$, instead of Eq.~(\ref{e:defSmunu}). 

The separation four-vector~$\eta^\mu$ is defined by comparing events only, independently from any metric. It is therefore invariant under conformal transformations. However, its projection over the Sachs basis changes (since the Sachs basis itself changes), indeed
\begin{equation}
\eta^a \equiv n^a_\mu \eta^\mu = a \tilde{n}^a_\mu \tilde{\eta}^\mu = a\,\tilde{\eta}^a.
\end{equation}
This implies that the Jacobi matrix transform as~\cite{Bonvin2010,Fleury2014}
\be\label{e.defDab}
{ \cal D}_{ab} = a\,\tilde{\cal D}_{ab}\,.
\ee
Hence, the angular distance $D_A$ in the universe described by a metric $g_{\mu\nu}$ is just $a \tilde D_A$, where $\tilde D_A$ is the angular distance in the universe described by the metric $\tilde g_{\mu\nu}$. At lowest order in perturbations, $\kappa$ is the relative perturbation of angular diameter distance whatever is the metric used. As for the reduced shear, it remains unaffected by the conformal transformation. In the remainder of this article, we will thus discard the effect of an overall scale factor, in order to simplify the computation. However it should be reminded that, as shown by Eq.~(\ref{e:cenergy}), a conformal transformation has an effect on the energetic aspects of light propagation, that is on the relation between the redshift and the affine parameter $U(v)$.

\subsubsection{General solution}

Using such a conformal transformation, it was shown in Ref.~\cite{Fleury2014} that the Sachs equation can be solved analytically in a Bianchi~$I$ universe. This solution relies on the fact that the Sachs equation can be rewritten as
\begin{equation}\label{e71}
  \frac{\dd^2\tilde{\cal D}_{ab}}{\dd\tilde{v}^2} = \tilde{\cal R}_{ac} \tilde{\cal D}_{cb},
\end{equation}
with the rescaled optical matrix given by
\begin{equation}\label{eq:optical_tidal_matrix}
 \tilde{\cal R}_{ab} = \tilde\omega^2 \left[ (\sigma_{ab})' + \sigma_{ac}\sigma_{cb} + \frac{\tilde\omega'}{\tilde\omega} \sigma_{ab} \right].
\end{equation}
The explicit solution of this equation is given in \S~VII.A-B of Ref.~\cite{Fleury2014}.

\section{Small shear limit}\label{sec5}

\subsection{Definition}\label{subsec5.1}

The current observational status of the  $\Lambda$CDM model shows that if the expansion is anisotropic, $\sigma/{\cal H}$ has to be small. Moreover, since any primordial anisotropy is washed out by the expansion of the universe, the term ${\cal C}^i_j$ in the evolution of the background shear is negligible compared to the integral term in Eq.~(\ref{e:sigij_sol}).

As discussed in the introduction, a late time anistropy may be generated during the acceleration of the universe, but the effect we are looking for needs to have an amplitude small enough to be below the detection threshold of ongoing observational surveys.  

In full generality a linear and gauge-invariant perturbative expansion around an anisotropic background should be performed. It was developped in Ref.~\cite{ppu1,ppu2} in the context of inflation and Appendix~\ref{secA} derives the full perturbation theory for a post-inflationary era. While a numerical integration of these equations can be performed, it is clear from the previous arguments that an analytical insight in the regime $\sigma/{\cal H}\ll1$ is sufficient.

We shall thus work in the {\em small shear limit} in which the background shear induced at late time by the anisotropic stress-energy tensor of the dark component is small, that is in the limit $\sigma/{\cal H}\ll1$.  More precisely, we assume that $\gamma_{ij}-\delta_{ij}\simeq 2 \beta_{i}\delta_{ij}$ is a small dimensionless perturbation, and $\sigma_{ij}/\HH$ is of the same order as this homogeneous perturbation. We shall thus consider the Bianchi~$I$ spacetime as a homogeneous perturbation around an isotropic Friedmann-Lema\^{\i}tre spacetime, hence ignoring non-linear  corrections in the background shear as well. In order to implement this approximation scheme, we introduce a  two-parameter perturbation scheme (see e.g. Ref.~\cite{Sopuerta:2003rg}) in which, besides the usual Scalar-Vector-Tensor (SVT) perturbations over a flat Friedmann-Lema\^{\i}tre background, the geometrical shear is considered as an extra perturbative degree. We refer to Ref.~\cite{Pontzen:2010eg} for a detailed description of general Bianchi spaces in this approach.

\subsection{Spacetime description}\label{subsec5.2}

\subsubsection{Metric}

We shall thus adopt the metric
\be\label{perturbedmetric}
   \dd{s}^{2}=a^{2}\left[-(1+2 \Phi)\dd\eta^{2}+2 \bar B_{i}\dd x^{i}\dd\eta+\left(\gamma_{ij}+h_{ij}\right)\dd x^{i}\dd x^{j}\right]\,,
\ee
where $h_{ij}$ is defined as [see Eqs.~(\ref{hij}) and (\ref{gauge-choice})]
\be
  h_{ij}=-2 \left(\gamma_{ij} +\frac{\sigma_{ij}}{\HH}\right)\Psi + 2 E_{ij}\,,
\ee
and $\gamma_{ij}$ is here understood as the Euclidian metric plus a small perturbation
\be
   \gamma_{ij}\simeq \delta_{ij}+2\int_{0}^{a}\frac{\sigma_{ij}}{\HH}\frac{\dd a'}{a'}\,,\quad\frac{\sigma_{ij}}{\HH}\ll1\,.
\ee
In order to simplify the notation, we also define the matrix 
\begin{equation}\label{e.defbetaij}
  \beta_{ij}\equiv{\rm diag}(\beta_i)
\end{equation}  
such that
\be\label{DefMatrixBeta}
\gamma_{ij} = \exp[2\beta]_{ij}\simeq \delta_{ij}+2 \beta_{ij}\,,\qquad \sigma_{\ti \tj} = \sigma^i_j=\beta'_{ij}\,.
\ee
Thus, $\beta_{ij}$ controls the homogeneous perturbation. Indices are now raised and lowered with the Euclidian metric $\delta_{ij}$ and $\delta^{ij}$, and the vector modes $B_i$ and tensor modes $E_{ij}$ satisfy $\partial^i B_i = \partial^i E_{ij}=E^i_i=0$. But since $\beta_{ij}$ is homogenous, everything happens as if we had usual cosmological perturbation, but also an infinite wavelength perturbations $2 \beta_{ij}$ to the spatial metric.

To control the perturbative series, we introduce the $\{n,p\}$ notation, where $n$ and $p$ indicate powers in $\beta$ and SVT variables, respectively. Thus, a term like $ \sigma_{ij}/\HH$ is of order $\oz$, terms like $\Psi$ and $\Phi$ are of order $\zo$, while a product like $\sigma^j_i\dbj\Psi/\HH$ is of order $\oo$. However, since vector and tensor modes only appear due to the coupling between the shear and scalar modes~\cite{ppu1}, vector perturbations $B_i$ and tensor perturbations $E_{ij}$ are also of order $\oo$. Hence, for any quantity $X$, one will consider the different quantities:
\begin{itemize}
 \item $X^{\{0,0\}}(\eta)$: the Friedmann-Lema\^itre background value;
 \item $X^{\{1,0\}}(\eta)$: the first order (homogeneous) scalar perturbed quantity in $\sigma/{\cal H}$;
 \item $X^{\{0,1\}}(\eta,{\bm x})$: the first order inhomogeneous perturbed quantity in $\Psi,\ldots$;
 \item $X^{\{1,1\}}(\eta,{\bm x})$: the first order inhomogeneous perturbed quantity in both  $\sigma/{\cal H}$ and $\Psi,\ldots$ and vector and tensor perturbations.
\end{itemize}

Before moving on we should make some general remarks about the adopted perturbative scheme. Indeed, one might be worried that adding
$\sigma_{ij}/\HH$ or $\beta_{ij}$ as a small homogeneous perturbation to the background metric would not have any significant observable effect, since the SVT decomposition was already designed to describe the most general perturbation over a flat Friedmann-Lema\^itre universe. Note however that SVT modes do not include a zero Fourier-mode in their spectrum (i.e., an infinite wavelength perturbation), since these modes will be isotropic by construction and hence merely rescale the background geometry. The tensor $\beta_{ij}$, on the other hand, is a homogeneous  (i.e., space-independent) field, which by definition corresponds to an anisotropic zero mode. Thus, its effect cannot be absorbed in a simple rescaling of the scale factor. Moreover, this field sources the background dynamics through Einstein's equations. 

\subsubsection{Tetrad basis}

Given this expansion scheme, the tetrad basis associated to the perturbed metric up to order $\oo$ is explicitely given by
\be\label{DefTetrad}
\begin{aligned}
\ThetradDU{\ti}{\,\,j} &\simeq (\delta_i^j -\beta_{ij}) (1+\Psi)+\frac{\beta'_{ij}}{\HH} \Psi  -E_i^j\,,\qquad &\ThetradDU{\ti}{\,\,0}&=0\,,\\
\ThetradUD{\ti}{\,\,j}  &\simeq (\delta^i_j +\beta_{ij} )(1-\Psi)-\frac{\beta'_{ij}}{\HH} \Psi+E^i_j\,,\qquad &\ThetradUD{\ti}{\,\,0}  &=\bar B^i\,,\\
\ThetradDU{\tz}{\,\,0} &= 1-\Phi\,,\qquad &\ThetradDU{\tz}{\,\,i} &=-\bar B^i\,,\\
\ThetradUD{\tz}{\,\,0}  &= 1+\Phi\,,\qquad  &\ThetradUD{\tz}{\,\,i}&=0\,,
\end{aligned}
\ee
where $\thetradDU{\ti}{\,\,j}$ refers to the background spatial triad defined in Eq.~(\ref{spatial_triad}). By chosing the observer to coincide with the timelike vector of the tetrad ($u^\mu = \ThetradDU{\tz}{\,\,\mu}$, $u_\mu = -\ThetradUD{\tz}{\,\,\mu}$) we obtain in general that the components of the direction vector $\bm{n}$ in the tetrad basis as
\be
\label{kivskz}
 k^\ti = -k^\tz n^\ti\,, \qquad n^\ti = \ThetradUD{\ti}{\mu} n^\mu\,.
\ee
At the position of the observer, the direction of the geodesic in the tetrad basis $n_\obs^\ti$ is also the direction in which the observation is made. Again, we remind that we are interested in the observables related to light propagation as expressed in function of this observed direction $n_\obs^\ti$.

Since we have introduced two types of tetrads ($\{\thetrad\}$ and  $\{\Thetrad\}$), there is an ambiguity whenever a tetrad index $\ti$ appears on a tensorial quantity.  First, for the geometric shear tensor, the tetrad index is defined with respect to the triad $\thetradDU{\ti}{\,\,\mu}$ and we remind that $\sigma_{\ti \tj} =\sigma^{\ti \tj}={\sigma^{\ti}}_{\tj} = {\rm diag}(\beta_i')=\beta'_{ij}$; see Eq.~(\ref{e.213}). Second, for partial derivatives the tetrad index corresponds also to the tetrad $\{ \thetrad\}$ and we define
\be\label{ditodi}
  \partial_{\ti} \equiv \thetradDU{\ti}{\,\,j}\partial_j \simeq \partial_i -\beta_{i}^{\,j}\partial_j\,.
\ee
It makes clear the difference between a derivative in the direction of a tetrad vector $\partial_{\ti}$ and the derivative in the direction of the vectors $\partial_i$ associated with the Cartesian coordinates. Since the vector perturbations $B_i$ and the tensor perturbations $E_{ij}$ are already of order $\oo$, there is absolutely no difference between their tetrad components $B_\ti$ and $E_{\ti\tj}$ at this order of perturbations and there is no need to be particularly careful. Everywhere else, a tetrad index refers to the tetrad $\{ \Thetrad\}$ defined in Eqs.~(\ref{DefTetrad}).

\subsection{Technical interlude}

Since we are interested in computing observables on the celestial sphere,
spherical coordinates are much more convenient than Cartesian
coordinates. This paragraph describes the use of such spherical
coordinates in real space and of the associated derivatives (radial and on the unit sphere). Several definitions of covariant derivatives have to be distinguished. We finish by relating them to each other and to the spin-raising operator of spherical harmonics.

\subsubsection{Spherical coordinates in real space}

Consider a tensor depending on Cartesian coordinates $T_{i_1 \dots i_n}(x^i)$ (with indices raised and lowered respectively with $\delta_{ij}$ and $\delta^{ij}$); it can always be constructed by considering the tetrad components of a given tensor. In spherical coordinates, one can then define from the partial derivative $\partial_i = \partial/\partial x^i$ a covariant derivative $D_i$ on the unit sphere and a radial derivative $\partial_r$. To be more precise, this requires the use of the projectors
\be
  S_{ij} \equiv \delta_{ij}-\hat x_i \hat x_j
  \,,\qquad 
  \hat x^i\equiv \frac{x^i}{r}\,,\qquad{\rm with}\qquad r^2 = \sum_{i=1}^3 (x^i)^2\,.
\ee
Remind that $S^{ij} =\delta^{ip} \delta^{jq} S_{pq}$ and $S_i^j =\delta^{ip} S_{pj}$. The covariant derivative on the unit sphere $S^2|_{\rm space}$ of the Cartesian coordinates centered on the observer is denoted by $D^\Rthreesymbol_i$ and is defined from the general projection
\be\label{DefRthreesymbol}
 \frac{1}{r }D^\Rthreesymbol_i T_{j_1\dots j_n} \equiv S_i^k S_{j_1}^{q_1} \dots
  S_{j_n}^{q_n}\frac{\partial}{\partial x^k} T_{q_1\dots q_n} \equiv P\left[\frac{\partial}{\partial x^i} T_{j_1\dots j_n}\right]
\ee
where $P[\dots]$ is to be understood as the projection of all free Cartesian indices with the projector $S_i^j$. This derivative only makes sense if
the tensor itself is a projected tensor, that is, if it satisfies $P[T_{j_1\dots j_n}] = T_{j_1\dots j_n}$. 

The radial derivative is then obtained simply by
\be
\partial_r T_{j_1 \dots j_n} \equiv \hat x^i \frac{\partial}{\partial x^i} T_{j_1 \dots j_n}\,.
\ee

Now, any combination of partial derivatives $\partial_i$ applied to some tensor, can be decomposed in terms of radial derivatives $\partial_r$ and covariant derivatives on the sphere $D^\Rthreesymbol_i$. The simplest such decomposition is
\be
 \partial_i f = \hat x^i \partial_r f+ \frac{1}{r} D_i^\Rthreesymbol f \,,
\ee
for any scalar function $f$. The decompositions for projected tensors of various ranks is detailed in appendix~\ref{App1plus2Cartesian}. For general tensors which are not necessarily projected, it is necessary to split them into their projected components on the sphere and their radial components before decomposing any derivative applying on them. Such decomposition for the vector and tensor modes is given in Eqs.~(\ref{FirstSplittingoftensor}). To finish, it is easy to check that 
\be
D^\Rthreesymbol_i S_{jk} = 0 \,,\qquad D^\Rthreesymbol_i \epsilon_{jk} =0\,, 
\ee
where the completely antisymmetric tensor on the sphere is
\be
 \epsilon_{ij} \equiv \epsilon_{ijk} \hat x^k\,.
\ee

\subsubsection{Covariant derivative on the tangent space}\label{SecDtangent}

For any spatial tensor constant in space, such as $\sigma_{\ti\tj}$, one can define scalar, vector and tensor fields on the unit sphere. First, one can define a scalar field on the unit sphere of observing directions, $S^2|_{\rm obs}$, by contracting  all free indices with the direction of observation,
\be\label{e.defSigma}
\Sigma\equiv \frac{1}{2}\sigma_{\ti \tk}n_\obs^\ti n_\obs^\tk\,.
\ee
Indeed, the observing direction can be considered as a point on $S^2|_{\rm obs}$, whose spherical coordinates are $(\theta_\obs,\varphi_\obs)$, and $\Sigma$ from the expression~\eqref{e.defSigma} is thus a function of $(\theta_\obs,\varphi_\obs)$, that is a scalar field on $S^2|_{\rm obs}$. Then, to define a vector field on the unit sphere, one needs to contract one index with the observing direction and project the remaining one on the sphere. Furthermore, in order to get a tensor field on the unit sphere, we shall project the two free indices on the sphere. These projections are obtained by contraction with the screen basis vectors $\bm{n}^\obs_a$ at the observer. For instance, the vector and tensor fields on the sphere build from the geometric
shear are simply
\be\label{e.defSigmaab}
 \Sigma_a\equiv {n^\obs_a}^\tj\sigma_{\tj \ti} n_\obs^\ti\,\qquad
 \Sigma_{ab}\equiv {n^\obs_a}^\ti{n^\obs_b}^\tk \sigma_{\ti \tk}\,.
\ee
We remark that $\Sigma_{ab}$, which is a symmetric $2\times2$ matrix, is not traceless. In fact, using the partition of the identity $\delta^{ij} = n^\ti n^\tj+n_1^\ti n_1^\tj+n_2^\ti n_2^\tj$, the trace is given by $\delta^{ab}{n^\obs_a}^\ti{n^\obs_b}^\tk \sigma_{\ti \tk} =- \sigma_{\ti \tk} n^\ti n^\tk =-2 \Sigma$. 
Alternatively, the vector and tensor fields~\eqref{e.defSigmaab} can be obtained by applying successively the covariant derivative on the unit sphere $D_a$ to $\Sigma(\theta_\obs,\varphi_\obs)$. Indeed, with this method, we find the relations
\be\label{def_DaS2}
\begin{aligned}
\Sigma_a &= D_a \Sigma \,,\qquad &\Sigma_{ab} &= D_{a} D_{b}\Sigma + 2
\delta_{ab} \Sigma\,,\\
 D_a D^a \Sigma &= -6 \Sigma\,,\qquad&\Sigma_{\langle ab\rangle}&= D_{\langle a} D_{b
  \rangle}\Sigma\,.
\end{aligned}
\ee
Note that the metric and the antisymmetric tensor on the sphere are obtained from
\be
\delta_{ab} = 2 n^{(+}_a n^{-)}_b\qquad \epsilon_{ab} = 2 \ii n^{[-}_a n^{+]}_b
\ee
and satisfy
\be
D_a \delta_{bc}=0\,, \qquad D_a \epsilon_{bc}=0\,.
\ee

\subsubsection{Background geodesics and identification of covariant derivatives}\label{IdentificationDD}

The covariant derivative $D_a$, related to the unit sphere in the observer's tangent space, and the derivative $D_i^\Rthreesymbol$, related to the unit sphere of Cartesian coordinates, are fundamentally different. But, they can be related in a simple way. Indeed, the solution to the background
geodesic at order $\{0,0\}$, that is, the geodesics of  the spatially flat Friedmann-Lema\^{\i}tre spacetime, is given by
\be
n^\ti = n_\obs^\ti\,,\qquad n_a^\ti = {n^\obs_a}^\ti\,,\qquad \frac{\dd x^{i\zz}}{\dd \chi}=n_\obs^\ti \quad \Rightarrow \quad
x^{i\zz} = \chi n_\obs^\ti\,,
\ee
where, we remind the reader, $n_\obs^\ti$ is the direction of the geodesic at the position of the observer in the tetrad basis. This is the direction of observation, since we have oriented the geodesic toward the past.  

There is thus a straightforward identification between the sphere of
the directions of observation, lying in the tangent space at the
observer (the set of directions $S^2|_{\rm obs}$ spanned by
$n_\obs^\ti$), and the set of points of $\mathds{R}^3$ reached at an
affine parameter $\chi$ (or $\eta$) on the background
geodesic. Indeed, the points spanned by the coordinates $x^{i\zz}$ at
a given affine parameter $\chi$ are such that
$$
\delta_{ij} x^{i\zz} x^{j\zz} = r^2(\chi)=\chi^2
$$ 
and form a sphere in the Cartesian coordinates. We can then subsequently identify this sphere of radius $\chi$ to the unit sphere $S^2|_{\rm space}$.

This means that we can identify  $n_0^\ti$ with $\hat x^i$ and then $D_a$ on $S^2|_{\rm obs}$  with ${n_a^\obs}^\ti D_i^\Rthreesymbol $, the projection onto the screen basis $\bm{n}_a^\obs$ being used only to switch from the extrinsic point of view of the derivative (the projection of the Cartesian derivative onto the sphere) to an intrinsic point of view on the sphere. In the rest of this article we thus replace the notation $D_i^\Rthreesymbol $ by $D_i$, ${n^\obs_a}^\ti D_i^\Rthreesymbol $ by $D_a$, and ${n^\obs_\pm}^\ti D_i^\Rthreesymbol $ by $D_\pm$.

\subsubsection{Link with spin-raising operator and spin-weighted spherical harmonics}\label{SecEdth}

The covariant derivative on the unit sphere is related to the usual spin-raising and spin-lowering operators. In spherical coordinates, these operators are defined for a spin-$s$ quantity by
\bea \label{def_spinder}
\edth X^s&=&-\sin^s\theta \left[\partial_\theta +\ii \frac{1}{\sin \theta} \partial_\varphi\right]\left(\sin^{-s} \theta X^s\right)\,,\\
\edthb X^s&=&-\sin^{-s}\theta \left[\partial_\theta -\ii \frac{1}{\sin \theta} \partial_\varphi\right]\left(\sin^{s} \theta X^s\right)\,.
\eea 
They are related to the covariant derivative through
\be
  \edth = -\sqrt{2} {n_{-}^{\obs}}^a D_a=-\sqrt{2} D_+\,,\qquad \edthb = -\sqrt{2} {n_{+}^{\obs}}^a D_a=-\sqrt{2}D_-\,,
\ee
the vector ${\bm n}_\pm^\obs$ being defined in Eq.~(\ref{e:helicity}). Hence, for a tensor field of spin $+|s|$ on the sphere, $X^{\nu_1\dots \nu_s}= X^s {n^\obs_+}^{\nu_1}\dots {n^\obs_+}^{\nu_s} $, and a tensor of spin $-|s|$, $Z^{\nu_1\dots \nu_s}= Z^{-s}  {n^\obs_-}^{\nu_1}\dots {n^\obs_-}^{\nu_s}$, we have 
\bea\label{edthDproperty}
-\sqrt{2}\nabla^\mu X^{\nu_1\dots \nu_s} &=& (\edth X^s) {n^\obs_+}^\mu {n^\obs_+}^{\nu_1}\dots {n^\obs_+}^{\nu_s}+(\edthb X^s) {n^\obs_-}^\mu {n^\obs_+}^{\nu_1}\dots {n^\obs_+}^{\nu_s}\,\\
-\sqrt{2}\nabla^\mu Z^{\nu_1\dots \nu_s} &=& (\edth Z^{-s}) {n^\obs_+}^\mu {n^\obs_-}^{\nu_1}\dots {n^\obs_-}^{\nu_s}+(\edthb Z^{-s}) {n^\obs_-}^\mu {n^\obs_-}^{\nu_1}\dots {n^\obs_-}^{\nu_s}\,.
\eea
Since the spin-weighted spherical harmonics satisfy the property
\be
Y^{s}_{\ell m} = \begin{cases}
\begin{aligned}
\sqrt{\frac{(\ell-s)!}{(\ell+s)!}} \edth^s Y_{\ell m}\qquad&{\rm if}\qquad 0\leq s\leq \ell\\
(-1)^s \sqrt{\frac{(\ell+s)!}{(\ell-s)!}} \edthb^{\,-s} Y_{\ell m}\qquad&{\rm if}\qquad -\ell \leq s \leq 0\,,
\end{aligned}
\end{cases}
\ee
any number of covariant derivatives applied on a spherical harmonic can be computed using the properties~\eqref{edthDproperty}.

As an application, consider the expansion of the variable $\Sigma$ in spherical harmonics
\be\label{def.Sig2m}
\Sigma(\chi,n_\obs^\ti) = \sum_{m=-2}^{+2} \Sigma_{2m}(\chi) Y_{2m}(n_\obs^\ti)\,.
\ee
If we align the azimuthal direction with an eigendirection of the geometric shear, the multipoles coefficients are then given by
\begin{equation}\label{EqSphericalComponents}
\Sigma_{20}(\chi) = -\sqrt{\frac{\pi}{5}}[\beta'_1(\chi)  + \hat \beta'_2(\chi) ] \,,\qquad \hat \Sigma_{2 \,\pm2}(\chi) = \sqrt{\frac{\pi}{30}} [\beta'_1(\chi)  - \beta'_2(\chi) ]\,.
\end{equation}
The most useful derivatives are then easily obtained to be 
\bea\label{def_Sigpm}
\Sigma_{\pm}(\chi,n_\obs^\ti)&=&{n^\obs_{\mp}}^\tk {n_\obs}^\tj\sigma_{\tk \tj}(\chi) =   D_{\pm}\Sigma(\chi,n_\obs^\ti) = \mp \sqrt{3}\sum_{m} \Sigma_{2m}(\chi) Y^{\pm1}_{2m}(n_\obs^\ti) \,,\\
\Sigma_{\pm \pm}(\chi,n_\obs^\ti)&=&{n^\obs_{\langle
    \mp}}^\tk {n^\obs_{\mp \rangle}}^\tj\sigma_{\tk \tj}(\chi) = D_{\pm} D_{\pm}\Sigma(\chi,n_\obs^\ti)= \sqrt{6}\sum_{m} \Sigma_{2m}(\chi) Y^{\pm2}_{2m}(n_\obs^\ti)\,.
\eea
Similarly, if we expand a scalar field $\varphi(\chi,x^i)$ in spherical harmonics
\be
\varphi(\chi,x^i) = \sum_{\ell,m} \varphi_{\ell m}(\chi,r) Y_{\ell m}(\hat x^i)\,,
\ee
then the most useful derivatives are 
\bea
D_{\pm} \varphi(\chi,x^i) &=& \mp \sqrt{\frac{\ell(\ell+1)}{2}}\sum_{\ell,m} \varphi_{\ell m}(\chi,r) Y^{\pm1}_{\ell m}(\hat x^i) \,,\\
D_{\pm}D_{\pm} \varphi(\chi,x^i) &=& \frac{1}{2}\sqrt{\frac{(\ell+2)!}{(\ell-2)!}}\sum_{\ell,m} \varphi_{\ell m}(\chi,r) Y^{\pm2}_{\ell m}(\hat x^i)\,,\\
2 D_{\pm}D_{\mp} \varphi(\chi,x^i) &=& D_a D^a \varphi(\chi,x^i) =  -\ell(\ell+1) \sum_{\ell,m} \varphi_{\ell m}(\chi,r) Y_{\ell m}(\hat x^i)\,.
\eea

\subsection{Geodesics and Sachs equations in term of the Friedmannian coordinates}\label{seczzzz}

In the approximation that we are considering, we can solve the perturbation equations and the Sachs equation up to order $\{1,1\}$. We shall define the distance down to the lightcone on the Friedmann-Lema\^{\i}tre background spacetime as
\be\label{echieta}
\chi \equiv \eta_0-\eta.
\ee
The geodesic equation~(\ref{e.geo3}) takes the form
\be\label{MasterGeodesic}
\frac{\dd k^{\nu}}{\dd v}+\Gamma_{\alpha\beta}^{\nu}k^{\alpha}k^{\beta}=0\,,\qquad k^0=-\frac{\dd \chi}{\dd v}\,,
\ee
and using~\eqref{kivskz} can be rewritten directly in terms of tetrad components as
\bea\label{MasterGeodesicTetrads}
 \frac{\dd k^{\ti}}{\dd v}&=&\frac{\dd k_{\ti}}{\dd v}=(k^\tz)^2\left(\omega_{\tk \tj \ti} n^\tk n^{\tj}+\omega_{\tz \tz \ti}-\omega_{\tj \tz \ti} n^\tj-\omega_{\tz \tj \ti} n^{\tj} \right) \,,\\
 \frac{\dd k^{\tz}}{\dd v}&=&-\frac{\dd k_{\tz}}{\dd v}=(k^\tz)^2\left(-\omega_{\ti \tj \tz} n^{\ti} n^\tj+\omega_{\tz \ti \tz} n^{\ti} \right) \,.
\eea
where the affine connections are defined in Appendix~\ref{AppAffineConnections}. Instead of the parameter $v$ we shall use the parameter $\chi$ since, once the wave-vector is integrated, we have
\be\label{MasterPosition}
\frac{\dd x^\mu}{\dd \chi} = -\frac{k^\mu}{k^0}\,.
\ee
The position on the geodesic then becomes a function of the parameter $\chi$ and the initial direction $n_\obs^\ti$.  Finally, the Sachs equation with the parameter $\chi$ reads~\cite{Bonvin2010}
\be\label{MasterBonvin}
\frac{\dd^{2}{\cal D}_{ab}}{\dd\chi^{2}}+\frac{1}{k^{0}}\frac{\dd k^{0}}{\dd\chi}\frac{\dd{\cal D}_{ab}}{\dd\chi}=\frac{1}{\left(k^{0}\right)^{2}}{\cal R}_{ac}{\cal D}_{cb}\,.
\ee

\section{Angular multipole correlations in anisotropic spaces}\label{SecAngularCorrelations}

As previously explained, we adopt an observer point of view in which all observable quantities are considered as functions of the direction of observation ${\bm n}_\obs$ and of the affine parameter $v$ or, equivalently, of the redshift $z$, keeping in mind that the later also depends on ${\bm n}_\obs$. All these quantities can be decomposed on a basis of spin-weighted spherical harmonics, $Y^s_{\ell m}$. The goal of this section is to derive a set of formal expressions concerning these expansions and to establish general results of the two-point correlation function valid in Bianchi $I$ geometries.

We consider that the universe  has undergone an early period of isotropic expansion followed by a late-time anisotropic phase. This is in sharp contrast with the approach of Ref.~\cite{Pullen:2007tu}, in which the universe is supposed to have an early inflationary stage followed by an isotropic evolution (so that geodesics are Friedmann-Lema\^{\i}tre geodesics and anisotropy is imprinted only in the source term).

The tools we shall develop are not specific to weak-lensing and can be used in other contextz, such as the study of the cosmic microwave background. We first describe, in \S~\ref{sec4.1}, the general expansion of spin $s$ quantities. This will allow us to express their angular power spectrum in \S~\ref{sec4.2}. We conclude by demonstrating that, while spatial parity symmetry implies that the $EB$ correlation matrix vanishes, some off-diagonal correlaions are necessarily non-vanishing and encode information on the geometrical shear.

\subsection{Multipolar expansions}\label{sec4.1}

The spin-$s$ polarization basis is defined as a tensor product  of $s$ polarization vectors as
\beanosub\label{def_polar}
\bm{m}_\obs^{s}\equiv
\begin{cases}
\bm{n}_\obs^{+}\otimes\dots\otimes\bm{n}_\obs^{+}
& \quad{\rm if}\quad s>0\,,\\
\bm{n}_\obs^{-}\otimes\dots\otimes\bm{n}_\obs^{-} & \quad{\rm if}\quad s<0\,.
\end{cases}
\eeanosub 
Under the action of an active rotation $R$, this basis transforms as
\be
 R[{\bm{m}}^s_\obs] \equiv R\cdot\bm{m}^{s}_\obs (R^{-1} \cdot\bm{n}_\obs)\,,
\ee
where $\bm{n}_\obs$ is the vector along the line of sight at the point of observation. Spin-weighted spherical harmonics transform under the same rotation as
\be
R [Y_{\ell m}^s(\bm{n}_\obs)\bm{m}^s_\obs] \equiv Y^s_{\ell
  m}(R^{-1} \cdot\bm{n}_\obs) R [\bm{m}^s_\obs]  = \sum_{m'}
Y_{\ell m'}^s(\bm{n}_\obs){\bm{m}}^s_\obs  D^\ell_{m' m}(R)\,,
\ee
where $D^\ell_{mm'}(R)$ are the components of the Wigner D-matrix. This means that they transform like {\em normal} spherical harmonics {\em provided} they are accompanied by the polarization basis to which they
are associated.\\

Now, any cosmological observable $X^s$ of spin $s$ can be expressed in the form
\be\label{e:GeneralLOS}
 X^s(\chi_{_{\rm S}}, \bm{n}_\obs) \bm{m}^{s}_{\obs} =
\int_0^{\chi_{_{\rm S}}} S^{X_s}(\chi_{_{\rm S}}, \chi,\bm{n}_\obs) \bm{m}^{s}_\obs \dd\chi\,,
\ee
where $\chi_{_{\rm S}}$ refers to the position of the source. Note that we are explicitly making use of the small shear expansion, since the source term is integrated along a geodesic of the Friedmann-Lema\^{\i}tre spacetime. This means that in order to compute $ X^s$ at order $\{n,p\}$ one needs to determine the source $S^{X_s}$ at the same order. The source term $S^{X_s}(\chi_{_{\rm S}}, \chi,\bm{n}_\obs)$ has to be understood as
\be
S^{X_s}(\chi_{_{\rm S}}, \chi,\bm{n}_\obs) = \left.S^{X_s}(\chi_{_{\rm S}}, \chi,x^i,\bm{n}_\obs)\right\vert_{x^i=\chi n_\obs^\ti}\,,
\ee
that is, evaluated on the background geodesic. Moreover, thanks to Eq.~(\ref{echieta}), the parameters $\chi$ and $\chi_{_{\rm S}}$ can be thought as time coordinates. The intrinsic angular dependence of $S^{X_s}$ on $\bm{n}_\obs$ is a consequence of the (possible) non scalar nature of the source. 
Moving forward, it is convenient to decompose $X^s$ into spherical harmonics as
\be
\label{X_ylm}
X^s(\chi_{_{\rm S}},\bm{n}_\obs) \bm{m}^{s}_{\obs} = \sum_{\ell,m} X^s_{\ell m}(\chi_{_{\rm S}}) Y^s_{\ell m}(\bm{n}_\obs) \bm{m}^{s}_{\obs}\,,
\ee
which will allows us to define multipolar correlations at unequal times of the form  $\langle X^s_{\ell m}(\chi_{_{\rm S1}}) X^{s\star}_{\ell' m'}(\chi_{_{\rm S2}}) \rangle$. In order to compute these angular correlators,  we first need to take the Fourier transform of the source~(\ref{e:GeneralLOS}) off the line of sight
\be
\label{source_fourier}
S^{X_s}(\chi_{_{\rm S}},\chi,x^i,\bm{n}_\obs)\,\bm{m}^{s}_{\obs} = \int \frac{\dd^3 \bm{k}}{(2 \pi)^{3/2}}
S^{X_s}(\chi_{_{\rm S}},\chi,\bm{k},\bm{n}_\obs) e^{\ii \bm{k}\cdot\bm{x}}\,\bm{m}^{s}_{\obs} \,, 
\ee
in the sense that we do not bind $x^i$ to $\chi$ by the relation $x^i=\chi n_\obs^\ti$ and $\chi$ has to be thought as a time coordinate thanks to Eq.~(\ref{echieta}). Then, the intrinsic dependence of the source on $\bm{n}_\obs$ is further expanded in terms of spherical harmonics, with the latter being defined with respect to an axis aligned with the Fourier mode $\bm{k}$. That is
\be
\label{source_fourier_ylm}
S^{X_s}(\chi_{_{\rm S}},\chi,,\bm{k},\bm{n}_\obs)  \bm{m}^{s}_{\obs} = \sum_{\ell,m} S^{X_s}_{\ell m}(\chi_{_{\rm S}},\chi,\bm{k}) \ii^\ell \sqrt{\frac{4 \pi}{2 \ell+1}} R_{\bm{k}}[Y^s_{\ell m}(\bm{n}_\obs) \bm{m}^{s}_{\obs}]\,,
\ee
where $R_{{\bm k}}$ is a rotation that transports the azimuthal direction to the direction of the Fourier mode ${\bm k}$
(see Appendix~\ref{SecFullLensing} for details about this notation).
The terms with $m=0,1,2$ correspond here to scalar, vector and tensor perturbations, respectively. If we now make use of the Rayleigh expansion
\be\label{Rayleigh}
e^{\ii \bm{k}\cdot\bm{x}} =4 \pi \sum_{\ell,m} \ii^\ell j_\ell(k r)
Y^\star_{\ell m}(\hat{\bm{k}}) Y_{\ell m}(\bm{n}_\obs) = \sum_\ell \ii^\ell
\sqrt{(4\pi)(2 \ell+1)} j_\ell(kr) R_{\bm{k}}[  Y_{\ell 0}(\bm{n}_\obs)]\,,
\ee
with $r=\delta_{ij}x^ix^j$, and insert the decomposition~(\ref{source_fourier_ylm}) into Eq.~(\ref{source_fourier}), we find, after comparing Eqs.~(\ref{X_ylm})  and~(\ref{source_fourier}), that
\be\label{X_s}
X^{s}_{\ell m}(\chi_{_{\rm S}}) = \sqrt{\frac{2}{\pi}}\int \dd^3 \bm{k}\int_0^{\chi_{_{\rm S}}} \dd\chi'\sum_{m'} D^\ell_{m m'}(R_{\bm{k}})\ii^\ell 
\sqrt{\frac{(2 \ell+1)}{4 \pi}} \sum_{\ell'} {}^s
j_{\ell}^{(\ell' m')}(k r) S^{X_s}_{\ell' m'}(\chi_{_{\rm S}},\chi,\bm{k})
\ee
where we have introduced the definitions
\be\label{DefGaunt}
{}^s
j_{\ell}^{(\ell' m')} (x) \equiv \sum_{L} {}^s C^{m' 0 m'}_{\ell L \ell'}
j_L(x)  \ii^{L+\ell'-\ell} \sqrt{\frac{(4 \pi)(2 L +1)}{(2 \ell + 1)(2
    \ell' +1)}}\,,
\ee
and
\be
\label{C_Gaunt}
{}^s C^{m_1 m_2 m_3}_{\ell_1  \ell_2 \ell_3} \equiv \int \dd^2 \Omega Y^{s,\star}_{\ell_1 m_1}(\bm{n}_\obs) 
Y_{\ell_2 m_2}(\bm{n}_\obs) Y^s_{\ell_3 m_3}(\bm{n}_\obs)\,.
\ee

The dynamical evolution and the initial conditions of the source can be split as
\be
\label{source_transfer}
S^{X_s}_{\ell m}(\chi_{_{\rm S}},\chi,\bm{k})= T^{X_s}_{\ell m}(\chi_{_{\rm S}},\chi,\bm{k})\Phi_i(\bm{k})\,,
\ee
where $T^{X_s}_{\ell m}$ is the (anisotropic) transfer function and $\Phi_{i}(\bm{k})$ is the primordial gravitational potential. Then, assuming that anisotropies are induced at late-time evolution only, the statistics of the primordial power spectrum must obey
\be
\label{iso_p_of_k}
\langle\Phi_i(\bm{k})\Phi^\star_i(\bm{q})\rangle = P(k)\delta^3(\bm{k}-\bm{q})\,,
\ee
with $P(k)$ being the (isotropic) primordial power spectrum. To account for the angular dependence of the transfer functions, we further decompose them
as
\be\label{e.def_ccc}
T^{X_s}_{\ell m}(\chi_{_{\rm S}},\chi,,\bm{k}) = \sum_{L,M} {}^{X_s}\!T^{LM}_{\ell m}(\chi_{_{\rm S}},\chi,k) Y_{LM}(\hat{\bm{k}})\,.
\ee

\subsection{Expression of the two-point angular correlators}\label{sec4.2}

These formulas can now be combined (using in particular Eq.~(\ref{gaunt_ylm}) to integrate out all spherical harmonics) to give an expression for the correlation between the multipoles of two different observables $X^{s_1}$ and $Z^{s_2}$. We find
\bea
\label{cov_matrix}
&&
\langle X^{s_1}_{\ell_1 m_1}(\chi_{_{\rm S1}}) Z^{s_2\,\star}_{\ell_2 m_2} (\chi_{_{\rm S2}})\rangle 
=
\frac{2}{\pi}(\ii)^{\ell_1}(-\ii)^{\ell_2}\int_0^\infty \dd k k^2 P(k)\nonumber\\
&&\qquad\qquad\int_0^{\chi_{_{\rm S1}}} \dd \chi_1 \int_0^{\chi_{_{\rm S2}}} \dd\chi_2 
\sum_{\ell,m}\sum_{\substack{\ell'_1,m'_1\\ \ell'_2,m'_2}}\sum_{\substack{L_1,M_1\\L_2,M_2}} (-1)^{m_1' +m_2'}
{}^{s_1}\!j_{\ell_1}^{(\ell_1' m_1')}(k\chi_1) \, {}^{s_2}\!j_{\ell_2}^{(\ell_2' m_2')\star}(k \chi_2)
\nonumber\\
&&
\qquad\qquad{}^{-m_1'} C^{m_1 m M_1}_{\ell_1 \ell L_1} \; {}^{-m_2'} C^{m_2 m M_2}_{\ell_2 \ell L_2}   
\;{}^{X_{s_1}}\!T^{L_1M_1}_{\ell_1 m_1}(\chi_{_{\rm S1}},\chi_1,k)
\;{}^{Z_{s_2}}\!T^{L_2M_2\,\star}_{\ell_2 m_2}(\chi_{_{\rm S2}},\chi_2,k)\,.
\eea

A central quantity in this description is the two-point correlation function of the $E$- and $B$-modes of a given spin-2 observable (as, for instance, the cosmic shear $\gamma$). This expression requires the decomposition of $X^{\pm s}$, ${}^{\pm s}j_{\ell}^{(\ell'm')}$ and $S^{X_s}_{\ell m}$ in their even/odd parity pieces as
\bea
X^{s}_{\ell m}(\chi) & = & E^{X_s}_{\ell m }(\chi)+\ii\,\hbox{sgn}(s) B^{X_s}_{\ell m}(\chi)\,, \label{dec_EB}\\
{}^{\pm s} j_{\ell}^{(\ell'm')}(x) & = & {}^{|s|}\epsilon^{(\ell'm')}_\ell(x)+\ii\,\hbox{sgn}(s){}^{|s|}\beta^{(\ell'm')}_\ell(x)\,, \\
S^{X_s}_{\ell m}(\chi_{_{\rm S}},\chi,\bm{k})& = & [T^{E^{X_s}}_{\ell m }(\chi_{_{\rm S}},\chi,\bm{k})+\ii\,\hbox{sgn}(s) T^{B^{X_s}}_{\ell m}(\chi_{_{\rm S}},\chi,\bm{k})]\Phi_i(\bm{k})\,.
\eea
Note that a spin $s=2$ field will have both $E$- and $B$-modes, while a scalar ($s=0$) field will only have the $E$ mode, so that ${}^{0}\beta^{(\ell'm')}_\ell=0$ and $B^{X_0}_{\ell m}=0$.

From these expressions and Eq.~(\ref{X_s}), one can verify that
\bea
E^{X_s}_{\ell m}(\chi_{_{\rm S}}) & =& \sqrt{\frac{2}{\pi}}\int \dd^3 \bm{k}\int_0^{\chi_{_{\rm S}}} \dd \chi
 \sum_{m'} D^\ell_{m m'}(R_{\bm{k}})\ii^\ell \sqrt{\frac{(2 \ell+1)}{4 \pi}} \\
 &&\qquad\sum_{\ell'} \left[{}^{|s|}\epsilon^{(\ell'm')}_m(k\chi)T^{E^{X_s}}_{\ell'm'}(\chi_{_{\rm S}},\chi,\bm{k})-{}^{|s|}\beta^{(\ell'm')}_m(k\chi)T^{B^{X_s}}_{\ell'm'}(\chi_{_{\rm S}},\chi,\bm{k})\right]\Phi_i(\bm{k})\,,  \nonumber\\
B^{X_s}_{\ell m}(\chi_{_{\rm S}}) & =& \sqrt{\frac{2}{\pi}}\int \dd^3 \bm{k}\int_0^{\chi_{_{\rm S}}} \dd\chi
\sum_{m'} D^\ell_{m m'}(R_{\bm{k}})\ii^\ell 
\sqrt{\frac{(2 \ell+1)}{4 \pi}}\\
&&\qquad \sum_{\ell'} 
\left[{}^{|s|}\epsilon^{(\ell'm')}_m(k\chi)T^{B^{X_s}}_{\ell'm'}(\chi_{_{\rm S}},\chi,\bm{k})+{}^{|s|}\beta^{(\ell'm')}_m(k\chi)T^{E^{X_s}}_{\ell'm'}(\chi_{_{\rm S}},\chi,\bm{k})\right]\Phi_i(\bm{k})\,. \nonumber
\eea
Then, the $EE$ and $BB$ covariance matrices can be computed by simply taking appropriate combinations of $X^{\pm s}_{\ell m}$. In order to simplify the notation we define
\bea
{\cal M}^{AC A'C'}_{\ell_1m_1\ell_2m_2}(\chi_{_{\rm S1}},\chi_{_{\rm S2}}) &\equiv& 
\frac{2}{\pi}(\ii)^{\ell_1}(-\ii)^{\ell_2}\int_0^\infty \dd k k^2 P(k)\int_0^{\chi_{_{\rm S1}}} \dd \chi_1 \int_0^{\chi_{_{\rm S2}}}  \dd\chi_2\\
&&
 \sum_{\ell,m}\sum_{\substack{\ell'_1,m'_1\\ \ell'_2,m'_2}}\sum_{\substack{L_1,M_1\\L_2,M_2}} (-1)^{m_1' +m_2'}
 \,{}^{A'} j_{\ell_1}^{(\ell_1' m_1')}(k\chi_1)
 \, {}^{C'} j_{\ell_2}^{(\ell_2' m_2')\star}(k\chi_2)
\nonumber \\
&&
{}^{-m_1'} C^{m_1 m M_1}_{\ell_1 \ell L_1} \;\, {}^{-m_2'} C^{m_2 m M_2}_{\ell_2 \ell L_2} 
\;{}^AT^{L _1M_1}_{\ell_1'  m_1'}(\chi_{_{\rm S1}},\chi_1,k)\,
{}^CT^{L_2M_2,\star}_{\ell_2'  m_2'}(\chi_{_{\rm S2}},\chi_2,k)\nonumber
\eea
together with the notation
\[
{}^{A}\!j_{\ell}^{(\ell'm')}=\begin{cases}
{}^{|s|}\epsilon_{\ell}^{(\ell'm')} & \quad\mbox{if}\quad A=E^{X_s}\,,\\
{}^{|s|}\beta_{\ell}^{(\ell'm')} & \quad\mbox{if}\quad A=B^{X_s}\,.
\end{cases}
\]
Thus, the $EE$ and $BB$ correlations become
\bea
\label{ee_cov_matrix}
\langle E^{X_s}_{\ell_1m_1}E^{X_s\star}_{\ell_2m_2}\rangle &=& \sum_{A,C} 
{\cal M}^{ACAC}_{\ell_1m_1\ell_2m_2}
\left[\delta^E_A\delta^E_C+\delta^B_A\delta^B_C-\delta^E_A\delta^B_C-\delta^B_A\delta^E_C\right]\,,\\
\langle B^{X_s}_{\ell_1m_1}B^{X_s\star}_{\ell_2m_2}\rangle &=& \sum_{A,C} 
{\cal M}^{AC\bar A\bar C}_{\ell_1m_1\ell_2m_2}
\left[\delta^E_A\delta^E_C+\delta^B_A\delta^B_C+\delta^E_A\delta^B_C+\delta^B_A\delta^E_C\right]\,,
\eea
where, in the last equality, we have introduced the notation according to which $\bar A$ equals $E^{X_s}$ (resp. $B^{X_s}$) whenever $A$ is equal to $B^{X_s}$ (resp. $E^{X_s}$); the same holding for $\bar C$. The $EB$ correlation can be computed using the same method.

The expression (\ref{ee_cov_matrix}) is quite general. Let us first focus on its diagonal part, which can be characterized by the following estimators of
\bea
C_\ell^{EE}(\chi_{_{\rm S1}},\chi_{_{\rm S2}}) &=& \frac{1}{2\ell+1}\sum_{m}\langle E^{X_s}_{\ell m}(\chi_{_{\rm S1}})E^{X_s\,\star}_{\ell m}(\chi_{_{\rm S2}})\rangle\label{def_CEE}\,,\\
C_\ell^{BB}(\chi_{_{\rm S1}},\chi_{_{\rm S2}}) &=& \frac{1}{2\ell+1}\sum_{m}\langle B^{X_s}_{\ell m}(\chi_{_{\rm S1}})\,B^{X_s\,\star}_{\ell m}(\chi_{_{\rm S2}})
\rangle\,.\label{def_CBB}
\eea
The angular power spectra are then given by simpler expressions
\bea
C^{EE}_{\ell}(\chi_{_{\rm S1}},\chi_{_{\rm S2}})  &= &\frac{2}{\pi}\int_0^\infty \frac{\dd k\,k^2}{4\pi} P(k) \sum_{m,L,M,A,C} 
\left[\delta^E_A\delta^E_C+\delta^B_A\delta^B_C-\delta^E_A\delta^B_C-\delta^B_A\delta^E_C\right]\nonumber\\
&&
\qquad\times\left[\sum_{\ell_1}\int_0^{\chi_{_{\rm S1}}}\dd\chi_1\,{}^AT^{LM}_{\ell_1  m}(\chi_{_{\rm S1}},\chi_{_{\rm S2}},k)  
\;{}^{A}\!j_{\ell}^{(\ell_1 m)}(k\chi_1)\right]\nonumber \\
&&\qquad \times\left[\sum_{\ell_2}\int_0^{\chi_{_{\rm S2}}}\dd\chi_2\,{}^CT^{LM}_{\ell_2  m} (\chi_{_{\rm S1}},\chi_{_{\rm S2}},k) 
\;{}^{C}\!j_{\ell}^{(\ell_2 m)}(k\chi_2)\right]^\star
\eea
with, again, a similar expression for the $B$-modes.

The case where the transfer functions are isotropic is easily recovered. To see that, let us consider the simpler situation where  $X^{s_1}=Z^{s_2}=\Theta$, with $\Theta$ being the CMB temperature fluctuations. Since $\Theta$ is a scalar, then it is a pure $E$-mode with no $B$-mode. In the previous formalism, we just need to set ${}^ET^{LM}_{\ell  m}={}^\Theta T^{LM}_{\ell  m}$ and ${}^BT^{LM}_{\ell  m}=0$.
Then, using ${}^\Theta \!T^{LM}_{\ell m} = \sqrt{4 \pi }T^\Theta_{\ell m} \delta^{L 0}\delta^{M 0}$, we get from Eq.~(\ref{cov_matrix}) the standard result
\be
C^{\Theta\Theta}_{\ell} = \frac{2}{\pi} \int_0^\infty k^2 \dd k  P(k) \sum_m \left|
  \int_0^\infty \dd \chi' \sum_{\ell'}T^\Theta_{\ell' m}(k,\chi') j_{\ell}^{(\ell' m)}(k \chi')\right|^2\,,
\ee
where in this specific case it must be understood that the visibility function is included in the transfer functions $T^\Theta_{\ell m}(k,\chi')$. If we consider only scalar sources, then only the $m=0$ mode contributes. Analogously, if we also have sources with no intrinsic direction (like, for example, no Doppler effect in the CMB), then we have 
$\ell'=0$.

\subsection{Implication of spatial parity}\label{sec4.3}

We would like to briefly elucidate the relationship between the symmetries of the underlying background spacetime and the cross-correlation functions of different observables. In particular, we want to show that (spatial) parity symmetry implies that the diagonal piece of the $EB$ correlation matrix is zero, while off-diagonal terms may not necessarily be.

We start by noticing that under a parity inversion $\{x,y,z\}\rightarrow\{-x,-y,-z\}$, or, equivalently, $\{\bm{n}_\obs,\bm{n}_\theta,\bm{n}_\phi\}\rightarrow\{-\bm{n}_\obs,-\bm{n}_\theta,\bm{n}_\phi\}$, the polarization vectors transform as
\be
\bm{n}^{\pm}_\obs(\bm{n}_\obs)\;\rightarrow\; - \bm{n}^{\mp}_\obs(\bm{n}_\obs)\,.
\ee
This implies that the polarization basis should transform under parity as
\be
\bm{m}^s_\obs(\bm{n}_\obs) \;\rightarrow\; (-1)^s\bm{m}^{-s}_\obs(\bm{n}_\obs)\,. 
\ee
Moreover, the sources transform as
\be\label{StoS}
S^{X_s}(\chi_{_{\rm S}},\chi,\bm{k},\bm{n}_\obs)\bm{m}^s_\obs(\bm{n}_\obs) 
\;\rightarrow\;
(-1)^s S^{X_s}(\chi_{_{\rm S}},\chi,-\bm{k},-\bm{n}_\obs)\bm{m}^{-s}_\obs(\bm{n}_\obs)\,.
\ee
We now demand that any physical quantity remains invariant under a
full parity inversion. That is, if at the same time we transform
$\bm{k}\rightarrow-\bm{k}$, $\bm{n}_\obs\rightarrow-\bm{n}_\obs$ and
$\bm{m}^s_\obs\rightarrow (-1)^s\bm{m}^{-s}_\obs$, the source
$S^{X_s}$ of a physical observable $X^s$ should remain invariant,
which from Eqs.~(\ref{source_fourier}) and~\eqref{StoS} implies the condition
\be
S^{X_s}(\chi_{_{\rm S}},\chi,\bm{k},\bm{n}_\obs) = (-1)^s S^{X_{-s}}(\chi_{_{\rm S}},\chi,-\bm{k},-\bm{n}_\obs)\,.
\ee
If we take into account the parity transformations of the Wigner
matrices and of the spherical harmonics -- see
Eqs.(\ref{parity1}-\ref{parity2}) -- then a comparison of the previous
expression with Eq.~(\ref{source_fourier_ylm}) shows that the parity
condition translates to
\be
\label{parity_source}
S^{X_s}_{\ell m}(\chi_{_{\rm S}},\chi,\bm{k}) = 
(-1)^{m+s}S^{X_{-s}}_{\ell,-m}(\chi_{_{\rm S}},\chi,-\bm{k})\,.
\ee
Then, we rewrite Eq.~(\ref{X_s}) as
\be\label{X_s_fourier}
X^s_{\ell m}(\chi_{_{\rm S}}) = \sqrt{\frac{2}{\pi}}\int\dd^3\bm{k}\,X^s_{\ell m}(\chi_{_{\rm S}},\bm{k})\Phi_i(\bm{k})\,.
\ee
The above expression should be seen as a definition of $X^s_{\ell m}(\chi_{_{\rm S}},\bm{k})$, and  corresponds to the contribution of each Fourier mode to the observable, but it is not its Fourier component. Its expression can be obtained by plugging Eq.~(\ref{source_transfer}) into Eq. (\ref{X_s}) and then comparing  with Eq.~(\ref{X_s_fourier}). Then, if we impose the symmetry (\ref{parity_source}) to Eq.~(\ref{X_s_fourier}), using again the parity transformation of the Wigner matrices, we finally find that
\be
X^{s}_{\ell m}(\chi_{_{\rm S}},\bm{k}) = (-1)^{\ell+s}X^{-s}_{\ell m}(\chi_{_{\rm S}},-\bm{k})\,.
\ee
From this expression, it is straightforward to check that the $E$- and $B$-modes of a spin-2 quantity transform under parity as
\bea
E^{X_2}_{\ell m}(\chi,\bm{k}) & = &\frac{1}{2}\left(X^{2}_{\ell m}(\chi,\bm{k})+X^{-2}_{\ell m}(\chi,\bm{k})\right) = (-1)^\ell E^{X_2}_{\ell m}(\chi,-\bm{k})\, \\
B^{X_2}_{\ell m}(\chi,\bm{k}) & =& \frac{-\ii}{2}\left(X^{2}_{\ell m}(\chi,\bm{k})-X^{-2}_{\ell m}(\chi,\bm{k})\right) = (-1)^{\ell+1} B^{X_2}_{\ell m}(\chi,-\bm{k})\,.
\eea
We can now establish our main conclusion: given the above symmetry, together with translational invariance of primordial fluctuations (see Eq.~(\ref{iso_p_of_k})), it follows that, for the $E$ and $B$ modes of a spin-2 field, we have
\begin{align}
\langle E^{X_2}_{\ell_1 m_1}(\chi)B^{X_2\,\star}_{\ell_2 m_2}(\chi)\rangle & = \frac{2}{\pi}\int \dd^3\bm{k}\,\langle E^{X_2}_{\ell_1m_1}(\chi,\bm{k})B^{X_2\,\star}_{\ell_2m_2}(\chi,\bm{k})\rangle P(k) \\
& = (-1)^{\ell_1+\ell_2+1}\,\frac{2}{\pi}\int \dd^3\bm{k}\,\langle E^{X_2}_{\ell_1m_1}(\chi,-\bm{k})B^{X_2\,\star}_{\ell_2m_2}(\chi,-\bm{k})\rangle P(k) \nonumber\\
& = (-1)^{\ell_1+\ell_2+1}\,\frac{2}{\pi}\int \dd^3\bm{k}\,\langle E^{X_2}_{\ell_1m_1}(\chi,\bm{k})B^{X_2\,\star}_{\ell_2m_2}(\chi,\bm{k})\rangle P(k) \nonumber\\
& = (-1)^{\ell_1+\ell_2+1}\,\langle E^{X_2}_{\ell_1 m_1}(\chi)B^{X_2\,\star}_{\ell_2 m_2}(\chi)\rangle\,,\nonumber
\end{align}
where, from the second to the third line, we have used $\int_{-\infty}^{+\infty}\dd^3\bm{k}=\int_{+\infty}^{-\infty}\dd^3(-\bm{k})$. Similarly, one can show that the $EE$ and $BB$ covariance matrices obey
\bea
\langle E^{X_2}_{\ell_1m_1}E^{X_2\,*}_{\ell_2m_2}\rangle & = &
(-1)^{\ell_1+\ell_2}\langle E^{X_2}_{\ell_1m_1}E^{X_2\,*}_{\ell_2m_2}\rangle\,, \\
\langle B^{X_2}_{\ell_1m_1}B^{X_2\,*}_{\ell_2m_2}\rangle & = &
(-1)^{\ell_1+\ell_2}\langle B^{X_2}_{\ell_1m_1}B^{X_2\,*}_{\ell_2m_2}\rangle\,.
\eea
We have thus proved that correlations between $E$ and $B$ modes will vanish whenever $\ell_1+\ell_2$ is an even number. In particular, the diagonal part of the $EB$ covariance matrix is always zero in spacetimes that respect parity (but not necesserily isotropy). Evidently, the same holds for the multipolar coefficients of a spin-0 quantity, such as the CMB temperature $a_{\ell m}$s (see e.g. Ref.~\cite{Abramo:2010gk}).

\section{Perturbation scheme in the small shear limit}\label{sec6}

\subsection{Expansion scheme}\label{subsec6.0}

The structure of the computation has been detailed in \S~\ref{subsec1.2}. Let us recall that order by order, we need to
\begin{enumerate}
\item solve the geodesic equation perturbatively in order to determine the displacement from the reference Friedmann-Lema\^{\i}tre geodesic, $x^{i\{n,p\}}(\chi,{\bm n}^\obs)$, and the local direction of propagation, ${\bm n}^{\{n,p\}}(\chi,{\bm n}^\obs)$. Note that $x^{i\{n,p\}}(\chi,{\bm n}^\obs)$ is split in a radial component, $\delta r^{\{n,p\}}(\chi,{\bm n}^\obs)$, and an orthoradial component which will be related to the deflection angle $\alpha^{a\{n,p\}}(\chi,{\bm n}^\obs)$;
 \item determine the transport of the Sachs basis, ${\bm n}_a^{\{n,p\}}(\chi,{\bm n}^\obs)$;
 \item expand the Sachs equation and determine the source terms for ${\cal D}_{ab}^{\{n,p\}}(\chi,{\bm n}^\obs)$;
 \item determine the evolution of the perturbations at the required order;
 \item perform the multipolar expansion in terms of the direction of observation ${\bm n}^\obs$.
\end{enumerate}

To avoid confusion, we shall use the notation that $X^{\{n,p\}}$ includes all terms up to order $\{n,p\}$ while $\delta X^{\{n,p\}}$ contains only the terms of order $\{n,p\}$.

\subsection{Order \{0,0\}}\label{subsec6.1}

Since on the background (i.e., a Friedmann-Lema\^{\i}tre spacetime), the metric is just the Minkowski metric, thanks to the overall conformal transformation described in \S~\ref{sec_conforme}, the (conformal) Riemann tensor vanishes, so that ${\cal R}_{ab}^\zz=0$.  Since the wave-vector is decomposed in accordance to Eq.~(\ref{e.34}), in which we can always choose to set $U_\obs=1$, one deduces that it is given by
\be 
k^{0\zz}=-1\,,\quad  k^{i\zz}=n_{\obs}^{\ti}\,
\ee
The Sachs equation~(\ref{MasterBonvin}) trivially reduces to
\be
\frac{\dd^{2}{\cal D}^\zz_{ab}}{\dd\chi^{2}}=0\,
\ee
so that the Jacobi matrix is given by
\be 
 {\cal D}_{ab}^\zz = D^\zz_A(\chi) I_{ab}, \qquad D_A^{\zz}(\chi)=\chi,
\ee
and its components reduce to
\be 
\kappa^{\zz} = \gamma_{ab}^{\zz}=V^{\zz}=0\,.
\ee
This completely specifies the property of the geodesic bundle at the background level.

\subsection{Order \{1,0\}}\label{subsec6Order10}

At this order, the spacetime remains homogeneous, but it now has an anisotropic perturbation described by the shear $\sigma_{ij}$, from which we can define a scalar field $\Sigma$ on the 2-sphere by
\be\label{DefSigmaB}
\Sigma(\chi) \equiv \frac{1}{2}\sigma_{\ti \tk}(\chi)n_{{\obs}}^{{\ti}} n^{{\tk}}_{{\obs}}\,.
\ee
We also introduce a new scalar function,
\be\label{DefSigmaBbis}
\MyB(\chi) \equiv \frac{1}{2}\beta_{ik}(\chi)n_{{\obs}}^{{\ti}} n^{{\tk}}_{{\obs}}\,,
\ee
where $\beta_{ij}$ is defined in Eqs.~(\ref{e.defbetaij}) and~(\ref{DefMatrixBeta}).

We will now show that all results at this order can be expressed in terms of these two fields on the unit 2-sphere and the covariant derivative $D_a$ defined in \S~\ref{SecDtangent}. In what follows we shall use the convention $\MyB_\obs=\MyB(\chi=0)$.

\subsubsection{Geodesic equation: tangent vector}

At this order, the 4-velocity of a fundamental observer is just $u_\mu = (\dd \eta)_\mu = \thetradUD{\tz}{\,\,\mu}$, so that 
\be
U =k^\mu u_\mu = k^0=k^\tz\,.
\ee
From Eq.~\eqref{MasterGeodesic}, and using the fact that, at first order, the only non-vanishing Christoffel symbols are~\cite{ppu1}
\be
\delta\Gamma_{ij}^{0\oz}= \sigma_{ij}\,,\qquad\delta\Gamma_{0j}^{i\oz}=\sigma_{\; j}^{i}\,,
\ee
we obtain that
\be\label{eq_k_oz}
\frac{\dd
  k^{\tz\oz}}{\dd\chi}=\frac{\dd
  k^{0\oz}}{\dd\chi}=-\sigma_{\ti \tk}n_{\obs}^{\ti}n_{\obs}^{\tk}\,.
\ee
It thus follows that
\be\label{eq_k_oz2}
k^{\tz\oz} =k^{0\oz}=-1+2 [\MyB(\chi)- \MyB_\obs]\equiv-1 + \delta k^{0\oz}\,.
\ee
This result is expected given, that for a Bianchi $I$ space-time, $k_i$ is a constant~\cite{Fleury2014}; see \S~\ref{subsec3.4a}. Alternatively, this result could have been obtained using Eq.~\eqref{MasterGeodesicTetrads}, with the $\oz$ order of the affine connections given in Appendix~\ref{AppAffineConnections}. Its physical interpretation is simple since the factor $2 [\MyB(\chi)- \MyB_\obs]$ can be identified to the Einstein effect between the events of emission and reception.

The evolution of the spatial components of the wave-vector is easily obtained using the tetrad components first. From Eq.~\eqref{MasterGeodesicTetrads} we obtain
\bea\label{e.ki010}
k^{\ti \oz}(\chi) &=& n_\obs^\ti + \int_0^\chi {\sigma^\ti}_\tj(\chi') n_\obs^\tj \dd \chi'=n_\obs^\ti+\left[\beta^i_{\,j}(0)-\beta^i_{\,j}(\chi)\right]n^\tj_\obs\\
k^{i \oz}(\chi) &=& k^{\ti \oz}(\chi)-\beta^i_{\,j}(\chi) n^\tj_\obs=n_\obs^\ti+\left[\beta^i_{\,j}(0)-2\beta^i_{\,j}(\chi)\right]n^\tj_\obs\\
k_{i}^{\oz}(\chi) &=& k_\ti^{\oz}(\chi)+\beta_{ij}(\chi) n^\tj_\obs=n_\obs^\ti+\beta^i_{\,j}(0)n^\tj_\obs
\eea
Again, this corresponds to the small shear limit of our previous general result~\cite{Fleury2014}, where $k_i=k_i(0)$ is used first.

\subsubsection{Geodesic equation: real space}

The parametric equation of the geodesic is obtained from the integration of Eq.~\eqref{MasterPosition} at order $\oz$. Separating the difference between the position at order $\oz$ and the position of the background geodesic into a radial displacement and an orthoradial displacement
according to
\be\label{defxi10}
{x^i}^{\oz}(\chi) =\chi  n_\obs^\ti + \delta x^{i\oz} \quad \hbox{with}\quad \delta x^{i\oz} =
n_\obs^\ti \delta r^\oz + \chi \alpha ^{a \oz} {n_a^\obs}^\ti,
\ee
which defines the deflection angle $\alpha ^{a \oz}$, and where we have used that ${x^i}^{\zz}(\chi) =\chi  n_\obs^\ti$, we get
\bea\label{xi10}
\delta r^\oz(\chi) &=&  -2 \int_0^\chi \MyB(\chi') \dd\chi'\,, \\
\alpha^{a \oz}(\chi)&=& D^a \MyB_\obs -\frac{2}{\chi} \int_0^\chi D^a \MyB(\chi') \dd \chi'\slabel{xi10bis}
\eea
in which $D^a \MyB_\obs$ stands for $(D^a \MyB)_{\chi=0}$, and where the last equalities of the equations above made use of an integration by parts. Note that 
\be
\lim_{\chi \to 0} \frac{x^{i\oz}}{\chi} =n_\obs^\ti - \beta_{ij}(0) n_\obs^\tj \,.
\ee
This can be interpreted simply, because it means that very close to the observer, everything happens as if $\beta_{ij}$ is constant and equal to $\beta_{ij}(0)$. Thus a constant change of coordinates $\tilde x^i \equiv x^i + \beta^i_j(0) x^j$ transforms the metric from $\gamma_{ij}=\delta_{ij}+2\beta_{ij}(0)$ to the Euclidian metric $\delta_{ij}$. The geodesic in these new coordinates is simply the Euclidian one, $\tilde x^{i}(\chi) = \chi n_\obs^\ti$.

\subsubsection{Evolution of the direction and screen vectors}

The infinitesimal change of a unit vector lies in the plan orthogonal to it. The perturbation of the direction vector is thus of the form
\be\label{def_varpi00}
n^{\ti \oz} = n_\obs^\ti + \myaleph^{a \oz}{n^\obs_a}^\ti\,.
\ee
From the previous results for $k^\ti$ and $k^\tz$, we get immediately 
\be
n^{\ti \oz} = n_\obs^\ti + \myaleph^{a \oz}{n^\obs_a}^\ti\,,\qquad \myaleph^{a \oz}=\int_0^\chi D^a \Sigma \,\dd \chi'\,.
\ee
The transport equation for the screen basis is in turn given by
\be\label{def_varpi}
n_a^{\ti \oz} ={n^\obs_a}^{\ti} -n_\obs^\ti \myaleph_a^{\oz} \,,
\ee
and it can be checked that the screen basis~(\ref{e:prop_n}) does remain orthogonal to the direction vector.

\subsubsection{Sachs equation}

At order $\oz$, given that ${\cal R}_{ab}^\oz=0$, the right hand side of Eq.~\eqref{MasterBonvin} reduces to
$$
{\cal R}_{ab}^\oz{\cal D}_{bc}^\zz={\cal  R}_{ab}^\oz \chi\delta_{bc} = \chi {\cal  R}_{ac}^\oz\,,
$$
so that the Sachs equation~\eqref{MasterBonvin} reduces to
\be \label{eq_jacobimap_zo}
\frac{\dd^{2}{\delta\cal D}_{ab}^\oz}{\dd \chi^{2}}= \frac{\dd \delta k^{0\oz}}{\dd\chi}\delta_{ab} + \chi{\cal R}_{ab}^{\oz}\,.
\ee
Its first integral yields
\be \label{eq_jacobimap_zo2}
\frac{\dd{\delta \cal D}_{ab}^\oz}{\dd \chi}= \delta_{ab}+ \int_0^\chi\left(\frac{\dd \delta k^{0\oz}}{\dd\tilde\chi}\delta_{ab} + \tilde\chi{\cal R}_{ab}^{\oz}\right)\dd\tilde\chi\,.
\ee
The first term gives $\delta_{ab}[1+\delta  k^{0\oz}(\chi)] = \delta_{ab}[2+  k^{0\oz}(\chi)]$, so that
\be
\label{jacobmatrix10}
\delta{\cal D}_{ab}^\oz(\chi,{\bm n}^\obs)= \delta_{ab} \int_{0}^{\chi}\left[2+k^{0\oz}(\tilde\chi)\right]\dd\tilde\chi
 +\int_0^\chi \dd\chi' \int_0^{\chi'} \tilde \chi{\cal R}_{ab}^\oz(\tilde\chi,{\bm n}^\obs)\dd\tilde \chi\,.
\ee
The double integral on the right hand side can be performed by means of an integration by parts. This gives
$$
\int_0^\chi \dd\chi' \int_0^{\chi'} \tilde \chi{\cal R}_{ab}^\oz(\tilde\chi,{\bm n}^\obs)\dd\tilde \chi = 
\int_0^\chi \tilde\chi(\chi-\tilde\chi){\cal R}_{ab}^\oz(\tilde\chi,{\bm n}^\obs)\dd\tilde\chi
$$
from where we finally conclude that
\be
\label{sol_jacobimap_zo}
\delta{\cal D}_{ab}^\oz(\chi,{\bm n}^\obs)=\int_{0}^{\chi}\left\lbrace\left[2+k^{0\oz}(\tilde\chi)\right]\delta_{ab}
+\left(\chi-\tilde \chi\right)\tilde \chi{\cal R}_{ab}^\oz(\tilde\chi,{\bm n}^\obs)\right\rbrace\dd\tilde \chi\,.
\ee

As detailed in Appendix~\ref{AppRiemannRicci}, the source term takes the form
\be
{\cal  R}_{ab}^{\oz}(\chi,{\bm n}^\obs)=-\frac{1}{2}(\sigma_{ij}'n_\obs^{i}n_\obs^{j})\delta_{ab}+n_{\langle
  a}^{\obs i}n_{b\rangle}^{\obs j}(\sigma_{ij})' = - \delta_{ab}\Sigma' + D_{\langle a} D_{b \rangle} \Sigma'\,.
\ee
By inserting the above in Eq. (\ref{sol_jacobimap_zo}) and using Eq.~(\ref{e.ki010}), we find the following expressions for the convergence and shear,
\bea
\kappa^{\oz}(\chi,{\bm n}^\obs)&=&-\int_0^\chi \frac{(\chi-\tilde{\chi})}{\chi}\left[2 \Sigma
  +\tilde{\chi}\Sigma'\right]\dd \tilde{\chi}\,,\\
\gamma^{\oz}_{ab}(\chi,{\bm n}^\obs)&=&\int_0^\chi \frac{(\chi-\tilde{\chi})\tilde{\chi}}{\chi}D_{\langle a} D_{b\rangle}
\Sigma' \dd\tilde{\chi}\,,
\eea
which simplify to
\bea
\kappa^{\oz}(\chi,{\bm n}^\obs)&=&-\MyB(\chi)-3\MyB_\obs+\frac{4}{\chi}\int_0^\chi \MyB(\tilde \chi)\dd \tilde{\chi}\,,\\
\gamma^{\oz}_{ab}(\chi,{\bm n}^\obs)&=&D_{\langle a} D_{b\rangle} \MyB_\obs +D_{\langle a} D_{b\rangle} \MyB -\frac{2}{\chi} \int_0^\chi D_{\langle a} D_{b\rangle} \MyB(\tilde \chi) \dd \tilde \chi\,.
\eea
Note that in the limit $\chi \to 0$, $\kappa^{\oz} \to 0$ and $\gamma^{\oz}_{ab} \to 0$ as it should be, given the initial condition~\eqref{initialconditions} for the Jacobi matrix.
Finally, since the Jacobi matrix~\eqref{jacobmatrix10} is symmetric, at this order we have
\be
 V^{\oz}=0\,.
\ee 
We have checked that these results match those found in~\cite{Fleury2014} when expanded in the small shear limit (where the special choice $\MyB_\obs=0$ is made). 

We can now perform the expansion of these observable quantities
in terms of spin-weighted spherical harmonics. Using the results of \S~\ref{SecEdth}, the shear can be
projected into the helicity basis so as to transform the covariant
derivatives into spin-raising and spin-lowering operators. The
spherical harmonics components of the convergence $\kappa$ and of the
cosmic shear $\gamma^\pm$ are then easily obtained at order
$\oz$. Both reduce to a quadrupolar contribution, inherited from the quadrupolar contribution of $\Sigma$, so that their only non-vanishing coefficients are
\bea
\kappa^{\oz}_{2 m}(\chi) &=& -3\MyB^\obs_{2m}-\MyB_{2m}+\frac{4}{\chi}\int_0^\chi \MyB_{2m}(\tilde \chi)\dd \tilde{\chi}\,,\\
\gamma^{\pm \oz}_{2 m}(\chi)&=&\sqrt{6}\left(\MyB^\obs_{2m}+\MyB_{2m} -\frac{2}{\chi} \int_0^\chi \MyB_{2m}(\tilde \chi) \dd \tilde \chi\right)\,.
\eea
We conclude that $B_{\ell m}^{\gamma\oz}=0$ and $E_{2 m}^{\gamma\oz}=\gamma^{\pm \oz}_{2 m}$.

\subsection{Order \{0,1\}}\label{subsec6.2}

We follow the same method for the order $\zo$ as for the order $\oz$. This corresponds to the standard approach to weak-lensing in the linear regime of cosmological perturbations. Our main goal is to rederive these standard results in our formalism, so as to serve as a basis for the study at order $\oo$. Note that, at this order, we only need to include scalar perturbations since, as stressed before, vectors and tensors modes are of order $\{1,1\}$.

\subsubsection{Geodesic equation: tangent vector}

Using the definition of the deflecting potential as $\varphi \equiv \Phi+\Psi$, the energy of a photon evolves according to
\be\label{defphipot}
\frac{\dd k^{\tz \zo}}{\dd \chi} = -\frac{\partial \varphi}{\partial
  \chi} + \frac{\dd \Phi}{\dd \chi}\,,\qquad \frac{\dd k^{0\zo}}{\dd \chi} = -\frac{\partial \varphi}{\partial
  \chi} + 2 \frac{\dd \Phi}{\dd \chi}\,,
\ee
the solution of which is
\be\label{dkodchi01}
 k^{0 \zo} = k^{\tz \zo}+\Phi = -1 + 2 \Phi-\Phi_\obs-\int_0^\chi \frac{\partial \varphi}{\partial \tilde \chi} \dd \tilde \chi\equiv-1+\delta k^{0 \zo}\,,
\ee
where it is understood the integrand is evaluated on the background
geodesic, i.e. that $x^i = \tilde \chi n_\obs^\ti$, at a time
associated with $\tilde \chi$. The notation is intentionally
simplified in this section, so, for example, $k^{0 ,\zo}$ means $k^{0\,
  \zo}(\chi,n_\obs^\ti)$,  $\Phi$ means $\Phi(\chi,x^i)$ with
$x^i=\chi n_\obs^\ti$, and so on. In other words, it is understood that everything is
evaluated on the background geodesic at parameter $\chi$. The only
exception is $\Phi_\obs$, which is the potential $\Phi$ evaluated at
the observer, that is, at $\chi=0$. Note that the total derivative $\dd/\dd \chi$, i.e. the total derivative along the background geodesic, satisfies
\be
\frac{\dd \varphi}{\dd\chi} = \frac{\partial \varphi}{\partial
  \chi}+\varphi_{,r}=-\varphi' + \hat x^i \partial_i\varphi \,,
\ee
since, from Eq.~(\ref{echieta}),
\be
\frac{\partial \varphi}{\partial \chi} = -\varphi'\,.
\ee

\subsubsection{Evolution of the direction and screen vectors}

The spatial component of the vector $k^\mu$ evolves according to
\be
\frac{\dd k^{\ti \zo}}{\dd \chi} = -\partial_i \varphi + n_\obs^\ti
\frac{\dd \Psi}{\dd \chi}\,,\qquad \frac{\dd k^{i \zo}}{\dd \chi} =  -\partial_i \varphi +2 n_\obs^\ti
\frac{\dd \Psi}{\dd \chi}\,,
\ee
the solution of which is
\be
k^{i \zo} =k^{\ti \zo}+n_\obs^\ti \Psi= n_\obs^\ti [1+2\Psi(\chi)-\Psi_\obs] - \int_0^\chi \partial_i \varphi \dd \tilde \chi\,.
\ee
Using Eq.~\eqref{kivskz}, we then deduce the evolution of the direction vector
\bea
\frac{\dd n^{\ti \zo}}{\dd \chi} &=&-S^{\ti \tj} \partial_j \varphi,
\eea
the solution of which is
\bea\label{varn01}
n^{\ti \zo}(\chi,n_\obs^\ti) = n_\obs^\ti + \myaleph^{a \zo}(\chi,n_\obs^\ti){n^\obs_a}^\ti,\qquad
\myaleph^{a \zo}(\chi,n_\obs^\ti) \equiv -\int_0^\chi D^a \varphi \dd \tilde \chi\,.
\eea
Similarly, the evolution of the screen projectors leads to
\be\label{varnsachs01}
n_a^{\ti \zo}(\chi,n_\obs^\ti) = {n^\obs_a}^\ti - \myaleph^{a \zo}(\chi,n_\obs^\ti){n_\obs}^\ti\,.
\ee

\subsubsection{Geodesic equation: real space}

We can then determine $x^{i\zo}$ from Eq.~(\ref{MasterPosition}) using
\be
\left(\frac{\dd x^i}{\dd \chi}\right)^{\zo} =
-\left(\frac{k^i}{k^0}\right)^{\zo} = n_\obs^\ti(1+\varphi)
-\int_0^\chi S^{\ti \tj} \partial_j \varphi \dd \tilde \chi
\ee 
and this leads to
\be\label{e.x01}
x^{i\zo} = \chi n_\obs^\ti + n_\obs^\ti \int_0^\chi 
\varphi \dd \tilde \chi-{n^\obs_a}^\ti\int_0^\chi \dd \tilde \chi \frac{(\chi-\tilde \chi)}{\tilde \chi}D^a \varphi\,.
\ee

\subsubsection{Sachs equation}

Finally, at order $\zo$, the right hand side of Eq.~\eqref{MasterBonvin} is simply  ${\cal  R}_{ab}^\zo {\cal D}_{bc}^\zz= \chi {\cal  R}_{ac}^\oz$. Thus, the Sachs equations becomes
\be
\frac{\dd^{2}{\delta\cal D}_{ab}^\zo}{\dd \chi^{2}}= \frac{\dd k^{0\zo}}{\dd\chi}\delta_{ab} + \chi{\cal R}_{ab}^{\zo}\,.
\ee
The solution of the Sachs equation follows formally the same steps as in the case $\oz$. That is, it can be integrated twice, and after an integration by parts for the double integral over the Riemann term, we get
\be\label{sol_jacobimap_zomm}
\delta{\cal D}_{ab}^\zo(\chi,{\bm n}^\obs)=\int_{0}^{\chi}\left\lbrace\left[2+k^{0\zo}(\tilde\chi)\right]\delta_{ab}
+\left(\chi-\tilde \chi\right)\tilde \chi{\cal R}_{ab}^\zo(\tilde\chi,{\bm n}^\obs)\right\rbrace\dd\tilde \chi\,.
\ee
Now, using the perturbed expression for ${\cal R}_{ab}$ found in the Appendix~\ref{AppRiemannRicci} (with $\sigma_{ij}=0$), 
\bea
\chi^2 {\cal  R}_{ab}^{\zo} &=& \chi^2 {n^\obs_a}^\ti{n^\obs_b}^\tj\left[-\partial_i \partial_j \varphi -\delta_{ij}\left(\Psi'' -2
    n^\ti \partial_i\Psi'  + n^\tp n^\tq \partial_p \partial_q \Psi \right) \right]\\
&=& -D_{\langle a} D_{b\rangle}\varphi -\delta_{ab}\chi^2\left[\frac{1}{2}\partial_i \partial^i \varphi + \Psi'' -2
    n^\ti \partial_i\Psi'  + \frac{1}{2}n^p n^q \partial_p \partial_q (\Psi-\Phi) \right]
\eea
and the expression ${\dd k^{0\oz}}/{\dd\chi}$ given in Eq.~(\ref{dkodchi01}), one obtains the formal solution of the Sachs equation~(\ref{MasterBonvin}) as~\cite{Bonvin2010}
\begin{eqnarray}
\label{DabOrder01}
\delta{\cal D}_{ab}^{\zo} &=&\chi \left[\delta_{ab}\left(1-\Psi(\chi)
  -\Phi_\obs+\frac{1}{\chi}\int_0^\chi \varphi(\tilde \chi) \dd \tilde \chi - \frac{1}{2}\int_0^\chi
  \frac{\chi-\tilde \chi}{\chi \tilde \chi} D_c D^c \varphi(\tilde \chi) \dd \tilde \chi\right)\nonumber\right.\\
&&\qquad \left.-\int_0^\chi \frac{\chi-\tilde \chi}{\chi\tilde \chi} D_{\langle a} D_{b \rangle}
\varphi(\tilde \chi) \dd \tilde \chi\right]
\end{eqnarray}
from which $\kappa$ and $\gamma_{ab}$ can be read directly from the expression in brackets in the first and second lines respectively; see our definitions in Eq.~\eqref{DecompositionDab}. Note that, since there is no antisymmetric part in $\delta{\cal D}^{\zo}_{ab}$, we conclude that $V^{\zo}=0$.

Dropping the (unobservable) monopole correction due to the local potential $\Phi_\obs$, we get their multipoles as
\bea
\kappa^{\zo}_{\ell m}&=&-\Psi_{\ell m}(\chi)+\frac{1}{\chi}\int_0^\chi \varphi_{\ell m}(\tilde \chi) \dd \tilde \chi + \frac{\ell(\ell+1)}{2}\int_0^\chi \frac{\chi-\tilde \chi}{\chi \tilde \chi} \varphi_{\ell m}(\tilde \chi) \dd \tilde \chi\,,\slabel{Eqkappa01}\\
\gamma^{\pm \zo}_{\ell m} &=& -\frac{1}{2}\sqrt{\frac{(\ell+2)!}{(\ell-2)!}}\int_0^\chi \frac{\chi-\tilde \chi}{\chi \tilde \chi}\varphi_{\ell m}(\tilde \chi) \dd \tilde \chi\,.
\eea
From which we conclude that $B_{\ell m}^{\gamma\oz}=0$ and $E_{\ell m}^{\gamma\oz}=\gamma^{\pm \oz}_{\ell m}$.

\subsubsection{Angular power spectra}\label{EmodesFL}

To determine the angular power spectrum of the convergence $\kappa$ and of the $E$-modes of the cosmic shear, we follow the procedure described in \S~\ref{SecAngularCorrelations}. At the order $\zo$, the transfer function is isotropic and there are only scalar sources. Consequently, only $E$-modes are generated. The power spectrum for the $E$-modes is then just given by
\be\label{e.ceezo}
C_\ell^{EE \zo} =\frac{2}{\pi}\int_0^\infty k^2 \dd k P(k) \left| \int_0^\infty \dd
  \chi \snumb(\chi) \int_0^{\chi} \dd \tilde \chi g^E_\ell(k,\chi,\tilde \chi)\right|^2 \,,
\ee
where the function $g^E$ is defined as
\be
g^E_\ell(k,\chi,\tilde \chi) = -\frac{\chi-\tilde \chi}{\chi \tilde \chi}j_\ell(k
\tilde\chi)\frac{1}{2}\sqrt{\frac{(\ell+2)!}{(\ell-2)!}} T^\varphi(k,\tilde\chi)\,.
\ee
In Eq.~(\ref{e.ceezo}), $\snumb(\chi)$ represents the distribution of
sources as a function of the radial distance $\chi$ defined such that
$\snumb(\chi)\dd\chi$ is the number of sources between $\chi$ and
$\chi+\dd\chi$. At order $\zo$, it is sufficient to consider the
homogeneous source distribution, so that the observed shear and convergence for sources distributed up to $\chi_+$ are then defined by
\be
\kappa_\obs(\chi_+,{\bm n}_\obs) = \int_0^{\chi_+} \snumb(\chi)\kappa(\chi,{\bm n}_\obs) \dd\chi,\qquad
\gamma^\pm_\obs(\chi_+,{\bm n}_\obs) = \int_0^{\chi_+} \snumb(\chi)\gamma^\pm(\chi,{\bm n}_\obs) \dd\chi.
\ee
Since here $\snumb$ depends on $\chi$ alone, this integration can be performed after the multipolar decomposition
so that we perform the replacement, e.g.
\be
E_{\ell m}(\chi) \rightarrow \int_0^{\chi_+} \snumb(\tilde\chi)E_{\ell m}(\tilde\chi) \dd\tilde\chi\,,
\ee
in order to build the cosmological observables. Let us emphasize that this derivation can actually be performed in a simpler way~\cite{Uzan-Peter-anglais}: since the source term derives from a potential, one could have simply used the Fourier transform directly in Eq.~\eqref{DabOrder01} and then expanded the exponential according to Eq.~(\ref{Rayleigh}). The present derivation is however more general when used to higher orders $\{n,p\}$.

On small angular scales, that is, in the limit $\ell\gg1$, it is possible to use the Limber approximation~\cite{LoVerde:2008re}. Such aproximation consists in using
\be\label{LimberApproximation}
\int_0^\infty \dd x f(x) j_\ell(x) \simeq \sqrt{\frac{\pi}{2 L}} f(L)
\ee
with $L\equiv \ell+1/2$. If we commute the time integrals according to
\be\label{CommuteIntegrals0}
\int_0^\infty \dd
  \chi \, \int_0^{\chi} \dd \tilde \chi f(\chi,\tilde \chi) =\int_0^\infty \dd
  \tilde \chi \int_{\tilde \chi}^\infty \dd \chi f(\chi,\tilde \chi)\,
\ee
we arrive at the simple expression
\be
C_\ell^{EE \zo} \simeq \frac{1}{4}\frac{(\ell+2)!}{(\ell-2)!}\,{\cal P}_{\ell} \,,
\ee
with
\be
{\cal P}_{\ell} \equiv \int_0^\infty \frac{\dd \tilde \chi}{\tilde \chi^2}P\left(\frac{L}{\tilde \chi}\right)\left|T^\varphi\left(\frac{L}{\tilde \chi},\tilde \chi\right)\int_{\tilde \chi}^\infty \dd\chi \snumb(\chi)\frac{(\chi-\tilde \chi)}{\chi\tilde \chi} \right|^2\,.
\ee

The angular power spectrum of the convergence $\kappa$ is obtained in a similar way. Indeed, if we consider only the dominant contribution of Eq.~\eqref{Eqkappa01} at small scales, it is sufficient to replace $g_\ell^E$ by 
\be
g^\kappa_\ell(k,\chi,\tilde \chi) = \frac{\ell(\ell+1)}{2}\frac{\chi-\tilde \chi}{\chi \tilde \chi}j_\ell(k
\tilde\chi)\ T^\varphi(k,\tilde\chi)\,
\ee
in the previous expressions to get $C_\ell^{\kappa\kappa \zo}$. Using the Limber approximation, we then obtain
\be
C_\ell^{\kappa\kappa \zo} \simeq \frac{\ell^2(\ell+1)^2}{4}{\cal P}_{\ell} \,,
\ee
and we check immediately that for large $\ell$, $C_\ell^{\kappa\kappa \zo}\simeq C_\ell^{EE \zo}$.

Finally, the angular power spectrum of the cross-correlations between the shear and the convergence is given by
\be
C_\ell^{\kappa E \zo} =\frac{2}{\pi}\int_0^\infty k^2 \dd k P(k) \left(\int_0^\infty \dd
  \chi \snumb(\chi) \int_0^{\chi} \dd \tilde \chi g^E_\ell(k,\chi,\tilde \chi) \right)\left(\int_0^\infty \dd
  \chi \snumb(\chi) \int_0^{\chi} \dd \tilde \chi g^\kappa_\ell(k,\chi,\tilde \chi)\right)\nonumber
\ee
for which the Limber approximation gives
\be
C_\ell^{\kappa E \zo}\simeq-\frac{\ell(\ell+1)}{4}\sqrt{\frac{(\ell+2)!}{(\ell-2)!}}{\cal P}_{\ell}\,.
\ee

\subsection{Order \{1,1\}}\label{subsec6.3}

\subsubsection{Geodesic equation}

In principle, we need to determine $k^{0\oo}$ from the geodesic equation and then $x^{i\oo}$. As we shall see, these terms are only needed for the expression of the convergence $\kappa^\oo$. We will  instead focus on the computation of the cosmic shear $\gamma_{ab}^\oo$ and also the rotation $V^\oo$, since they give the leading order of the $B$-mode and the rotation. Fortunately, that computation does not require the solution of the geodesic equation up to order $\oo$.

\subsubsection{Sachs basis}

In order to get a definite expression involving only covariant and radial derivatives, we need to expand the direction vector $n^\ti$ around its background value $n^\ti_\obs$, so as to use the definition of \S~\ref{IdentificationDD}, taking into account the contributions of order $\zo$ and $\oz$, and similarly for the projection vectors $n_a^\ti$. We must use
\bea\label{Projrules}
n^\ti(\chi,n_\obs^\ti)&=&n_\obs^\ti+\left[\myaleph^{a \zo}(\chi,n_\obs^\ti)+\myaleph^{a \oz}(\chi,n_\obs^\ti)\right] {n_a^\obs}^\ti\,,\\
n_a^\ti(\chi,n_\obs^\ti)&=&{n^\obs_a}^\ti-\left[\myaleph^{\zo}_a(\chi,n_\obs^\ti)+\myaleph^{\oz}_a(\chi,n_\obs^\ti)\right] n_\obs^\ti\,.
\eea
It turns out that only the expression for the projection vectors is needed since the direction vector $n^\ti$ appears only in terms which are already of order $\oo$. 
Additionally, we must convert the derivative along the tetrads $\thetrad_{\ti}$ noted by $\partial_\ti$ to derivatives along the Cartesian coordinates, and these are related from Eq.~\eqref{ditodi}. This correction  is only relevant for the term $\partial_\ti \partial_\tj \varphi$ because the other terms are already of order $\oo$. We thus use
\bea\label{Propbeta}
\partial_\ti \partial_\tj \varphi &=&\partial_i \partial_j \varphi - 2 \beta^k_{\,\,(i}\partial_{j)} \partial_k \varphi \,,\\
\beta_{i j} &=& D_i D_j \MyB + 2 \MyB S_{ij} + 2 D_{(i} \MyB n^\obs_{\tj)} + \MyB n^\obs_\ti n^\obs_\tj\,.
\eea

\subsubsection{General form}

Since in Eq.~\eqref{MasterBonvin} the two terms $\frac{1}{k^{0}}\frac{\dd k^{0}}{\dd\chi}\frac{\dd{\cal D}_{ab}}{\dd\chi}$ and $\frac{1}{\left(k^{0}\right)^{2}}{\cal R}_{ac}{\cal D}_{cb}$ do not contain ${\cal D}^\oo_{ab}$ (because $\dd k^0/\dd\chi$ and ${\cal R}_{ab}$ vanish at order $\zz$), it can be integrated to give
\be\label{e.hjk}
\frac{\dd{\delta\cal D}^\oo_{ab}}{\dd\chi} = \delta_{ab} + \int_0^\chi\left[
-\frac{\dd \ln k^{0}}{\dd\chi}\frac{\dd{\cal D}_{ab}}{\dd\chi}+\frac{1}{\left(k^{0}\right)^{2}}{\cal R}_{ac}{\cal D}_{cb}
\right]\dd\chi'\,.
\ee
(We remind the reader of our convention, in which we split $k^0$ and ${\cal D}_{ab}$ respectively as $k^0=-1+\delta k^0$ and ${\cal D}_{ab}=\chi\delta_{ab} + \delta {\cal D}_{ab}$.)
In the first term of the integral, given that $\dd k^0/\dd \chi$ is at least of order $\oz+\zo$, the term ${\cal D}'_{ab}$ can be expressed using the formulas found in the two previous sections, that is
\be
\frac{\dd{\delta\cal D}^{\oz/\zo}_{ab}}{\dd\chi} =\int_0^\chi\left(
\frac{\dd k^{0,\oz/\zo}}{\dd\chi}\delta_{ab}+
\tilde\chi{\cal R}_{ab}^{\oz/\zo}
\right)\dd\tilde\chi.
\ee
Equation~(\ref{e.hjk}) can then be integrated as
\be
\delta{\cal D}_{ab}^{\oo}(\chi) =
\int_0^{\chi}\frac{\chi - \tilde \chi}{\tilde \chi} S^{\oo}_{ab}(\tilde \chi)\dd \tilde \chi \,,
\ee
where $S^{\oo}_{ab}$ contains all source terms of order $\oo$. It is explicitely given by
\bea
S^{\oo}_{ab}(\chi)&=&\frac{\chi^2 {\cal R}_{ab}}{(k^0)^2} +\chi {\cal R}_{ac} \delta {\cal D}_{cb}-\frac{\dd \ln k^0}{\dd \ln \chi}(2+k^0)\delta_{ab}+\frac{\dd k^0}{\dd \ln \chi}\int_0^\chi \dd \tilde \chi \tilde \chi {\cal R}_{ab} + \chi^2 \delta x^i \partial_i {\cal R}_{ab}\nonumber\,,
\eea
evaluated at order $\oo$, and where the last term arises from the fact that, at this order, there is a correction to be considered since we
have to go beyond the Born approximation. That is, we cannot just integrate on the Friedmann-Lema\^{\i}tre geodesic; instead we integrate on the geodesic $\tilde x^i(\chi,{\bm n}_\obs)=\chi n_\obs^i + \delta x^i(\chi,{\bm n}_\obs)$, so that the source term is 
\be
S_{ab}(\tilde x^i(\chi,{\bm n}_\obs)) = S_{ab}(\chi,n_\obs^\ti) +   \delta x^i(\chi,{\bm n}_\obs)\partial_jS_{ab},
\ee
which implies that
\be
S^\oo_{ab}(\tilde x^i(\chi,{\bm n}_\obs)) = S^\oo_{ab}(\chi,n_\obs^\ti) +   \delta x^{j,\oz}(\chi,{\bm n}_\obs)\partial_jS^{\zo}_{ab},
\ee
since $\partial_i S_{ab}^\zz=\partial_i S_{ab}^\oz=0$. It follows that the source term is explicitely given by
\beanosub\label{source_dev}
S^{\oo}_{ab}(\chi)&=& \chi^2\left\{\frac{{\cal R}_{ab}}{(k^0)^2}\right\}^\oo +\chi\frac{\dd\delta k^{0\oo}}{\dd\chi}\delta_{ab}\nonumber\\
 &+&\chi\left( {\cal R}_{ac}^\oz\delta{\cal D}_{cb}^\zo +  {\cal R}_{ac}^\zo\delta{\cal D}_{cb}^\oz
 +\frac{\dd \delta k^{0\zo}}{\delta\chi}\int_0^\chi \tilde\chi {\cal R}_{ab}^\oz  \dd\tilde\chi
  +\frac{\dd \delta k^{0\oz}}{\delta\chi}\int_0^\chi \tilde\chi {\cal R}_{ab}^\zo  \dd\tilde\chi
 \right)
   \nonumber\\
  &+& \chi\delta_{ab}\left[ \frac{\dd\left( \delta k^{0\oz}\delta k^{0\zo}\right)}{\dd\chi} + \delta k^{0\oz}\delta k^{0\zo}
  \right]\nonumber\\
  &+& \chi^2 \delta x^{j,\oz}\partial_j{\cal R}^{\zo}_{ab}.
\eeanosub
We see on this expression that the general source $S^{\oo}_{ab}(\chi)$ has several contributions. First, it has contributions from the vector and tensor modes $\bar B_i$ and $E_{ij}$ (respectively noted $S^{\oo V}$ and $S^{\oo T}$) which are at least of order $\oo$ since they vanish in the pure FL case; they enter the terms ${\cal R}_{ab}^\oo$ and $\delta k^{0\oo}$. Then, all the other contributions are formally products of the scalar perturbations by the geometrical shear; they appear as products of $\oz\times\zo$ terms. To compute explicitely these terms, we decompose the source term as
\be\label{DefVTquadSources}
S^{\oo}_{ab}(\chi) = S^{\oo V}_{ab}(\chi)+S^{\oo T}_{ab}(\chi)+S^{\oo {\rm quad}}_{ab}(\chi)\,.
\ee
Each contribution can be further decomposed into its trace, symmetric traceless and antisymmetric parts as
\be
S_{ab} = \delta_{ab} S+ S_{\langle a b \rangle}+S_{[ab]}\,.
\ee
Since our goal is to compute the effect of an anisotropic phase on the cosmic shear, and not on the convergence, we are mostly interested only in the symmetric traceless part. We shall thus not report the computation of the trace contribution to the trace part, except for the contribution coming from vectors and tensors, so as to be able to compare our results with the standard results in the literature, in the cases where the vector and tensor modes are considered even around a Friedmann-Lema\^{\i}tre background. A full computation may be useful in order to cross-correlate weak-lensing with the magnitude of supernovae.

\subsubsection{Vector and tensor modes contributions}

The vector and tensor contributions are easily found from the literature~\cite{Yamauchi:2013fra,Bonvin2010}. Splitting the vector mode into a radial and orthoradial parts as
\be
\bar B_i = \tilde B_i + \hat r_i B_r\,,\qquad E_{ij} = \tilde E_{ij} + 2
\hat r_{(i} \tilde E_{j)} + E_r \hat r_i \hat r_j\,,
\ee
the expression for the Riemann tensor given in Appendix~\ref{AppRiemannRicci} for vector and tensor modes gives
\bea
{\cal R}_{ab}^{\oo T} &=& {n_a^\obs}^\ti {n_b^\obs}^\tj\left[E_{ij}''-4 n^q \partial_{[q} E_{i]j}' +n^p n^q (\partial_i \partial_j E_{pq} +\partial_p\partial_q E_{ij})\! -2 n^p n^q \partial_q \partial_{(i} E_{j)p}\right]\\
{\cal R}_{ab}^{\oo V} &=& {n_a^\obs}^\ti {n_b^\obs}^\tj\left(-\partial_{(i} \bar B_{j)}' +n^q\partial_q \partial_{(i} \bar B_{j)} -n^q \partial_i\partial_j \bar B_q \right)\,.
\eea
Using the projections of partial derivatives into radial and covariant derivatives (see Appendix~\ref{App1plus2Cartesian}), we deduce that the vector and tensor contributions to the sources \eqref{DefVTquadSources} are 
\bea\label{SooVT}
S^{\oo V}_{ab}(\chi) &=& \chi \delta_{ab}B_{r,r}+\frac{\delta_{ab}}{2}\left[-D_c D^c B_r
  +\frac{1}{\chi^2}\frac{\dd }{\dd \chi}\left(\chi^3 D^c \tilde
  B_c\right) +\frac{2}{\chi}\frac{\dd}{\dd \chi} (\chi^2 B_r)\right] \nonumber\\
&+& \frac{\dd}{\dd \chi}\left(\chi D_{\langle a}
\tilde B_{b \rangle} \right)-D_{\langle a} D_{b \rangle} B_r\,,\\
S^{\oo T}_{ab}(\chi) &=& \chi \delta_{ab}\frac{\partial
    E_r}{\partial \chi}+\frac{\delta_{ab}}{2}\left[D^c D_c E_r -\chi^2
  \frac{\dd^2}{\dd \chi^2}E_r-6\frac{\dd }{\dd \chi}(\chi
  E_r)-\frac{2}{\chi}\frac{\dd}{\dd \chi}(\chi^2 D^c \tilde E_c)\right]\nonumber\\
&+&\chi \frac{\dd^2}{\dd \chi^2}(\chi \tilde E_{\langle ab \rangle}) +
D_{\langle a} D_{b \rangle} E_r -2 \frac{\dd}{\dd \chi}(\chi
D_{\langle a} \tilde E_{b \rangle})\,,
\eea
with the notation for the radial derivative  ${}_{,r} \equiv \hat x^i\partial_i$. The first terms of each expression are respectively the V and T contribution of the term in $\delta k^{0\oo}$ in Eq.~(\ref{source_dev}). For each of these two expressions, the first line contributes to the trace of the Jacobi matrix, that is, to convergence $\kappa^\oo$. The second line contributes to the cosmic shear $\gamma_{ab}^\oo$, since it is symmetric and traceless. By construction there is no antisymmetric part, so the vectors and tensors do not contribute to the rotation $V^\oo$.

In order to compare and recover the results of Refs.~\cite{Yamauchi:2013fra,Bonvin2010}, we must use the fact that vector modes are transverse, and that tensor modes are transverse and traceless. This allows us to get the relations (see also appendix~\ref{App1plus2Cartesian})
\bea
0&=&D^a \tilde B_a + \chi (B_r)_{,\chi} + 2 B_r \,,\\
0&=&D^a \tilde E_{ab} + 3 \tilde E_{b} + \chi (\tilde E_b)_{,\chi}\,,\\
0&=&D^a \tilde E_a + 3 E_r + \chi (E_r)_{,\chi} \,.
\eea

\subsubsection{Trace free part of the quadratic contributions}

Starting from the general expression~(\ref{source_dev}), the only contribution of the terms of order $\oo$ to the trace free part is $\chi^2{\cal R}_{\langle ab\rangle}^\oo$. Then, the terms $\delta{\cal D}_{ab}^{\oz/\zo}$ are decomposed as
\be
\delta{\cal D}_{ab}^{\oz/\zo} = \chi\kappa^{\oz/\zo} \delta_{ab} + \chi\gamma_{\langle ab\rangle}^{\oz/\zo} 
\ee
since, as we have just seen, there is no rotation at order ${\oz}$ and $\zo$. To finish, it is obvious that
\be
\frac{\dd\delta k^{0\oz/\zo}}{\dd\chi} = \left(\frac{\dd k^{0}}{\dd\chi}\right)^{\oz/\zo}.
\ee
It follows that the trace-free part of the quadratic contribution of the source term is given by
\beanosub\label{exhaustiveterms}
S^{\oo {\rm quad}}_{\langle ab \rangle}(\chi) &=& \chi^2 \left\{\frac{{\cal R}_{ab}}{(k^0)^2}\right\}^\oo+ \chi^2 {\cal R}^\oz_{\langle ab
  \rangle} \kappa^{\zo} + \chi^2 {\cal R}^\zo_{\langle ab
  \rangle} \kappa^{\oz} \label{exhaustivetermsa}\\
&&+ \chi^2 {\cal R}^\oz \gamma^{\zo}_{\langle ab \rangle}  + \chi^2 {\cal
  R}^\zo \gamma^{\oz}_{\langle ab \rangle}  \nonumber\\
&&+\chi \left(\frac{\dd k^0}{\dd \chi}\right)^{\zo} \int_0^\chi \tilde \chi {\cal R}_{\langle
  ab \rangle}^\oz\dd \tilde \chi +\chi \left(\frac{\dd k^0}{\dd \chi}\right)^{\oz} \int_0^\chi \tilde \chi {\cal R}_{\langle
  ab \rangle}^\zo\dd \tilde \chi \nonumber\nonumber\\
&&+ \chi^2 \delta x^{i\oz} \partial_i {\cal
  R}_{\langle ab \rangle}^\zo\,.\nonumber
\eeanosub
All terms, except the first one, involve products of quantities which
are $\zo$ and $\oz$, and have been already computed.  Note that the
first term is kept in the form $({\cal R}_{\langle
  ab\rangle})/(k^0)^2)^\oo$, and its detailed expression must be found
using the perturbed Riemann tensor given in Appendix~\ref{AppRiemannRicci}. This is indeed more convenient since
we shall express everything in terms of the tetrad basis, and we will
just need to use the fact that
\be
k^{0} =k^\tz \ThetradDU{\tz}{\,\,0}  =(1-\Phi)k^{\tz}\,.
\ee
We find for this first term
\beanosub\label{S11s}
&&\left(\frac{ {\cal R}_{\langle ab \rangle }}{(k^0)^2}\right)^\oo =  n_{\langle a}^\ti n_{b \rangle}^\tj\left[ 
-\sigma_{\ti \tj}(\varphi' + 2 \varphi_{,r})+ 2
  \sigma_{\tk (\ti} n^\tk \partial_{j)} \varphi-\partial_\ti \partial_\tj \varphi -\left(\frac{\sigma_{ij} \Psi}{\HH}\right)''\right.\\
  &&\qquad \qquad \qquad \left. -2\left(\frac{\sigma_{ij}
      \Psi_{,r}}{\HH}\right)'+2\left(\frac{\sigma_{ik}n^k}{\HH} \partial_j
    \Psi\right)'-\frac{\sigma_{ij}}{\HH} \Psi_{,rr} -\frac{\sigma_{kl}n^k
  n^l}{\HH} \partial_i \partial_j\Psi +2 \frac{\sigma_{ik}
  n^k}{\HH} \partial_j \Psi_{,r} \right]\,.\nonumber
\eeanosub
Then, we can split all partial derivatives into covariant and radial derivatives, using the expressions of Appendix~\ref{App1plus2Cartesian}. This term is then given by
\beanosub\label{Sderoo}
&&\left(\frac{\chi^2
    {\cal R}_{\langle ab \rangle }}{(k^0)^2}\right)^{\oo} = 
    -\chi^2 D_{\langle a} D_{b
  \rangle}\Sigma \left(2\varphi_{,r} +\varphi'\right)+2 \chi
D_{\langle a} \Sigma D_{b \rangle} \varphi \\
&&+ 2\chi^2 D_{\langle a} \MyB D_{b \rangle} \left(\frac{\varphi}{r}\right)_{,r}+2 D_c D_{\langle
  a} \MyB D_{b \rangle} D^c \varphi-(1-2\MyB )D_{\langle a} D_{b
  \rangle}\varphi\nonumber\\
&&-\chi^2 \left(\frac{\Psi D_{\langle a} D_{b
  \rangle}\Sigma}{\HH}\right)'' -2\chi^2 \left( \frac{\Psi_{,r} D_{\langle a} D_{b
  \rangle}\Sigma}{\HH} \right)' +2\chi^2  \left(\frac{D_{\langle a} \Sigma
D_{b \rangle} \Psi}{\HH}\right)'-\chi^2 \frac{D_{\langle a}D_{b \rangle}
\Sigma}{\HH}\Psi_{,rr}\nonumber\\
&&{-\frac{\Sigma}{\HH} D_{\langle a}D_{b
    \rangle}\Psi + \frac{2\chi^2 }{\HH} D_{\langle a} \Sigma D_{b \rangle}
  \left(\frac{\Psi}{r}\right)_{,r} } -2 \chi^2 D_{\langle a} \Sigma'\myaleph_{b \rangle}^{\zo} + 2 \chi^2 \myaleph_{\langle a}^{\oz}D_{b \rangle} \left(\frac{\varphi}{r}\right)_{,r}\,.\nonumber
\eeanosub

Finally, the last term of Eq.~\eqref{exhaustiveterms} needs to be evaluated. It can be read directly from the previous results at order $\oz$ and $\zo$.  We need only to split it into radial and covariant derivatives using the formulas of Appendix~\ref{App1plus2Cartesian}. We find that its contribution to the traceless part of the Jacobi matrix is given by
\beanosub\label{MasterLensing}
\chi^2 \delta x^{i\oz} \partial_i {\cal
  R}_{\langle ab \rangle}^\zo &=& -\delta r^{\oz} D_{\langle a} D_{b \rangle}\left(\varphi_{,r}-2 \frac{\varphi}{\chi}\right)\\
&&- \alpha^{c \oz} D_c D_{\langle a} D_{b
  \rangle}\varphi - 2 \chi^2 \,\alpha_{\langle a}^{ \oz}D_{b \rangle}\left(\frac{\varphi}{r}\right)_{,r}\nonumber\,.
\eeanosub
Let us emphasize that, when $\alpha^a\not=\varpi^a$, the source is partially seen on its side.

To conclude, the source term~(\ref{source_dev}) is obtained by
combining the two terms~(\ref{SooVT}) for the vector and tensor
contribution to the $\oo$ part,  the term~(\ref{S11s}) for the
quadratic scalar contribution and the term~(\ref{MasterLensing}) for
the post-Born approximation, to which we need to add the 6 terms which
are products $\oz\times\zo$ in~(\ref{exhaustiveterms}), obtained from the expressions of the former paragraphs. In principle, once all these contributions to the sources of the Sachs equation are identified and decomposed into radial and covariant derivatives, one should apply the formalism detailed in Appendix~\ref{SecAngularCorrelations}, and expand each term in spherical harmonics for both the angular dependence and the Fourier dependence. 

This procedure is however extremely long and includes a large number of terms. We will not detail it here but instead just identify the dominant contribution and compute its effect on the Jacobi matrix in order to derive the leading contribution to the $B$-modes in the next Section. Indeed, once we convert the covariant derivatives into spin-raising and spin-lowering operators, each covariant derivative is clearly associated with a factor $\ell$. In the flat sky approximation, that is, in the small angle approximation, the dominant contribution arises from the first term on the second line of Eq.~\eqref{MasterLensing}, 
\be
\delta\gamma_{ab}^\oo(\chi,{\bm n}_\obs) \simeq- \int_0^\chi
\frac{\chi-\tilde\chi}{\tilde\chi} \alpha^{c \oz}(\tilde\chi,{\bm n}_\obs) D_c D_{\langle a}
D_{b \rangle}\varphi(\tilde\chi,{\bm n}_\obs) \dd \tilde \chi
\ee
as it enjoys three covariant derivatives.

\subsubsection{Trace part of the quadratic contributions}

As discussed in the previous paragraph, the computation of the trace of $\delta{\cal D}_{ab}^\oo$ involves a lot of terms such as $k^{0,\oo}$ and the fourth line of Eq.~(\ref{source_dev}). This tedious computation can indeed be performed with all the details given in this article. It will however give only a small correction to $\kappa$, the leading order of which is the standard convergence $\kappa^{\zo}$.

We thus decide not to include this computation here since we are mostly interested by the lowest order dominant effect related to the anisotropic expansion.

\subsubsection{Rotation quadratic contributions}

As we have seen, the rotation vanishes at orders $\zo$ and $\oz$ so that its leading order contribution appears at order $\oo$. Since ${\cal R}_{ab}^\oo$ is symmetric, its contribution arises simply from the two first terms of the second line of Eq.~(\ref{source_dev}), that is from the source term
\be
 S_{[ab]}^\oo(\chi) = \chi\left( {\cal R}_{[a| c}^\oz\delta{\cal
     D}_{c|b]}^\zo +  {\cal R}_{[a|c}^\zo\delta{\cal D}_{c|b]}^\oz
 \right)\equiv \epsilon_{ab} S^{\oo}_{\rm rot}(\chi)\,.
\ee
Using the expression of the previous sections, it is explicitely given by
\bea
S_{\rm rot}^\oo(\chi,{\bm n}_\obs) &=& -\frac{\ii}{2} D_+D_+ \Sigma'
  \int_0^\chi \frac{\chi(\chi-\tilde \chi)}{\tilde \chi} D_-D_-
  \varphi(\tilde \chi) \dd \tilde \chi -  (-\leftrightarrow+)\nonumber \\
&&-\frac{\ii}{2}  D_+D_+ \varphi \int_0^\chi
\frac{(\chi-\tilde \chi) \tilde \chi}{\chi} D_-D_- \Sigma' (\tilde \chi) \dd \tilde \chi -  (-\leftrightarrow+)\,.
\eea
The general expression for the rotation is then obtained through
\be
V^\oo(\chi,{\bm n}_\obs) = \int_0^\chi \frac{\chi-\tilde \chi}{\tilde \chi} S_{\rm rot}^\oo(\tilde \chi,{\bm n}_\obs) \dd \tilde \chi\,.
\ee
The rotation is thus sourced by the coupling between the usual cosmic shear of the standard scalar perturbation around a Friedmann-Lema\^{\i}tre spacetime ($D_+D_+ \varphi$) and the quadrupolar contribution due to the geometric shear ($D_- D_- \MyB''=D_- D_-\Sigma' $).

\subsubsection{Integration over the source distribution}

The last point that needs to be discussed before turning to the multipolar decomposition and the computation of the angular power spectra is the source distribution.

The source distribution represents the mean number of object normalized to the mean density observed in a solid angle $\dd\Omega_\obs$, that is
\be
 \frac{\dd N}{\dd\Omega_0\dd\chi} \rightarrow \snumb\qquad\,.
\ee
In the Friedmann-Lema\^{\i}tre and Bianchi $I$ background spacetimes, which are both homogeneous, $\snumb$ is constant on any constant time hypersurface, which means that it depends on $\chi$ alone. Thus
\be
\snumb(\chi,{\bm n}_\obs) = \snumb(\chi) + \snumb^{\zo}(\chi,x^i) + \snumb^{\oz}(\chi)
\ee
where the second term is the standard fluctuation of the number density due to the large scale cosmological perturbations and for which it is understood that the position $x^i$ is evaluated on the background geodesic, that is $x^i= \chi n_\obs^\ti$.

Note however than when one turns to redshift space, on which the observations are actually performed, one needs to take into account that $z$ is a function of $\chi$ and the direction of observation ${\bm n}_\obs$, so that we should rather use
\be
\snumb(z,{\bm n}_\obs) = \snumb(z) + \snumb^{\zo}(z,x^i) + \snumb^{\oz}(z,{\bm n}_\obs)\,,
\ee
where again the position is evaluated on the background geodesic with $x^i= \chi n_\obs^\ti$. It follows that, when computing the observed quantities,
$$
\gamma_{ab}^\oo(\chi,{\bm n}_\obs) = \int \dd\chi \left[ \snumb(\chi)\gamma_{ab}^\oo 
+ \snumb^\zo(\chi,x^i)\gamma_{ab}^\oz + \snumb^\oz(\chi)\gamma_{ab}^\zo 
\right]\,.
$$
The second term is the standard correlation between the fluctuations
of the source distribution and the cosmic shear. It inherits a
directional dependence from the spatial dependence $\snumb^\zo(\chi,x^i)$
given that it is evaluated on the background line of sight, that is
with $x^i=\chi n_\obs^\ti$. Because of the coupling to the pure
$E$-mode $\gamma_{ab}^\oz$ it will induce $B$-modes in the source
averaged cosmic shear. This component is expected to be important on
large angular scales. The third term is a correction that arises from
the fact that the formation of structure differs a priori in the presence of a geometrical shear, but it does not contribute the $B$-modes since it does not have a directional dependence. However, it induces a correction for the $E$-modes and for the convergence.

Now, in redshift space, one needs to be more careful since
$$
\gamma_{ab}^\oo(z,{\bm n}_\obs) = \int \dd \tilde z \left[ \snumb(\tilde z)\gamma_{ab}^\oo 
+ \snumb^\zo(\tilde z,x^i)\gamma_{ab}^\oz + \snumb^\oz(\tilde z,{\bm n}_\obs)\gamma_{ab}^\zo 
\right].
$$
Both the second and the third term depend now explicitely on the direction of observation, so that the convolution by the source distribution has to be performed before the decomposition in spherical harmonics, and both terms will generate $B$-modes out of the $E$-modes of $\gamma_{ab}^\oz$ and $\gamma_{ab}^\zo$ respectively. However, these effects should not dominate for small angular scales and we shall thus neglect them.

\section{Orders of magnitude}\label{sec7}

The previous sections have provided all the elements needed to compute the contribution of the $B$-modes at order $\oo$ and their correlations with the $E$-modes and the cosmic shear. It is obvious that any further computation has to be performed numerically. It is however important to exhibit the dominant contribution.

\subsection{Dominant effects}

Once the covariant derivatives are expressed in terms of spin-raising and spin-lowering operators, it is rather straightforward to realize that any covariant derivative is associated with a factor $\ell$ in multipole space. The dominant terms contributing to the shear are thus the ones with the highest number of covariant derivatives applied to $\varphi$.

For instance, at order $\zo$, the convergence is dominated by the last term of Eq.~\eqref{Eqkappa01} on small scales. That is
$$
\kappa^{\zo}_{\ell m}\sim \frac{\ell(\ell+1)}{2}\int_0^\chi \frac{\chi-\tilde \chi}{\chi \tilde \chi} \varphi_{\ell m}(\tilde \chi) \dd \tilde \chi\,,
$$
simply because of the geometrical factor $\ell^2$. It is indeed the term which is usually presented in textbooks. This term dominates over the second one even at small $\ell$, i.e. for $\ell>2-3$, that is for all practical purposes.

When applying this small scale approximation scheme at order $\oo$, we realize that there is just one dominant term -- the first one on the second line of Eq.~\eqref{MasterLensing} -- which possesses three covariant derivatives, that is
$$
\delta\gamma_{ab}^\oo(\chi,{\bm n}_\obs) \simeq- \int_0^\chi
\frac{\chi-\tilde\chi}{\tilde\chi} \alpha^{c \oz}(\tilde\chi,{\bm n}_\obs) D_c D_{\langle a}
D_{b \rangle}\varphi(\tilde\chi,{\bm n}_\obs) \dd \tilde \chi\,.
$$
Physically it corresponds to the orthoradial displacement of the central geodesic on which the Sachs equation is evaluated, when compared with the background geodesic. It is as if the sources of order $\zo$ contributing to the Jacobi map had been lensed by the orthoradial displacement of order $\oz$, resulting in an order $\oo$ effect. This is similar to the lensing of first order sources of CMB around the last-scattering surface by first order gravitational potential in the foreground, resulting in a second order lensing effect in the CMB. 

The first consequence of this is that the formalism used to compute the CMB lensing can also be applied to obtain the resulting Jacobi map due to this leading order term. However, there is a slight difference. Indeed, for the CMB the sources are all located in a background around the last-scattering surface, for which there is a deflection due to the gravitational potential crossed between emission and reception. For the general solution giving the Jacobi map, however, the sources are distributed from the observer up to the maximum redshift of the survey. For each source there is a different deflection angle as it depends on the trajectory between the source and the observer. 

Finally, we must remind that the treatment of CMB lensing by a gradient expansion~\cite{Hu2000} holds only until the deflection angle is comparable to the angle of structures in the CMB. Beyond that scale, this method underestimates the effect of lensing and one has to resort to a full-lensing method as exposed in Ref.~\cite{Challinor:2005jy,Lewis2006}.  Since we are interested in an order of magnitude estimate of the effect of geometrical shear on the cosmic shear, we will present in the next section a gradient expansion method based on Ref.~\cite{Hu2000}, but one should be aware that for any amplitude of the geometric shear, there must exist a scale $\ell_*$ beyond which this treatment is inaccurate. The method for the full-lensing method is exposed briefly in Appendix~\ref{SecFullLensing}.

\subsection{Lensing of the central geodesic}

\subsubsection{General formalism of the gradient expansion}

Any observable at a given affine parameter $\chi$ in a given direction $\bm{n}_\obs$ is formally obtained from an integration on the background geodesic over its sources given by Eq.~(\ref{e:GeneralLOS}), that is
\be
X^s(\chi, \bm{n}_\obs) \bm{m}^{s}_{\obs} =
\int_0^\chi S^{X}(\chi,\tilde \chi,\bm{n}_\obs) \bm{m}^{s}_\obs \dd \tilde \chi\,.
\ee
However, and as discussed above, a true observable like the cosmic shear is obtained by averaging over the true normalized profile $\snumb(\chi)$ of sources as
\be\label{Integralchisources}
X^s(\bm{n}_\obs) \bm{m}^{s}_{\obs}=\int_0^\infty \dd \chi\, \snumb(\chi) X^s(\chi, \bm{n}_\obs) \bm{m}^{s}_{\obs} \dd \chi=\int_0^\infty \snumb(\chi) \dd \chi\int_0^\chi S^{X}(\chi,\tilde \chi,\bm{n}_\obs) \bm{m}^{s}_\obs \dd \tilde \chi \,.
\ee
Note that the integrals can be interchanged using
\be
\int_0^\infty \dd
  \chi \,\snumb(\chi) \int_0^{\chi} \dd \tilde \chi f(\chi,\tilde \chi) =\int_0^\infty \dd
  \tilde \chi \int_{\tilde \chi}^\infty \dd \chi \,\snumb(\chi) f(\chi,\tilde \chi)\,. 
\ee
We consider only the effect of the dominant term in Eq.\eqref{MasterLensing}, which corresponds to the lensing of the sources, that is, it transforms the sources according to a parallel transport along the lensing vector ${\bm \alpha}$. A lensed observable $\tilde X^s$ is then obtained from an integration over the lensed sources. If the lensing effect is small, it is sufficient to use a Taylor expansion of the lensed sources to express them in terms of the unlensed sources, the small parameter being the lensing vector ${\bm \alpha}$. Furthermore, if the lensed vector can be written as the gradient of a scalar, as $\alpha_a = D_a \alpha$, then at lowest order in the Taylor expansion, we get for the lensed source
\be
\tilde S^{X^s}(\chi,\tilde \chi, \bm{n}_\obs) \bm{m}^{s}_{\obs} = S^{X^s}(\chi,\tilde \chi, \bm{n}_\obs) \bm{m}^{s}_{\obs} +  D^a \alpha(\tilde \chi,\bm{n}_\obs) D_a \left[S^{X^s}(\chi,\tilde \chi,\bm{n}_\obs) \bm{m}^{s}_\obs \right]\,.
\ee
Using Eq.~\eqref{IHU}, the multipoles are easily extracted as
\begin{equation}\label{HuI}
\tilde S^{X^s}_{\ell m}(\chi,\tilde \chi) = S^{X^s}_{\ell m}(\chi,\tilde \chi) + \sum_{m_1, \ell_2, m_2} \alpha_{2\,m_1}(\tilde \chi)S^{X^s}_{\ell_2 m_2}(\chi,\tilde \chi) {}_{s}I_{\ell\, 2 \,\ell_2}^{m\,m_1\,m_2}\,,
\end{equation}
where the $\alpha_{\ell m}$ are the multipoles of the lensing scalar when decomposed into spherical harmonics, and the coefficients ${}_s I_{\ell\, \ell_1\,\ell_2}^{m\,m_1\,m_2}$ are defined in Eq.~(\ref{PropertyIHU}).

\subsubsection{Multipoles of the lensing vector}

The previous expression depends on the multipoles of the lensing scalar, that can actually be obtained very easily. First, following the definitions~(\ref{DefSigmaB}-\ref{DefSigmaBbis}) we define a matrix $\alpha_{ij}$ such that
\be
{ \alpha}_\pm(\bm{n}_0,\chi) \equiv D_\pm \alpha(\bm{n}_\obs,\chi) =
D_\pm \left[\frac{1}{2}\alpha_{ij}(\chi) n_\obs^i n_\obs^j\right] \,.
\ee
Given Eqs.~\eqref{xi10bis} and \eqref{MasterLensing}, the components of $\alpha_{ij}(\chi)$ are just 
\bea
\alpha_{ij}(\chi) &=&-\beta_{ij}(0)+ 2\int_0^\chi \dd \chi'
\frac{\chi-\chi'}{\chi} \sigma_{ij}(\chi')\\
\label{alpha_shear}
&=&\beta_{ij}(0)-\frac{2}{\chi}\int_0^\chi \dd \chi' \beta_{ij}(\chi')\,.
\eea
Then, similarly to the computation of the coefficients $\Sigma_{2m}$ in Eq.~\eqref{EqSphericalComponents}, the multipoles of $\alpha(\bm{n}_\obs,\chi)$, defined by $\alpha = \sum_m \alpha_{2m} Y_{2m}$ reduce to a quadrupole and are explicitely given by
\begin{equation}
\label{alpha2m}
\alpha_{20}(\chi) = -\sqrt{\frac{\pi}{5}}[\alpha_{xx}(\chi)  + \alpha_{yy}(\chi) ] \,,\qquad \alpha_{2 \,\pm2}(\chi) = \sqrt{\frac{\pi}{30}} [\alpha_{xx}(\chi)  - \alpha_{yy}(\chi) ]\,,
\end{equation}
if the coordinates system is adapted to the eigendirections of the geometrical shear.

\subsubsection{Extracting the spatial shear components from off-diagonal correlations}

A byproduct of the formalism just developed is that we can extract information about the geometric shear $\sigma_{ij}$ from cross-correlations between the $E$- and $B$-modes multipoles of the cosmic shear, $E_{\ell m}$ and $B_{\ell m}$, and the multipoles $\kappa_{\ell m}$ of the convergence $\kappa$, that would otherwise vanish in the pure Friedmann-Lema\^{\i}tre case. Indeed, even if the $B$-modes are not sourced initially, as is the case of a Friedmann-Lema\^{\i}tre background, at the perturbative level there will be a lensed $B$-mode term sourced by the $E$-modes of the background shear. In order to extract this effect we decompose the (lensed) $E$- and $B$-modes of the source as
\be
\widetilde{S}^{\gamma^{\pm}}_{\ell m}(\chi,\tilde{\chi}) = \widetilde{S}^{E}_{\ell m}(\chi,\tilde{\chi})\pm\ii
\widetilde{S}^{B}_{\ell m}(\chi,\tilde{\chi})
\ee
with a similar decomposition for the (unlensed) $S^{X}_{\ell m}$. Then, using the properties (\ref{PropertyIHU}), it follows that
\bea
\label{StildeB}
\widetilde{S}^{B}_{\ell m}(\chi,\tilde{\chi}) &=& -\ii\sum_{\substack{m_1,m_2 \\ \ell_2=\ell\pm1}}
\alpha_{2 m_1}(\tilde \chi)S^{E}_{\ell_2m_2}(\chi,\tilde{\chi})\,{}_{+2}I^{m m_1 m_2}_{\ell 2 \ell_2}\,,\\
\widetilde{S}^{E}_{\ell m}(\chi,\tilde{\chi}) &=& {S}^{E}_{\ell m}(\chi,\tilde{\chi})+\sum_{\substack{m_1,m_2 \\ \ell_2=\ell,\ell\pm2}}
\alpha_{2 m_1}(\tilde \chi)S^{E}_{\ell_2m_2}(\chi,\tilde{\chi})\,{}_{+2}I^{m m_1 m_2}_{\ell 2 \ell_2}\,.
\eea
We remind that there is no tilde on $S^{E}_{\ell m}$ on the right-hand side of the above equation, since it corresponds to the unlensed sources. Since the convergence is a spin $0$ quantity, then from Eq.~\eqref{HuI}, its sources are transformed under lensing as
\be
\widetilde{S}^{\kappa}_{\ell m}(\chi,\tilde{\chi}) = {S}^{\kappa}_{\ell m}(\chi,\tilde{\chi})+\sum_{\substack{m_1,m_2 \\ \ell_2=\ell,\ell\pm2}}
\alpha_{2 m_1}(\tilde \chi)S^{\kappa}_{\ell_2m_2}(\chi,\tilde{\chi})\,I^{m m_1 m_2}_{\ell 2 \ell_2}\,.
\ee
From these expressions, it is clear that the off-diagonal terms coming from the $EB$, $EE$, $\kappa\kappa$, $\kappa E$, $\kappa B$ cross-correlation matrices allow us to put constraints on $\alpha_{2m}$ and, consequently, on the geometric shear components $\sigma_{ij}$ by means of Eqs.~(\ref{alpha_shear}). To see how that is possible, we must remember that the sources should be convolved with $\snumb(\chi)$ by means of Eq.~(\ref{Integralchisources}). 

In Section~\ref{EmodesFL}, the $EE$, $\kappa \kappa$, and $\kappa E$ correlations at order $\zo$ (that is, without the effect of lensing by the geometric shear) have been computed and they are of the form
\be
C_\ell^{XZ} = \int_0^\infty \dd
  \tilde \chi_1 \int_{\tilde \chi_1}^\infty \dd \chi_1 \,\snumb(\chi_1) \int_0^\infty \dd  \tilde \chi_2 \int_{\tilde \chi_2}^\infty \dd \chi_2 \,\snumb(\chi_2)C_\ell^{XZ}(\chi_1,\chi_2,\tilde \chi_1,\tilde \chi_2) 
\ee
where the indices $X$ and $Z$ take the values $\kappa,E$, and with the source correlations
\bea
\langle S^X_{\ell_1 m_1}(\chi_1,\tilde{\chi}_1)S^{Z}_{\ell_2 m_2}(\chi_2,\tilde{\chi}_2)\rangle&=& \delta_{\ell_1 \ell_2} \delta_{m_1 m_2}C_{\ell_1}^{XZ}(\chi_1,\chi_2,\tilde \chi_1,\tilde \chi_2) \,,\\
C_\ell^{XZ}(\chi_1,\chi_2,\tilde \chi_1,\tilde \chi_2) &\equiv& \frac{2}{\pi}\int_0^\infty k^2 \dd k P(k) g^X_{\ell}(k,\chi_1,\tilde \chi_1) g^{Z}_{\ell}(k,\chi_2,\tilde \chi_2)\,.
\eea
For the lensed observables, we define similarly
\bea
{\cal P}_{\ell M}^{XZ}&\equiv&\int_0^\infty \dd
  \tilde \chi_1 \int_{\tilde \chi_1}^\infty \dd \chi_1 \,\snumb(\chi_1)\nonumber
\int_0^\infty \dd
  \tilde \chi_2 \int_{\tilde \chi_2}^\infty \dd \chi_2 \,\snumb(\chi_2) \alpha_{2\,M}(\tilde{\chi}_1)C_{\ell}^{XZ}(\chi_1,\chi_2,\tilde{\chi}_1,\tilde{\chi}_2)\,\\
  &=&\frac{2}{\pi}\int_0^\infty k^2 \dd k P(k)\left(\int_0^\infty \dd
  \tilde \chi_1 \int_{\tilde \chi_1}^\infty \dd \chi_1 \,\snumb(\chi_1)\alpha_{2\,M}(\tilde{\chi}_1)g^X_{\ell}(k,\chi_1,\tilde \chi_1)\right)\,\nonumber\\
  && \qquad \qquad \times\left(\int_0^\infty \dd
  \tilde \chi_2 \int_{\tilde \chi_2}^\infty \dd \chi_2 \,\snumb(\chi_2)g^Z_{\ell}(k,\chi_2,\tilde \chi_2)\right)\,,
\eea
such that the following non-vanishing correlations are expressed as
\bea
\langle\widetilde{B}_{\ell m}E^{\star}_{\ell\pm1 \,m-M}\rangle
&=&-\ii \,\,{}_{+2}I^{m \,M\, (m-M)}_{\ell\, 2\, \ell\pm 1}\,\,{\cal P}_{\ell\pm1\,M}^{EE}\,,\\
\langle\widetilde{B}_{\ell m}\kappa^{\star}_{\ell\pm1 \,m-M}\rangle
&=&-\ii \,\,{}_{+2}I^{m \,M\, (m-M)}_{\ell\, 2\, \ell\pm 1}\,\,{\cal P}_{\ell\pm1\,M}^{E\kappa}\,.
\eea
Not only do we get off-diagonal contributions for $B$-modes with the $E$-modes and the convergence, but we also get off-diagonal correlations between $\kappa$ and $E$-modes. They read
\bea
\langle\widetilde{E}_{\ell m} \widetilde{E}^{\star}_{\ell\pm2 \,m-M}\rangle
&=&{}_{+2}I^{m \,M\, (m-M)}_{\ell\, 2\, \ell\pm 2}{\cal
  P}_{\ell\pm2\,M}^{EE}+(-1)^M {}_{+2}I^{(m-M)\,-M\,m}_{\ell\pm 2\,2\,\ell}{\cal P}_{\ell\,M}^{EE}\,,\\
\langle\widetilde{\kappa}_{\ell m} \widetilde{\kappa}^{\star}_{\ell\pm2 \,m-M}\rangle
&=&I^{m \,M\, (m-M)}_{\ell\, 2\, \ell\pm 2}{\cal
  P}_{\ell\pm2\,M}^{\kappa \kappa}+(-1)^M I^{(m-M)\,-M\,m}_{\ell\pm 2\,2\,\ell}{\cal P}_{\ell\,M}^{\kappa \kappa}\,,\\
\langle\widetilde{E}_{\ell m} \widetilde{\kappa}^{\star}_{\ell\pm2 \,m-M}\rangle
&=&{}_{+2}I^{m \,M\, (m-M)}_{\ell\, 2\, \ell\pm 2}\,\, {\cal
  P}_{\ell\pm2\,M}^{E\kappa }+(-1)^M I^{(m-M)\,-M\,m}_{\ell\pm 2\,2\,\ell}\,\, {\cal P}_{\ell\,M}^{\kappa E}\,.
\eea
Note that in all these cross-correlators, $M$ ranges from $-2$ to $2$, thus spanning the five degrees of freedom of the lensing potential $\alpha_{2M}$, and consequently of the underlying Bianchi geometrical shear $\sigma_{ij}$. These expressions for the correlators are however not ideal to relate the correlations to the lensing potential, and thus to the components of $\sigma_{ij}$. Instead, we define appropriate combinations of the correlators by resumming them as~\cite{Biposh}
\be
{}^{XZ}{\cal A}^{2M}_{\ell_1\,\ell_2}  \equiv  \sum_{m}\sqrt{5}(-1)^{m+\ell_1+\ell_2}\troisj{\ell_1}{2}{\ell_2}{-m}{M}{m-M}\langle\widetilde{X}_{\ell_1 m}Z^{\star}_{\ell_2,m-M}\rangle\,.
\ee
For instance, for the $EB$ and $EE$ correlations, we get
\bea
{}^{BE}{\cal A}^{2M}_{\ell\,\ell\pm1}  &\equiv & \sum_{m}\sqrt{5}(-1)^{m+1}\troisj{\ell}{2}{\ell\pm1}{-m}{M}{m-M}\langle\widetilde{B}_{\ell m}E^{\star}_{\ell\pm1,m-M}\rangle\,,\\
{}^{EE}{\cal A}^{2M}_{\ell\,\ell\pm2}  &\equiv & \sum_{m}\sqrt{5}(-1)^{m}\troisj{\ell}{2}{\ell\pm2}{-m}{M}{m-M}\langle\widetilde{E}_{\ell m}\widetilde{E}^{\star}_{\ell\pm2,m-M}\rangle\,.
\eea
Then, by using the definition of the symbols ${}_{+2}I^{mMm'}_{\ell 2\ell\pm1}$ and $I^{mMm'}_{\ell 2\ell\pm1}$ and the orthogonality relations of the Wigner 3j symbols, we get 
\bea\label{FiveCorrelators}
{}^{BE}{\cal A}^{2M}_{\ell\ell\pm1}  &=&  \ii \frac{{}_2F_{\ell2\ell\pm1}}{\sqrt{5}}\,\,  \,{\cal P}_{\ell\pm1 M}^{EE} \,,\\
{}^{B\kappa}{\cal A}^{2M}_{\ell\ell\pm1}  &=&  \ii \frac{F_{\ell2\ell\pm1}}{\sqrt{5}}\,\,  \,{\cal P}_{\ell\pm1 M}^{E\kappa } \,,\\
{}^{EE}{\cal A}^{2M}_{\ell\ell\pm2}  &=& \frac{{}_2F_{\ell2\ell\pm2}}{\sqrt{5}}{\cal P}_{\ell\pm2 M}^{EE}+\frac{{}_2F_{\ell\pm2\,2\,\ell}}{\sqrt{5}}{\cal P}_{\ell\,M}^{EE}\,,\\
{}^{\kappa \kappa}{\cal A}^{2M}_{\ell\ell\pm2}  &=& \frac{F_{\ell2\ell\pm2}}{\sqrt{5}}{\cal P}_{\ell\pm2 M}^{\kappa \kappa}+\frac{F_{\ell\pm2\,2\,\ell}}{\sqrt{5}} {\cal P}_{\ell\,M}^{\kappa \kappa}\,,\\
{}^{E\kappa}{\cal A}^{2M}_{\ell\ell\pm2}  &=& \frac{{}_2F_{\ell2\ell\pm2}}{\sqrt{5}}\,\,  {\cal P}_{\ell\pm2 M}^{E\kappa}+\frac{F_{\ell\pm2\,2\,\ell}}{\sqrt{5}}\,\, {\cal P}_{\ell\,M}^{\kappa E}\,,
\eea
where the symbols ${}_2F_{\ell \ell_1 \ell_2}$ are defined in Appendix~\ref{3j_identities}. 

Approximate expressions of this correlators can be obtained in the Limber approximation~\eqref{LimberApproximation} and read
\bea
{\cal P}_{\ell\,M}^{\kappa \kappa} &\simeq & \frac{\ell^2(\ell+1)^2}{4}{\cal P}_{\ell\,M}\,,\\
{\cal P}_{\ell\,M}^{\kappa E} &\simeq &{\cal P}_{\ell\,M}^{E\kappa} =\frac{\ell(\ell+1)}{4}\sqrt{\frac{(\ell+2)!}{(\ell-2)!}}\,\,{\cal P}_{\ell\,M}\,,\\
{\cal P}_{\ell\,M}^{\kappa \kappa} &\simeq & \frac{1}{4}\frac{(\ell+2)!}{(\ell-2)!}\,{\cal P}_{\ell\,M}\,,
\eea
with the function ${\cal P}_{\ell\,M}$ given by
\be
{\cal P}_{\ell\,M} \equiv \int_0^\infty
\frac{\dd \tilde \chi}{\tilde \chi^2}P\left(\frac{L}{\tilde \chi}\right)\alpha_{2 M}(\tilde \chi)\left|T^\varphi\left(\frac{L}{\tilde \chi},\tilde \chi\right)\int_{\tilde \chi}^\infty \dd\chi \snumb(\chi)\frac{(\chi-\tilde \chi)}{\chi\tilde \chi} \right|^2\,,
\ee
where we used the notation $L\equiv \ell+1/2$.

This provides the expressions of the $5$ (off-diagonal) correlators~(\ref{FiveCorrelators}), each having $5$ components, and all being linear in $\sigma_{ij}$. We stress that the measurement of these quantities from further surveys will allow us to get stronger constraints on the spatial isotropy of the universe, thus pushing forward the ``beyond $\Lambda$CDM'' program.

\subsubsection{Autocorrelations of B-modes from the lensing of the central geodesic}

The previous off-diagonal correlators are the most direct consequence of a late time geometrical shear on weak-lensing. However, experiments are mostly designed to measure the diagonal part. In this section we compute the autocorrelation of $B$-modes induced by the dominant lensing term. This angular power spectrum will thus be quadratic in $\sigma_{ij}$. Contrary to the previous estimators, it does not allow us to reconstruct the full geometrical shear $\sigma_{ij}$, but can be used to set constraints on $\sigma^2$.

Using  the properties of the Wigner 3j symbols given in Appendix~\ref{3j_identities} and starting from the lens sources~\eqref{StildeB}, we obtain that the $B$-modes angular power spectrum of weak-lensing cosmic shear generated by the lensing of the central geodesic is
\beanosub
C_\ell^{BB \oo} &=& \int_0^{\infty} \dd \chi_1 \int_0^{\infty} \dd
\chi_2\,\snumb(\chi_1)\snumb(\chi_2)\int_0^{\chi_1}\dd \chi_1' \int_0^{\chi_2}\dd
\chi_2' \nonumber\\
&&\times \sum_{s=\pm 1, m} \frac{\alpha_{2 m}(\chi'_1) \alpha_{2 m}^\star(\chi'_2) }{5}C_{\ell+s}^{EE}(\chi_1,\chi_2,\chi_1',\chi_2')  \frac{({}_2 F_{\ell\,2\,\ell+s})^2}{2 \ell+1}\,.
\eeanosub
If we now factorize the time integrals we simply get
\be
C_\ell^{BB \oo} = \frac{2}{5\pi}\int_0^\infty k^2 \dd k P(k) \sum_{s=\pm 1}\frac{({}_2
   F_{\ell\,2\,\ell+s})^2}{2 \ell+1}  \sum_m \left|\int_0^\infty \dd
   \chi \snumb(\chi) \int_0^\chi \dd \chi' \alpha_{2 m}(\chi')  g_{\ell+s}(k,\chi,\chi') \right|^2\,.
\ee
Note that for large $\ell$, the $F$-coefficients behave as
\be
\lim_{\ell\to\infty}\frac{({}_2   F_{\ell\,2\,\ell+1})^2}{2 \ell+1} =\frac{15}{2 \pi} \,,\qquad\lim_{\ell\to\infty}\frac{({}_2   F_{\ell\,2\,\ell-1})^2}{2 \ell+1}= \frac{15}{2 \pi}\,.
\ee
Apart from the six terms (sum over $m=-2,0,+2$ and over $s=\pm1$), this is numerically as fast as computing the correlation $C_\ell^{EE}$ at order $\zo$. Then, using the Limber approximation~\eqref{LimberApproximation}, with the definitions $L_s = L+s=\ell+1/2+s$ and $\ell_s=\ell+s$, it leads to
\beanosub
C_\ell^{BB \oo} &\simeq& \frac{1}{20}\int_0^\infty
\frac{\dd \tilde \chi }{\tilde \chi^2}\sum_{s=\pm1} \frac{(\ell_s+2)!}{(\ell_s-2)!} P\left(\frac{L_s}{\tilde\chi}\right) \frac{({}_2
   F_{\ell\,2\,\ell+s})^2}{2 \ell+1}\nonumber\\
&&\times
\sum_m \left|T^\varphi\left(\frac{L_s}{\tilde \chi},\tilde \chi\right) \alpha_{2\,m}(\tilde \chi)\int_{\tilde \chi}^\infty \dd \chi \snumb(\chi) \frac{(\chi-\tilde \chi)}{\chi \tilde \chi}\right|^2\,.
\eeanosub
Finally, using~\eqref{alpha2m} to get
\be
\sum_m \left| \alpha_{2\,m}(\chi)\right|^2 = \frac{2\pi}{15}
\alpha_{ij}(\chi) \alpha^{ij}(\chi)\equiv \frac{2\pi}{15}|\alpha(\chi)|^2
\ee
we obtain a very compact expression, valid only for large $\ell$, which is
\be
C_\ell^{BB \oo} \simeq\frac{\ell^4}{10}\int_0^\infty
\frac{\dd \tilde \chi}{\tilde \chi^2}P\left(\frac{\ell}{\tilde \chi}\right) \left|T^\varphi\left(\frac{\ell}{\tilde \chi},\tilde \chi\right) \alpha(\tilde \chi)
\int_{\tilde \chi}^\infty \dd \chi \snumb(\chi)\frac{(\chi-\tilde \chi)}{\chi \tilde \chi}\right|^2\,.
\ee
A numerical analysis of some simple anisotropic phenomenological models, together with observational constraints from Euclid~\cite{euclid} and SKA~\cite{ska} surveys, will appear in a companion paper~\cite{ppulett15}.

\section{Discussion}\label{sec8}

In this article we have derived the observational signature of a late time anisotropic expansion on the weak-lensing observables. To that purpose, we have provided all the technical tools, including the evolution of the background spacetime, the perturbation theory, the description of the evolution of a geodesic bundle and the manipulation of observables on the celestial sphere. 

Our strategy is to adopt an approach based on the observer point of view, in which all observables are expressed in terms of the direction of observation at the observer. Since a full solution to the problem cannot be attained straightforwardly, and given that CMB observations suggest that spatial anisotropy cannot be too large, we have developed a small shear approximation scheme. It allowed us to identify the following contributions compared to the standard Friedmann-Lema\^{\i}tre case:
\begin{enumerate}
\item the tensor and vector contributions to the source of the Sachs equation, which starts at order $\oo$, and the contribution of the scalar modes coupled to the geometrical shear, which is of order $\oo$ as well;
\item the evolution of all the perturbative modes, that is, of the transfer functions, which are decomposed as $T^{X_s}({\bm k},t)$, where the dependence with the direction of $\bm{k}$ comes from the coupling with the geometrical shear in the Einstein equation (for that, see the Appendix~\ref{secA});
\item the fact that the geodesic deviates from is Friedmannian form and which leads to post-Born corrections;
\item the effect of the source distributions which are affected by the background shear or the scalar perturbations -- that are respectively at orders $\oz$ and $\zo$ -- and for which we would in principle need a theory of structure formation;
\end{enumerate}
We have then argued that the dominant term is related to the orthoradial displacement of the central geodesic on which the Sachs equation is evaluated, when compared with the background geodesic. 

While we have provided all the elements to perform the full computation, we have focused on this dominant term and demonstrated that there exist 5 off-diagonal correlators between $E_{\ell m}$, $B_{\ell m}$ and $\kappa_{\ell m}$ each of which has 5 independent components and thus allow in principle to fully reconstruct the geometrical shear $\sigma_{ij}$. All of them are linear in $\sigma/\HH$ and only two of them involve the $B$-modes. We advocate that their measurements in future surveys such as Euclid and SKA, on scales where the linear regime holds, can set strong constraints on the anisotropy. The amplitude for these two surveys is estimated in our companion article~\cite{ppulett15}.

The existence of non-vanishing $B$-modes also reflects iteslf in the existence of an angular power spectrum that is quadratic in $\sigma/\HH$. While probably easier to measure, it does not allow one to fully reconstruct the shear $\sigma_{ij}$.

This analysis sets the ground for stronger constraints on an anisotropic expansion, and possibly on the anisotropic stress on the dark energy sector. The new estimators that we proposed will also allow the control of systematics, and are new in the weak-lensing literature. 

\begin{acknowledgments} 
We thank  Pierre Fleury and Yannick Mellier for discussions. This work was made in the ILP LABEX (under reference ANR-10-LABX-63) and was supported by French state funds managed by the ANR within the Investissements d'Avenir programme under reference ANR--11--IDEX--0004--02, the Programme National Cosmologie et Galaxies, and the ANR THALES (ANR-10-BLAN-0507-01-02). Thiago Pereira thanks the Institut d'Astrophysique de Paris for hospitality during his visits. Cyril Pitrou thanks the State University of Londrina for its friendly reception during the conclusion of this work.
  \end{acknowledgments}



\pagebreak
\appendix

\section{Perturbation theory in Bianchi~$I$ universes}\label{secA}

This section summarizes the general framework of linear perturbation theory in Bianchi~$I$ universes. Our approach is an extension of the formalism we introduced in Ref.~\cite{ppu1}, where perturbation theory was developed in the context of an early anisotropic stage. Here, we adapt this formalism for the physics of the late-time universe. Before we introduce the parameterization of the perturbations and the whole machinery of gauge-ivariant linear perturbation theory, we summarize some results regarding the appropriate Fourier transform in anisotropic spacetimes, and show how they can be used to extract the dynamics of scalar, vector and tensor modes from Einstein's equations. We then use these tools to decompose the background shear and anisotropic stress in a general basis adapted to our coordinate system.

\subsection{Mode decomposition}\label{sec:FourierMode}

\subsubsection{Fourier and SVT decomposition}

In order to correctly describe the dynamics of perturbative modes one needs a complete set of spatial eigenfunctions adapted to the symmetries of the spacetime one is dealling with. Since Bianchi~$I$ universes are spatially flat, at each constant time hypersurface these eigenfunctions are standard plane waves. Therefore, any scalar function of the comoving coordinates $\{x^i\}$ and time can be Fourier decomposed as
\be
f(x^j,\eta)=\int\frac{{\rm d}^3k_i}{(2\pi)^{3/2}}\hat{f}(k_i,\eta)e^{\ii k_ix^i}\,.
\ee
with the obvious inverse transfomation. Due to the lack of rotational symmetry, the direction of a wave vector will vary with time. In particular, since $k_i$ is constant, $k^i=\gamma^{ij}k_j$ vary with time -- its rate of change being given by:
\be\label{kiprime}
   (k^i)'=-2\sigma^{ij}k_j\,.
\ee
Note however that $k_ix^i=k^ix_i$ remains constant. From the above expression, one can easily deduce the time evolution of the modulus $k^2\equiv k^ik_j$ and unit vector $\kiu\equiv k^i/k$ as
\be
\label{khatprime}
\frac{k'}{k}=-\sigma^{ij}\kid\kjd\,,\qquad (\kiu)'=(\sigma^{jl}\kjd\kld)\kiu-2\sigma^{ij}\kjd\,.
\ee
As we are going to see, these expressions are crucial for extracting different perturbative modes from Einstein equations.

Once equiped with a Fourier transform, we can proceed and decompose any 3-dimensional geometrical object in terms of its scalar, vector and tensor pieces. We start by decomposing any (3-dimensional) vector $V_i$ in its longitudinal and transverse pieces as $V_i=\dbi V+\bar{V}_i$, with $\dhi\bar{V}_i=0$. In Fourier space, this decomposition is equivalent to\footnote{Note that we can always reabsorb $\ii$ factors in the terms of the decomposition.}
\be
V_i=\kid V+\bar{V}_i\,,\qquad \kiu\bar{V}_i=0\,.
\ee
Since $\bar{V}_i$ is orthogonal to $\kiu$, it can be further decomposed as
\be
\label{vbar}
\bar{V}_i=\sum_{a=1,2}V_a(\kiu,\eta)e^a(\kid)
\ee
where $\{e^a_i\}$ represents a 2-dimensional basis defined so that
\be
e^a_ik^i=0\,,\qquad e^a_ie^b_i\gamma^{ij}=\delta^{ab}\,.
\ee

Analogously, any (3-dimensional) symmetric tensor $V_{ij}$ can be decomposed into a trace plus a traceless part as 
$V_{ij}=T\gamma_{ij}+\Delta_{ij}S+2\partial_{(i}\bar{V}_{j)}+2\bar{V}_{ij}$, where $\Delta_{ij}\equiv\dbi\dbj-\gamma_{ij}\Delta/3$, $\bar{V}_i$ is transverse and $\bar{V}_{ij}$ is transverse and traceless. In Fourier space, such decomposition becomes
\be
V_{ij}=T\gamma_{ij}+\left(\kid\kjd-\frac{\gamma_{ij}}{3}\right)S+2\hat{k}_{(i}\bar{V}_{j)}+\bar{V}_{ij}
\ee
where $\bar{V}_i$ is given by Eq.~(\ref{vbar}). $\bar{V}_{ij}$ is a transverse and traceless tensor decomposed as
\be
\bar{V}_{ij}=\sum_{\lambda=+,\times}V_\lambda(\kiu,\eta)\varepsilon^\lambda_{ij}(\kiu)
\ee
with the traceless ($\varepsilon^\lambda_{ij}\gamma^{ij}=0$), transverse ($\varepsilon^\lambda_{ij}\kiu=0$) and perpendicular ($\varepsilon^\lambda_{ij}\varepsilon_\mu^{ij}=\delta^\lambda_\mu$) polarization tensor being defined as:
\be
\varepsilon^\lambda_{ij}=\frac{e^1_ie^1_j-e^2_ie^2_j}{\sqrt{2}}\delta^\lambda_{+}+
\frac{e^1_ie^2_j+e^2_ie^1_j}{\sqrt{2}}\delta^\lambda_\times\,.
\ee

Given the above decomposition, the correspondence between SVT components of any geometrical equation can be extracted uniquelly. For example, the scalar part of any vectorial equation of the form $V_i=0$ can be extracted by projecting it along $\kiu$, whereas its vector part can be extracted with the help of the projector
\be\label{projectors1}
P_{ij}\equiv\gamma_{ij}-\hat{k}_i\hat{k}_j=e^1_ie^1_j+e^2_ie^2_j\,.
\ee
Likewise, the scalar components of any tensorial equation like $V_{ij}=0$ can be extracted by projecting it along 
$\gamma_{ij}$ and $T_{ij}$, with the later projector defined as
\be\label{projectors2}
T_{ij}\equiv\hat{k}_i\hat{k}_j-\frac{1}{3}\gamma_{ij}\,.
\ee
The remaining vector and tensor pieces can be extracted with the help of $P^i_l\kju$ and $\Lambda_{ij}^{ab}$, respectively, where
\be\label{projectors3}
\Lambda_{ij}^{ab}=P^a_iP^b_j-\frac{1}{2}P_{ij}P^{ab}\,.
\ee
In conclusion, the SVT degrees of freedom of any tri-dimensional vector and tensor are given explicitly by
\bea
V_i&=&\left(\kju V_j\right)\kid+P^j_iV_j\,,\nonumber\\
V_{ij}&=&\left(\frac{1}{3}\gamma^{kl}V_{kl}\right)\gamma_{ij}+\left(\frac{3}{2}T^{kl}V_{kl}\right)T_{ij}
+2\hat{k}_{(i}\left[P_{j)}^{m}\hat{k}^n V_{mn}\right]+\Lambda^{mn}_{ij}V_{mn}\,.\nonumber
\eea

\subsubsection{Mode evolution}

The SVT decomposition introduced above is based on the properties of a tensor according to the rotation symmetries of the background spacetime. As such, in Friedmann-Lema\^{\i}tre spacetimes this decomposition is always possible and will hold during the entire cosmic evolution. In the Bianchi~$I$ case, on the other hand, this decomposition will hold, strictly speaking, only on a given constant-time hypersurface. Due to the anisotropic evolution of space, SVT modes which are initially decoupled will couple nontrivialy as time evolves, implying in a set of coupled dynamical equations already at linear order in perturbations. Therefore, it is important to have expressions for the time evolution of basis vectors and polarization tensors, which will be directly dependent on the spacetime shear. We have already met the time evolution of 
$\kiu$; Eq.~(\ref{khatprime}). For completeness, we also give the time evolution of the vector $e^a_i$ and polarization tensor $\varepsilon^\lambda_{ij}$ \cite{ppu1}
\bea
(\kiu)'&=&(\sigma^{jl}\hat{k}_j\hat{k}_l)\kiu-2\sigma^{ij}\kjd\,,\slabel{modeevok}\\
\lp e^i_a\rp'&=&-\sum_{b}(\sigma_{jl}e_a^je_b^l)e^i_b\,,\slabel{modeevoe}\\
\lp \varepsilon^\lambda_{ij}\rp'&=&-\lp\sigma^{kl}\varepsilon_{kl}^\lambda\rp P_{ij}-\lp\sigma^{kl}P_{kl}\rp\varepsilon^\lambda_{ij}
+4\sigma^k_{(i}\varepsilon^\lambda_{j)k}\,.\slabel{modeevoepsilon}
\eea

Special care is needed when extracting SVT modes from Einstein equations, for the projections of SVT modes do not commute with time evolution anymore. As an illustration, let us consider the extraction of the scalar component of an equation like $(\bar{V}_i)'+\HH\bar{V}_i=0$, where $\bar{V}_i$ is any transverse tensor. In FLRW this equation does not have a scalar component, since $\kiu\bar{V}_i=0$. However, due to Eq.~(\ref{khatprime}) we now get
\be
\kiu\lcr(\bar{V}_i)'+\HH\bar{V}_i\rcr=(\kiu\bar{V}_i)'-(\kiu)'\bar{V}_i=2\sigma^{ij}\kjd\bar{V}_i
\ee
which is only zero when $\sigma_{ij}=0$. Further mode-extracting relations can be easily found in an analogous manner. For a comprehensive list of relations the reader can check Ref.~\cite{ppu1}.

\subsubsection{Background shear and anisotropic stress}

Both the (background) spatial shear $\sigma_{ij}$ and spatial anisotropic stress $\Pi_{ij}$ are transverse and traceless tensors. As such, each of them is described by five independent degrees of freedom, which are best described in the basis $\{\kid,e^1_i,e^2_i\}$ adapted to the modes we are considering. In this basis, these two tensors can be written in terms of 10 new scalar functions as
\bea
\label{decshear}
 \sigma_{ij} &=& \frac{3}{2}\left(\hat k_i\hat k_j-\frac{1}{3}\gamma_{ij}\right)\spar
 + 2\sum_{a=1,2}\sva \,\hat k_{(i}e^a_{j)}
 + \sum_{\lambda=+,\times}\stl\,\varepsilon^\lambda_{ij}\,,\\
\label{decstress}
 \Pi_{ij} &=& \frac{3}{2}\left(\hat k_i\hat k_j-\frac{1}{3}\gamma_{ij}\right)\ppar
 + 2\sum_{a=1,2}\pva \,\hat k_{(i}e^a_{j)}
 + \sum_{\lambda=+,\times}\ptl\,\varepsilon^\lambda_{ij}\,.
\eea
It is important to note that these new functions, which depend of both $\kid$ and time, are not the Fourier transform of their respective tensors, which in fact  are homogeneous and depend only on time. In other words, the dependence of $(\spar,\sva,\stl)$ and $(\ppar,\pva,\ptl)$ with $\kid$ arises solely from the local anisotropy of space.

With the help of background Einstein equations~(\ref{g00}) and~(\ref{gTT})  written in conformal time and the mode evolution described by Eqs.~(\ref{modeevok}), one can show with a bit of algebra that
\bea
 \sparp &\equiv& \spar'+2\HH\spar +2\sum_a\sva^2 = \kappa a^2\ppar\,, \label{background_sigma}\\
 \svap &\equiv& \sva' +2\HH\sva -\frac{3}{2}\sva\spar +\sum_{b,\lambda}\svb\stl\mathcal{M}_{ab}^\lambda 
  = \kappa a^2\pva\,, \\
 \stlp &\equiv& \stl'+2\HH\stl -2\sum_{a,b}\mathcal{M}_{ab}^\lambda\sva\svb = \kappa a^2\ptl 
 \label{background_sigma3}\,,
\eea
where $\mathcal{M}^\lambda_{ab}$ is defined as \cite{ppu1} 
\[
{\cal M}_{ab}^{\lambda}=\frac{1}{\sqrt{2}}\left(\begin{array}{cc}
1 & 0\\
0 & -1
\end{array}\right)\delta_{+}^{\lambda}+\frac{1}{\sqrt{2}}\left(\begin{array}{cc}
0 & 1\\
1 & 0
\end{array}\right)\delta_{\times}^{\lambda}\,.
\]

\subsection{Gauge-invariant variables}\label{secA.2}

\subsubsection{Geometry}

The most general linearly perturbed metric over a Bianchi~$I$ spacetime can be parameterized as follows
\be
\label{perturbedmetric2}
\dd s^2 = a^2[-(1+2A)\dd\eta^2+2B_i\dd x^i\dd\eta+(\gamma_{ij}+h_{ij})\dd x^i\dd x^j]
\ee
where $A$ is a free scalar function and
\bea
\label{vectorB} 
B_i & \equiv & \dbi B + \bar{B}_i\,,\\
\label{hij}
h_{ij} & \equiv & 2C \left(\gamma_{ij}+\frac{\sigma_{ij}}{\HH}\right)+
2\dbi\dbj E + 2\partial_{(i}E_{j)}+2E_{ij}\,,
\eea
defined together with the usual transversality and trace-free conditions:
\be
\dbi \bar{B}^i=0=\dbi E^i,\qquad E^i_i=0=\dbi E^{ij}\,.
\ee

Under an active coordinate transformation, the coordinates of any point will change according to
\be
\label{gt}
x^\mu\rightarrow\tilde{x}^\mu=x^\mu-\xi^\mu(x^\nu)\,,
\ee
where the gauge vector $\xi^\mu$ is itself decomposed as
\be
\xi^0=T\,,\qquad\xi^i=\dhi L + L^i
\ee
with $\dbi L^i=0$. Under the transformation of Eq.~(\ref{gt}), the perturbations of the metric will transform 
as
\be
\delta g_{\mu\nu}\rightarrow\delta g_{\mu\nu}+{\cal L}_\xi\bar{g}_{\mu\nu}
\ee
where ${\cal L}_\xi\bar{g}_{\mu\nu}$ is the Lie derivative of the background metric along $\xi$. Using the above parameterization and the mode decomposition introduced in \S~\ref{sec:FourierMode}, it is straightforward to show that the scalar and vector metric potentials transform respectively as
\bea
A &\rightarrow& A + T' + \mathcal{H}T\,, \\
B &\rightarrow& B - T + \frac{(k^2L)'}{k^2}\,, \\
C &\rightarrow& C + \mathcal{H}T\,, \\
E &\rightarrow& E + L \,, \\
\bar{B}_i &\rightarrow& \bar{B}_i + \gamma_{ij}(L^j)' -2{\rm i}k^j\sigma_{lj}P^l_iL\,,\\
E_i &\rightarrow& E_i + L_i \,,
\eea
whereas $E_{ij}$ is automatically gauge invariant. Based on these transformations, we can construct
the following gauge-invariant variables:
\bea\label{eq.A25}
\Phi &=& A + \frac{1}{a}\left\{a\left[B-\frac{(k^2E)'}{k^2}\right]\right\}'\,, \\
\Psi &=& -C -\mathcal{H}\left[B-\frac{(k^2E)'}{k^2}\right]\,, \\ 
\Phi_i &=& \bar{B}_i - \gamma_{ij}(E^j)'+2{\rm i}k^j\sigma_{lj}P^l_iE\,.
\eea
It is easily verifiable that whenever $\sigma_{ij}=0$ the Fourier wave vector $k$ will be constant and the above variables become the standard Bardeen variables for a Friedmann-Lema\^{\i}tre universe.

\subsubsection{Matter sector}

Moving forward, we now parameterize the perturbations of the energy-momentum tensor defined in Eq.~(\ref{e:Tmunu_de}), which can be decomposed as
\be
T^\mu_\nu=\bar{T}^\mu_\nu+\delta T^\mu_\nu\,.
\ee
From the normalization condition of the fluid total four velocity we can write
\be
\delta u^\mu = \frac{1}{a}(-A,v^i)\,,\qquad v^i=\dhi v+\bar{v}^i\,.
\ee
with $\dbi\bar{v}^i=0$, as usual. Likewise, the perturbations to the energy density ($\delta\rho$), pressure ($\delta P$) and anisotropic stress ($\delta\pi^i_j$) are introduced as follows:
\bea
\delta T^0_0 &=& - \delta\rho\,, \\
\delta T^0_i &=& [\rho(1+w)\gamma_{ij}+\Pi_{ij}](v^j+B^j)\,, \\
\delta T^i_0 &=& -\rho(1+w)v^i+\gamma^{ij}\delta\pi_{j0}\,, \\
\delta T^i_j &=& \delta P\delta^i_j+\gamma^{il}\delta\pi_{lj}-\Pi_{jk}h^{ki}\,,
\eea
where $B_i$ and $h_{ij}$ were defined in Eqs.~(\ref{vectorB}). Special care to the notation is in order here because, as one can check,
$\bar{g}^{\mu\lambda}\delta\pi_{\lambda\nu}\neq\delta\pi^\mu_\nu\equiv 
\bar{g}^{\mu\lambda}\delta\pi_{\lambda\nu}+\Pi_{\lambda\nu}\delta g^{\lambda\mu}$.

We also need to parametrize the perturbed anisotropic stress tensor $\delta\pi^i_j$. From the transversality condition 
$(u^\mu+\delta u^\mu)(\Pi_{\mu\nu}+\delta\pi_{\mu\nu})=0$, we conclude that
\be
\delta\pi_{00}=0\,,\quad\delta\pi_{0i}=-\Pi_{ij}v^j\,.
\ee
Note however that these conditions do not fix $\delta\pi_{ij}$. We therefore further decompose $\delta\pi_{ij}$ as
\be
\label{perturbed-stress}
\delta\pi_{ij}=2\left[\pi^T\gamma_{ij}+\dbi\dbj\pi^S+\partial_{(i}\pi^V_{j)}+\pi^T_{ij}\right]
\ee
where $T$ in $\pi^{T}$ stands for ``trace'' and where, as usual, we have
\be
\dhi\pi^V_i=0=\dhi\pi^T_{ij}\,,\qquad\pi^{Ti}_{\phantom{T}i}=0\,.
\ee 
Moreover, note that 
$\bar{g}^{\mu\nu}\delta\pi_{\mu\nu}=-\Pi_{\mu\nu}\delta g^{\mu\nu}\neq0$, which is why the 
above decomposition tensor has a trace.

Under the gauge transformation (\ref{gt}) and using again the appropriate Fourier decomposition, 
the above variables transform as
\bea	
\delta\rho &\rightarrow& \delta\rho+\rho'T\,, \\
\delta P &\rightarrow& \delta P +P'T\,, \\
v &\rightarrow& v-\frac{(k^2L)'}{k^2}\,, \\
\bar{v}^i &\rightarrow& \bar{v}^i-(L^i)'+2{\rm i}k^j\sigma_{lj}P^{li}L\,\,.
\eea
These transformations suggest the introduction of the following gauge-invariant variables
\bea\label{eq.A33}
\delta\hat{\rho} &\equiv& \delta\rho+\rho'\left[B-\frac{(k^2E)'}{k^2}\right]\,, \\
\delta\hat{P} &\equiv& \delta P+P'\left[B-\frac{(k^2E)'}{k^2}\right]\,, \\
\hat{v} &\equiv& v+\frac{(k^2E)'}{k^2}\,, \\
\hat{\bar{v}}^i &\equiv& \bar{v}^i+\bar{B}^i\,.
\eea

The perturbed variables in Eq.~(\ref{perturbed-stress}), on the other hand, do not have simple transformations as above, essentially 
because there is no simplifying relation between the background tensor $\Pi_{\mu\nu}$ and the wave vector $k^i$. 
Using
\[
\mathcal{L}_\xi\Pi_{ij}=\Pi_{ij}'T+\Pi_{il}\xi^l_{\,,\,j}+\Pi_{jl}\xi^l_{\,,\,i}
\]
we find that
\bea
\pi^T &\rightarrow& \pi^T+\left(-\frac{1}{4}T^{ij}\Pi_{ij}'+\frac{1}{3}\sigma^{ij}\Pi_{ij}\right)T\,,\\
\pi^S &\rightarrow& \pi^S - \frac{3}{4k^2}T^{ij}\Pi_{ij}'\,T
+\Pi_{ij}\hat{k}^i\hat{k}^jL-{\rm i}\Pi_{ij}\frac{\hat{k}^i}{k}L^j\,,\\
\pi^V_i &\rightarrow& \pi^V_i-\frac{\rm i}{k}P^j_i\hat{k}^l\Pi_{jl}'\,T
+{\rm i}kP^j_i\Pi_{jl}\klu L + P^j_i\Pi_{jl}L^l\,,\\
\pi^T_{ij}&\rightarrow&\pi^T_{ij}+\frac{1}{2}\Lambda^{lm}_{ij}\Pi_{lm}'\,T \,,
\eea 
where $P_{ij}$, $T_{ij}$ and $\Lambda^{ij}_{kl}$ were defined in Eqs.~(\ref{projectors1}),~(\ref{projectors2}) 
and~(\ref{projectors3}). From the variables above we construct the following new variables
\bea \label{eq.A35}
\hat{\pi}^T &\equiv& \pi^T+\left[-\frac{1}{4}T^{ij}\Pi_{ij}'+\frac{1}{3}\sigma^{ij}\Pi_{ij}\right]
\left(B-\frac{(k^2E)'}{k^2}\right)\,,\\
\hat{\pi}^S &\equiv& \pi^S-\frac{3}{4k^2}T^{ij}\Pi_{ij}'\left(B-\frac{(k^2E)'}{k^2}\right)
-\Pi_{ij}\hat{k}^i\hat{k}^jE+{\rm i}\Pi_{ij}\frac{\hat{k}^i}{k}E^j\,,\\
\hat{\pi}^V_{i}&\equiv& \pi^V_{i}-\frac{\rm i}{k}P^j_i\hat{k}^l\Pi_{jl}'\left(B-\frac{(k^2E)'}{k^2}\right)-
{\rm i}kP^j_i\Pi_{jl}\klu E - P^j_i\Pi_{jl}E^l\,,\\
\hat{\pi}^T_{ij}&\equiv& \pi^T_{ij}+\frac{1}{2}\Lambda^{lm}_{ij}\Pi_{lm}'\left(B-\frac{(k^2E)'}{k^2}\right)\,,
\eea
which, as one can easily check, are gauge-invariant.

\subsubsection{Gauge choice}

From the construction of gauge-invariant variables presented above, it is clear that an enormous simplification will 
be achieved if we work in a gauge where
\be
\label{gauge-choice}
B=E=0=E^i\,.
\ee
In this gauge the scalar modes become
\be
\Phi=A\,,\quad \Psi=-C\,,\quad \delta\hat{\rho}=\delta\rho\,,\quad \delta\hat{P}=\delta P\,,\quad
\hat{v}=v\,,\quad \hat{\pi}^T=\pi^T\,,\quad \hat{\pi}^S=\pi^S,
\ee
whereas the vector and tensor variables become
\be
\bar{\Phi}_i=\bar{B}_i\,,\quad \hat{\bar{v}}_i=\bar{v}_i+\bar{B}_i\,,\quad
\hat{\pi}^V_i=\pi^V_i\,,\quad \hat{\pi}^T_{ij}=\pi^T_{ij}\,.
\ee
Apart from the spatial velocity $\hat{\bar{v}}_i$, in this gauge the gauge-invariant variables coincide with the original potentials. In other words, by working in this gauge the final equations can be trivially (again, apart 
from $\hat{\bar{v}}_i$) replaced with the same equations satisfied by gauge-invariant variables. Note that this choice fixes the gauge completely, and is slightly different from the choice made in \cite{ppu1}.

\subsection{Perturbed Einstein's equations in Bianchi $I$}\label{secA.3}

We have now everything needed to obtain the fully mode-projected and gauge-invariant Einstein equations. This is a tedious but straightforward procedure which requires careful comutation of time derivatives and the Fourier vectors through the use of Eqs.~(\ref{modeevok}-\ref{modeevoepsilon}). We note that the main difference with the approach followed in Ref.~\cite{ppu1} is that the trick bellow Eq. (3.21) in \cite{ppu1} cannot be used when $\Pi_{ij}$ is non-zero.

\subsubsection{Scalar modes}

Einstein equations give four equations for the evolution of the scalar modes. The first of them comes from $\delta G^0_0=\kappa\delta T^0_0$, and is given by
\bea
&& k^2\Psi+3\HH(\Psi'+\HH\Phi)-\frac{1}{2}\sigma^2\left[X-3\Psi\right]-\frac{k^2}{2\HH}\spar\Psi
-\frac{1}{2}\ii k\sum_a\sva\Phi_a-\frac{1}{4\HH}\Psi\left[(\sigma^2)'+4\HH\sigma^2\right]\nonumber\\
&& +\frac{1}{2}\sum_\lambda\left[E_\lambda'\stl+2E_\lambda\sum_{a,b}\sva\svb\mathcal{M}^\lambda_{a,b}\right]
=-a^2\frac{\kappa}{2}\delta\rho\,.
\eea
where, for simplicity, we have introduced the new variable
\[
X\equiv\Psi+\Phi+\left(\frac{\Psi}{\HH}\right)'\,.
\]
The second scalar equation can be extracted from $\delta G^0_i=\kappa\delta T^0_i$ by projecting it along the vector  $\hat{k}^i$ and is given by
\be
\Psi'+\HH\Phi-\frac{\spar}{2}X+(\sigma^2-\sparp)\frac{\Psi}{2\HH}-\frac{1}{2}\sum_\lambda E_\lambda\stl
=-\frac{a^2}{2}\kappa\left[\rho(1+w)v+\ppar v-\frac{\ii}{k}\sum_a\pva v_a\right]\,.
\ee
The third and fourth equations come from trace and traceless part of $\delta G^i_j=\kappa\delta T^i_j$. They are
\bea
&& \Psi''+2\HH\Psi'+\HH\Phi'+(2\HH'+\HH^2)\Phi-\frac{1}{3}k^2(\Phi-\Psi)-\frac{1}{6}\frac{k^2}{\HH}\spar\Psi
+\frac{1}{2}\sigma^2(X-3\Psi) \nonumber\\
&& -\frac{1}{2}\sum_\lambda\left(E_\lambda'\stl+2E_\lambda\sum_{a,b}\sva\svb\mathcal{M}^\lambda_{ab}\right)
+\frac{1}{2}\ii k\sum_a\sva\Phi_a+\frac{\Psi}{4\HH}\left[(\sigma^2)'+4\HH\sigma^2\right]\nonumber\\
&& = a^2\kappa\left[\frac{1}{2}\delta P+\pi^T-\frac{1}{3}k^2\pi^S+
\frac{1}{3}\left(\frac{\Psi}{\HH}\Pi^{ij}\sigma_{ij}-\sum_\lambda E_\lambda\ptl\right)
\right]
\eea
and
\bea
\label{eq:scalar_modes}
&& \frac{2}{3}k^2(\Phi-\Psi)-\spar\left[X'-\frac{k^2}{3}\frac{\Psi}{\HH}\right]-2\ii k\sum_a\sva\Phi_a
-2X\sparp-\frac{\Psi}{\HH}\left(\sparp\right)'
+4\sum_{a,b,\lambda}\sva\svb E_\lambda\mathcal{M}^\lambda_{ab}\nonumber\\
&&= a^2\kappa\left[-\frac{4}{3}k^2\pi^S+\frac{\Psi}{\HH}\ppar\spar
+\frac{8}{3}\frac{\Psi}{\HH}\sum_a\pva\sva-\frac{2}{3}\frac{\Psi}{\HH}\sum_{\lambda}\stl\ptl
+\frac{2}{3}\sum_\lambda E_\lambda\ptl\right]
\eea
respectively. Note that, despite the appearence of $\rm{i}$ factors, these equations are real.

\subsubsection{Vector modes}

The two equations for the vector modes can be obtained through the combinations  $e^i_a(\delta G^0_i-\kappa\delta T^0_i)=0$ and  $\kid e^j_a(\delta G^i_j-\kappa\delta T^i_j)=0$. They are given respectively by
\bea
&& \Phi_a-\frac{2\ii}{k}\sva X + \frac{4\ii}{k}\sum_{b,\lambda}E_\lambda\svb\mathcal{M}^\lambda_{ab}\nonumber\\
&& = \frac{-2a^2\kappa}{k^2}\left[\rho(1+w)v_a+\ii k\pva v -\frac{1}{2}\ppar v_a+
\sum_{b,\lambda}\mathcal{M}^\lambda_{ab}\ptl v_b-\ii k\frac{\Psi}{\HH}\pva\right]\,.
\eea
and
\bea
\label{eq:vector_modes}
&& \Phi_a'+2\HH\Phi_a-\frac{5}{2}\spar\Phi_a
+\sum_{b,\lambda}\Phi_b\stl\mathcal{M}^{\lambda}_{ab}-\frac{2\rm{i}}{k}\sva X'\nonumber \\
&& + \frac{2\rm{i}\Psi}{k\HH}\left[3\spar\svap -3\sva\sparp
+2\sum_{b,\lambda}\mathcal{M}^\lambda_{ab}\left(\svb\stlp-\stl\svbp\right)\right]\nonumber \\
&& +\frac{4\rm{i}}{k}\sum_{b,\lambda}E_\lambda\left[\mathcal{M}^\lambda_{ab}
\svbp+\mathcal{N}_{ab}\svb(\stplus\delta^\times_\lambda-\stcross\delta^+_\lambda)\right]
+\frac{4\rm{i}}{k}\sum_{b,\lambda}E'_\lambda\svb\mathcal{M}^\lambda_{ab}\nonumber \\
&& -\frac{4\rm{i}}{k}X\svap-\frac{2\rm{i}\Psi}{k\HH}\left[(\svap)'-\frac{3\spar}{2}\svap
-2\HH\svap+\sum_{b,\lambda}\mathcal{M}^\lambda_{ab}\svbp\right]\nonumber\\
&& =\frac{2{\rm i}}{k}a^2\kappa\left[{\rm{i}}k\pi^V_a+2\Psi\pva+2\frac{\Psi}{\HH}\left(
\spar\pva-\frac{1}{2}\sva\spar+\sum_{b,\lambda}\svb\ptl\mathcal{M}^\lambda_{ab}\right)\right]\,.
\eea

\subsubsection{Tensor modes}

There is only one dynamical equation for the tensor modes. This equation follows from the projection $\varepsilon^{\lambda\,j}_i(\delta G^i_j-\kappa\delta T^i_j)=0$, which gives
\bea
\label{eq:tensor_modes}
&& E_\lambda''+2\HH E_\lambda'+k^2E_\lambda-2E_\lambda\sum_a\sva^2-2E_\lambda\stoneminusl^2
+2E_{(1-\lambda)}\stplus\stcross \nonumber \\
&& -\stl\left[k^2\left(\frac{\Psi}{\HH}\right)+X'\right]-2X\stlp
+2{\rm i}k\sum_{a,b}\svb\Phi_a\mathcal{M}^\lambda_{ab}-\frac{\Psi}{\HH}(\stlp)'\nonumber\\
&& =a^2\kappa\left[2\pi^T_\lambda+\ppar E_\lambda
-\frac{\Psi}{\HH}\left(\spar\ptl+\stl\ppar+\sum_{a,b}\pva\svb\mathcal{M}^\lambda_{ab}\right)\right].
\eea

\subsection{Perturbed Fluid equations}\label{secA.4}

The perturbed conservation equation follows from
\be
\label{pertconseq}
(\delta\nabla_\mu)T^\mu_\nu+\nabla_\mu(\delta T^\mu_\nu)=0\,.
\ee
Working in the gauge (\ref{gauge-choice}), the time component ($\nu=0$) of the above expression gives the perturbed continuity equation
\bea
&&\delta\rho'+\rho(1+w)\nabla^{2}v+3\mathcal{H}(\delta\rho+\delta P)-(\rho+P)3\Psi'=
\nonumber \\
&&\partial^{j}\delta\pi_{j0}-\mathcal{H}\gamma^{ij}\delta\pi_{ij}+\mathcal{H}\Pi_{ij}h^{ij}
-\sigma_{j}^{i}\delta\pi_{il}\gamma^{lj}+\sigma_{j}^{i}\Pi_{il}h^{lj}-\frac{1}{2}(h_{i}^{j})'\Pi_{j}^{i}\,,
\eea
where we remind that $h_{ij}$ was introduced in Eq.~(\ref{hij}). Likewise, the perturbed Euler equation follows from spatial part ($\nu=i$) of Eq.~(\ref{pertconseq}). We find
\bea
&&\frac{\partial}{\partial\eta}\{[\rho(1+w)\gamma_{ij}+\Pi_{ij}]v^j\}+\partial_i\delta P 
+\partial^l\delta\pi_{li}+(1+w)\rho\partial_i\Phi+\Pi^j_i\partial_j\Phi\nonumber\\
&&+4\HH\{[\rho(1+w)\gamma_{ij}+\Pi_{ij}]v^j\}-\Pi_{il}\partial_j h^{lj}
+\frac{1}{2}\Pi^j_i\partial_j h^l_l - \frac{1}{2}\Pi^{jk}\partial_i h_{jk}=0\,.
\eea

Despite their generality, the above equations are not very useful since they are implemented in real space. In order to obtain their Fourier counterparts we need to project these equations along the scalar $(\hat{k}^i)$ and vector $(e^i_a)$ modes. This mode extraction procedure is tedious but straightforward, and requires special attention to the use of the evolution Eqs.~(\ref{modeevok}a-\ref{modeevoepsilon}b) of the Fourier wave vectors.

\subsubsection{Scalar modes}

Both continuity and Euler equations lead to conservation equations for the scalar modes. They are given 
respectively by

\bea
\label{pertconteq}
&&\delta'+3\HH[(c_s^2-w)\delta+\omega\Gamma]-(1+w)(k^2v+3\Psi')
-\frac{\delta}{\rho}\sigma^{ij}\Pi_ {ij}= 
\frac{1}{\rho}\left\{\left[\left(\frac{\Psi}{\HH}\right)'\sigma_{ij}-\frac{\Psi}{\HH}(\sigma_{ij})'
\right]\Pi^{ij}\right.\nonumber \\
&&\left. +2a^2\kappa\frac{\Psi}{\HH}\Pi^{ij}\Pi_{ij}-8\Psi\Pi^{ij}\sigma_{ij}+k^2\ppar v
-\ii k\sum_a\pva(v_a-\Phi_a)-6\HH\pi^T+2k^2\pi^S(\HH+\spar)\nonumber\right.\\
&& \left. -2{\rm i}k\sum_a\sva\pi^V_a -2\sum_\lambda\stl\pi^T_\lambda
-\sum_\lambda E_\lambda(\stl\ppar+\spar\ptl)
+\sum_\lambda\Pi_\lambda(2\HH E_\lambda- E_\lambda')\right\}
\eea
and
\bea
\label{perteulereq}
&& v'+\HH(1-3c^2_s)v+\Phi+\frac{1}{1+w}\left[c_s^2\delta+w\Gamma + 
\frac{2}{\rho}(\pi^T-k^2\pi^S)\right]-\frac{2 \ii}{k}\sum_a\sva v_a
-\frac{\sigma^{ij}\Pi_{ij}}{\rho}v\left(\frac{1+c^2_s}{1+w}\right) =\nonumber \\
&& \frac{-1}{\ii k(1+w)\rho}\left\{\ii k(\ppar v)' + \ii k\ppar\Phi 
+4\HH(\ii k\ppar v +\sum_a\pva v_a) + \ii k\Psi\ppar\left(\frac{2\spar}{\HH}-1\right)
-\ii k\sum_\lambda \ptl E_\lambda\nonumber\right.\\
&&\left.+\left(2\sum_a\pva\sva+\Pi^{ij}\sigma_{ij}\right)\frac{\ii k\Psi}{\HH}
+\sum_a\left[(\pva v_a)'+\spar\pva v_a
+ 2\ii k\sva\pva v\right]\right.\nonumber\\
&&\left.-\sum_a\left(\ppar\sva v_a + 2\sum_{b,\lambda}\sva v_b\mathcal{M}^\lambda_{ab}\ptl\right)\right\}\,,
\eea
where we have made use of the equation
\be
w'=-\left[3\HH(1+w)+\frac{\Pi^{ij}\sigma_{ij}}{\rho}\right](c^2_s-w)\,.
\ee
and of the definition \cite{Uzan-Peter-anglais}
\be
\delta=\frac{\delta\rho}{\rho}\,,\quad\delta P = c^2_s\delta\rho+w\rho\Gamma\,,
\ee
where $c^2_s$ and $\Gamma$ are the sound speed and the entropy perturbation, respectively.

\subsubsection{Vector Modes}

There is only one conservation equation for the vector modes, which follows from the vector projection of Eq.~(\ref{perteulereq}). This equation is given by
\bea
&& v_a'+\HH(1-3c^2_s)v_a-\frac{k^2\pi_a^V}{\rho(1+\omega)}
-\frac{\Pi^{ij}\sigma_{ij}}{\rho}\left(\frac{1+c^2_s}{1+\omega}\right)v_a-\frac{1}{2}\spar v_a
+\sum_{b,\lambda}\mathcal{M}^\lambda_{ab}\stl v_b=\nonumber\\
&& -\frac{1}{(1+\omega)\rho}\left[\partial_\eta(\ii k\pva v - \sum_b \mathcal{V}_{ab}v_b)
-\ii k \sum_b \mathcal{U}_{ab}\pvb v + \sum_{b,c} \mathcal{U}_{ab}\mathcal{V}_{bc}v_c
+4\HH(\ii k\pva v-\sum_b \mathcal{V}_{ab}v_b)\right.\nonumber \\
&& \left.\ii k\pva\Phi +\ii k\left(2\sum_{\lambda,b}\ptl\svb\mathcal{M}^\lambda_{ab}
+2\pva\spar-\ppar\sva -\HH\pva \right)\frac{\Psi}{\HH}\right]\,.
\eea
where we have introduced
\be
\mathcal{U}_{ab}\equiv -\sigma_{ij}e^i_a e^j_b\,,\quad \mathcal{V}_{ab}\equiv-\Pi_{ij}e^i_a e^j_b\,.
\ee

\subsubsection{Friedmannian limit}

It is a straightforward exercise to verify that the above equations have a well-defined Friedman-Lema\^{\i}tre limit. Redefining $\delta\pi_{ij}\rightarrow P\delta\pi_{ij}$ and $\pi^T\rightarrow k^2\pi^S/3$ to compare with Ref.~\cite{Uzan-Peter-anglais}, we find immediately that
\bea
\delta'+3\HH[(c^2_s-\omega)\delta+\omega\Gamma]&=&(1+\omega)(k^2 v+3\Psi')\,\\
v'+\HH(1-3c^2_s)v+\Phi&=&-\frac{c^2_s}{1+\omega}\delta
-\frac{\omega}{1+\omega}\left[\Gamma-\frac{2}{3}(2k^2\pi^S)\right]\,,\\
v_a'+\HH(1-3c^2_s)v_a&=&\frac{\omega}{1+\omega}k^2\pi^V_a\,,
\eea
which are the expected equations.

\section{Perturbed geometric quantities}\label{secB}

This section gathers the expression of the geometrical quantities at order $\{1,1\}$, as needed for the computation of this article.

\subsection{Connections}\label{AppAffineConnections} 

Using the commutators of the tetrad field
\be
{\gamma^\tc}_{\ta\tb} \equiv \ThetradUD{\tc}{\,\,\mu} [\Thetrad_\ta, \Thetrad_\tb]^\mu\,,\qquad
{\gamma}_{\tc \ta\tb} \equiv \eta_{\tc\td} {\gamma^\td}_{\ta\tb}\, 
\ee
the Ricci rotation coefficients are obtained through
\be
\omega_{\ta\tb\tc} \equiv
\eta_{\tb\td}\ThetradUD{\td}{\,\,\nu}\ThetradDU{\ta}{\,\,\mu}\nabla_\mu\ThetradDU{\tc}{\,\,\nu} \,,\qquad\omega_{\ta\tb\tc} = \frac{1}{2}\left(\gamma_{\ta \tb \tc} +
  \gamma_{\tc \tb \ta} - \gamma_{\tb \tc \ta} \right)\,.
\ee
Up to order $\oo$, the commutators are
\bea
{\gamma^{\tz}}_{\tz\tz}&=&{\gamma^{\ti}}_{\tz\tz}=0\,,\\
{\gamma^{\tz}}_{\tz\ti}&=&-{\gamma^{\tz}}_{\ti\tz}=\partial_\ti \Phi = \partial_i\Phi-\beta_i^j\partial_j \Phi\,,\\
{\gamma^{\tj}}_{\tz\ti}&=&-{\gamma^{\tj}}_{\ti\tz}=-E_{ij}'+\left(\frac{\sigma_{\ti\tj}\Psi}{\HH}\right)'+\partial_i \bar B_j-\sigma_{\ti \tj}(1-\Phi)+\Psi' \delta_{i}^j\,,\\
{\gamma^{\tk}}_{\ti\tj}&=&-2\partial_{[i} E_{j]k} + \frac{2}{\HH}\partial_{[i} \Psi \sigma_{\tj]\tk} + 2\partial_{[i} \Psi\delta_{j]}^k-2\beta_{[i}^q \partial_q\Psi\delta_{j]}^k\,,
\eea
and the Ricci rotation coefficients are thus
\bea
\omega_{\tz \tz \ti} &=& -\omega_{\tz \ti \tz} =
-{\exp[-\beta]}_{ij}\partial_j \Phi\,,\\
\omega_{\ti \tz \tj} &=& -\omega_{\ti \tj \tz} = \delta_{ij} \Psi'+\left(\frac{\sigma_{ij}}{\HH}\Psi\right)'
-\sigma_{\ti \tj} (1-\Phi)-E_{ij}'+\partial_{( i} \bar B_{j)}\,,\\
\omega_{\tz \ti \tj} &=&-\omega_{\tz \tj \ti}= \partial_{[j} \bar B_{i]}\,,\\
\omega_{\ti \tj \tk} &=&-\omega_{\ti \tk \tj}=2 \delta_{i [k}\partial_{\tj]} \Psi+2 \frac{\sigma_{i [k}\partial_{\tj]} \Psi}{\HH}-2 \partial_{[j} E_{k] i}\,.
\eea

\subsection{Riemann and Ricci tensors}\label{AppRiemannRicci}

We report  the Riemann and Ricci tensor components of the metric~\eqref{perturbedmetric}, where the overall scale factor $a^2$ has been removed by a conformal transformation, up to order $\oo$. We first give their components in the coordinated basis (with the use of the package {\it xPand}~\cite{xPand}), and then in the tetrad basis $\{\Thetrad\}$. 

In the coordinate basis, the non-vanishing components are given by
\bea
R_{0i0j} &=& \gamma_{ij}\Psi'' +\left(\frac{\sigma_{ij}}{\HH}\Psi\right)'' + \sigma_{ij} (\Phi' + 2
\Psi') - \sigma_{ij}'(1-2 \Psi) \nonumber\\
&& + \partial_i \partial_j
\Phi-E_{ij}''+\partial_{(i} \bar B_{j)}'\\
R_{0ijk} &=& 2 \gamma_{i[j} {\sigma_{k]}}^q \partial_q
\Psi - 2 \sigma_{i [j} \partial_{k]} \varphi -2 \left[\left( \gamma_{i[j}  +\frac{\sigma_{i[j} }{\HH}
\right) \partial_{k]}\Psi \right]'\nonumber\\
&&+\partial_i \partial_{[j} \bar B_{k]} + 2 \partial_{[k} E_{j]i}'\\
R_{ijpq} &=& -4 \gamma_{[i [p} \sigma_{j ] q]} \Psi' +4
\left(\gamma_{[i[p} + \frac{\sigma_{[i[p}}{\HH}\right) \partial_{j]} \partial_{q]}\Psi\nonumber\\
&&-\partial_{q}\partial_j E_{ip}-\partial_{i}\partial_p E_{jq}+\partial_{p}\partial_j E_{iq}+\partial_{q}\partial_i E_{jp}
\eea
and
\bea
R_{00} &=& 3 \Psi'' + \gamma^{ij}\partial_j \partial_i \Phi\\
R_{0i} &=& 2 \partial_i \Psi' - \sigma_i^j \partial_j
(\Phi+3 \Psi) - \frac{1}{2}\Delta \bar B_i  {-
  \left[\frac{\sigma_i^j\partial_j \Psi}{\HH}\right]'}\\
R_{ij} &=& \sigma_{ij}'(1-2\Phi-2\Psi) - \sigma_{ij} (\Phi' + 3\Psi')
+\gamma_{ij} \left[ \gamma^{kq}\partial_q \partial_k \Psi -\Psi''
\right]\nonumber\\
&&+\partial_i \partial_j (\Psi-\Phi) +E_{ij}''-\Delta E_{ij}
-\partial_{(\ti} \bar B_{\tj )}'\nonumber\\
&&{ -\left(\frac{\sigma_{ij} \Psi}{\HH}\right)''
  +\frac{\sigma_{ij}}{\HH} \partial^k \partial_k \Psi -2
  \frac{\sigma_{(i}^k}{\HH} \partial_{j )} \partial_k \Psi}\,.
\eea

Projecting using the tetrad \eqref{DefTetrad} leads to the components in the tetrad basis
\bea
R_{\tz\ti\tz\tj} &=& \delta_{\ti \tj}\Psi'' +\left(\frac{\sigma_{\ti \tj}}{\HH}\Psi\right)'' + \sigma_{\ti \tj} (\Phi' + 2
\Psi') - \sigma_{\ti \tj}' (1-2\Phi)\nonumber\\
&& + \partial_\ti \partial_\tj
\Phi-E_{ij}''+\partial_{(i} \bar B_{j)}'\\
R_{\tz\ti \tj \tk} &=& 2 \delta_{\ti[\tj} {\sigma_{\tk]}}^\tq \partial_\tq
\Psi - 2 \sigma_{\ti [\tj} \partial_{\tk]} \varphi -2 \delta_{\ti
  [\tj}\partial_{\tk]}\Psi'-2\left(\frac{\sigma_{\ti[\tj}\partial_{\tk]}\Psi }{\HH}  \right)'\nonumber\\
&&+\partial_i \partial_{[j} \bar B_{k]} + 2 \partial_{[k} E_{j]i}'\\
R_{\ti \tj \tp \tq} &=& -4 \delta_{[\ti [\tp} \sigma_{\tj ] \tq]} \Psi' +4
\left(\delta_{[\ti[\tp} {+
    \sigma_{[\ti[\tp}/\HH}\right) \partial_{\tj]} \partial_{\tq]}
\Psi\nonumber\\
&&-\partial_{q}\partial_j E_{ip}-\partial_{i}\partial_p E_{jq}+\partial_{p}\partial_j E_{iq}+\partial_{q}\partial_i E_{jp}
\eea
and
\bea
R_{\tz \tz} &=& 3 \Psi'' + \partial^{\ti} \partial_\ti\Phi\\
R_{\tz \ti} &=& 2 \partial_\ti \Psi' - \sigma_\ti^\tj \partial_\tj
(\Phi+3 \Psi) - \frac{1}{2}\Delta \bar B_i  {-
  \left[\frac{\sigma_\ti^\tj\partial_\tj \Psi}{\HH}\right]'}\\
R_{\ti \tj} &=& \sigma_{\ti \tj}'(1-2\Phi) - \sigma_{\ti \tj} (\Phi' + 3\Psi')
+\delta_{\ti \tj} \left[ \partial^\tk \partial_\tk \Psi -\Psi''
\right]\nonumber\\
&&+\partial_\ti \partial_\tj (\Psi-\Phi) +E_{ij}''-\Delta E_{ij}
-\partial_{(\ti} \bar B_{\tj )}'\nonumber\\
&&{ -\left(\frac{\sigma_{\ti \tj} \Psi}{\HH}\right)''
  +\frac{\sigma_{\ti \tj}}{\HH} \partial^\tk \partial_\tk \Psi -2
  \frac{\sigma_{(\ti}^\tk}{\HH} \partial_{\tj )} \partial_\tk \Psi}\,.
\eea

\section{Glimpse on the full lensing method}\label{SecFullLensing}

This section details how a tensor field on the sphere $\bm{X}$ is lensed by a vector field ${\bm \alpha}$, the lensing being defined as the result of a parallel transport with respect to this vector field.

First, for any direction on the sphere, there exists a rotation which connects the azimuthal direction with this particular direction $\bm{n}$. If this direction has spherical coordinates $(\theta,\varphi)$  this is simply
\be
\bm{n} \equiv  R_{\bm{n}} \cdot \bm{e}_z \qquad R_{\bm{n}} = R(\varphi,\theta,0)=R_z(\varphi)\cdot R_y(\theta)\cdot R_z(0)
\ee
where $R(\alpha,\beta,\gamma)$ is a general rotation parameterized by Euler angles. Now, if we want to define the helicity basis at a given direction $\bm{n}$ as a result of this rotation applied to the helicity basis at the north pole, we have to face the fact that the helicity basis at the north pole is not well defined, since $\bm{e}_\varphi$ is not defined at this point. We choose that at the north pole $\bm{n}^\pm(\bm{e}_z) \equiv \frac{1}{\sqrt{2}}(\bm{e}_x\mp \ii \bm{e}_y)$, since this ensures that the helicity basis at any point is obtained from the one at the north pole through a rotation, that is
\be
\bm{n}^\pm (\bm{n}) = R_{\bm{n}} \cdot \bm{n}^\pm(\bm{e}_z) \,.
\ee
A spin $s$ tensor is defined as ${\bm X}(\bm{n}) \equiv  X^{s}(\bm{n}) \bm{m}^s(\bm{n}) = [X^{s} \bm{m}^s](\bm{n})$. Its component on the polarization basis are simply obtained by projection
\be
X^s(\bm{n}) = \bm{m}^{-s}(\bm{n}) \cdot {\bm X}(\bm{n}) =
\bm{m}^{-s}(\bm{e}_z)\cdot [R^{-1}_{\bm{n}} . {\bm X}(\bm{n}) ] = \bm{m}^{-s}(\bm{e}_z)\cdot [R^{-1}_{\bm{n}}{\bm X}](\bm{e}_z)\,.
\ee
This means that instead of projecting a tensor on the polarization basis at a point $\bm{n}$ we can equivalently rotate it so that the point
which initially in $\bm{n}$ becomes located on the azimuthal direction. Then we can evaluate its components on the polarization basis at this azimuthal direction. The azimuthal direction can thus be used as a common reference for all points on the sphere since for each point there is a unique natural rotation to transport from this point to the azimuthal direction.

\begin{figure}[ht!]
\centering
\includegraphics[width=8.5cm]{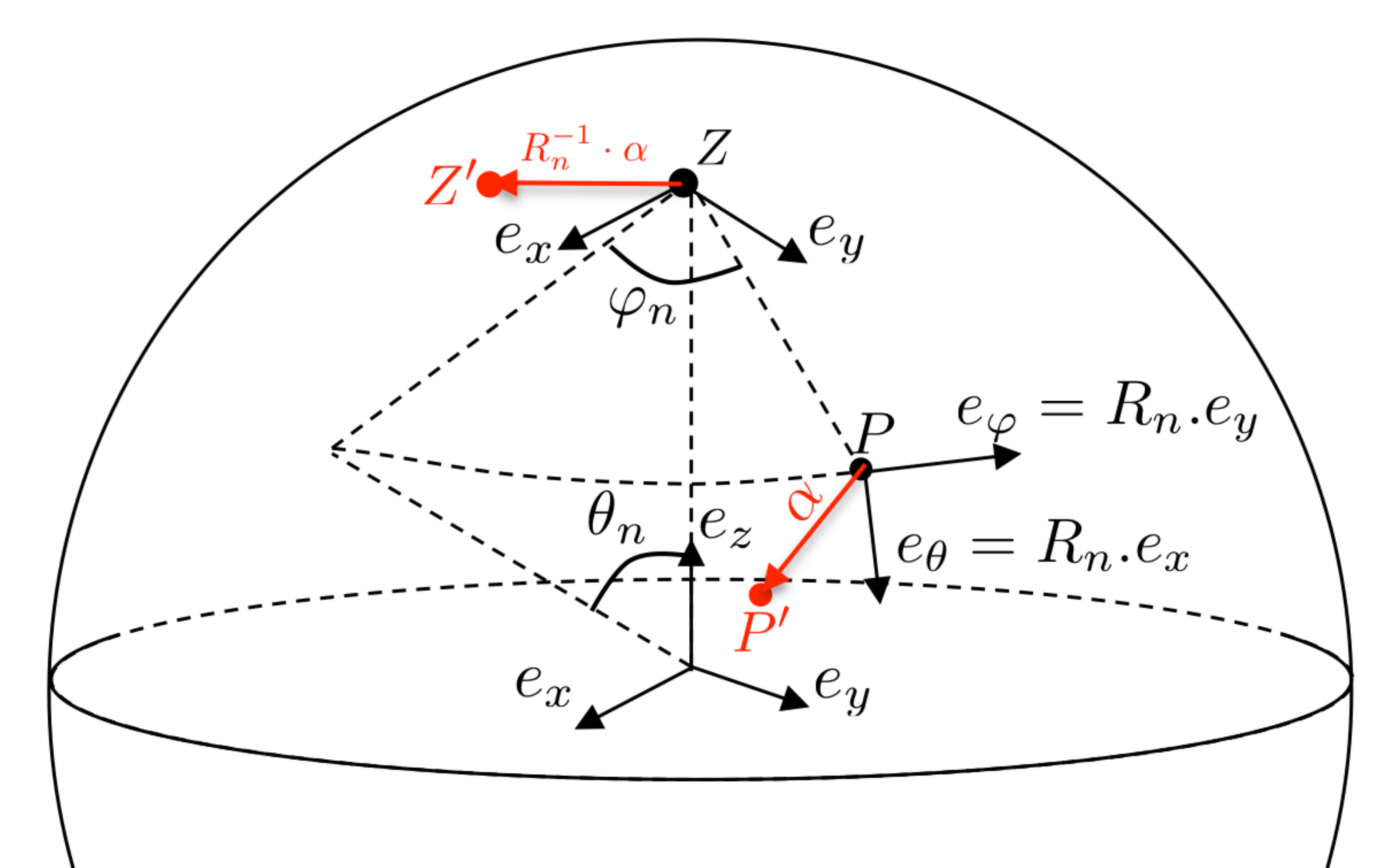}\includegraphics[width=8.5cm]{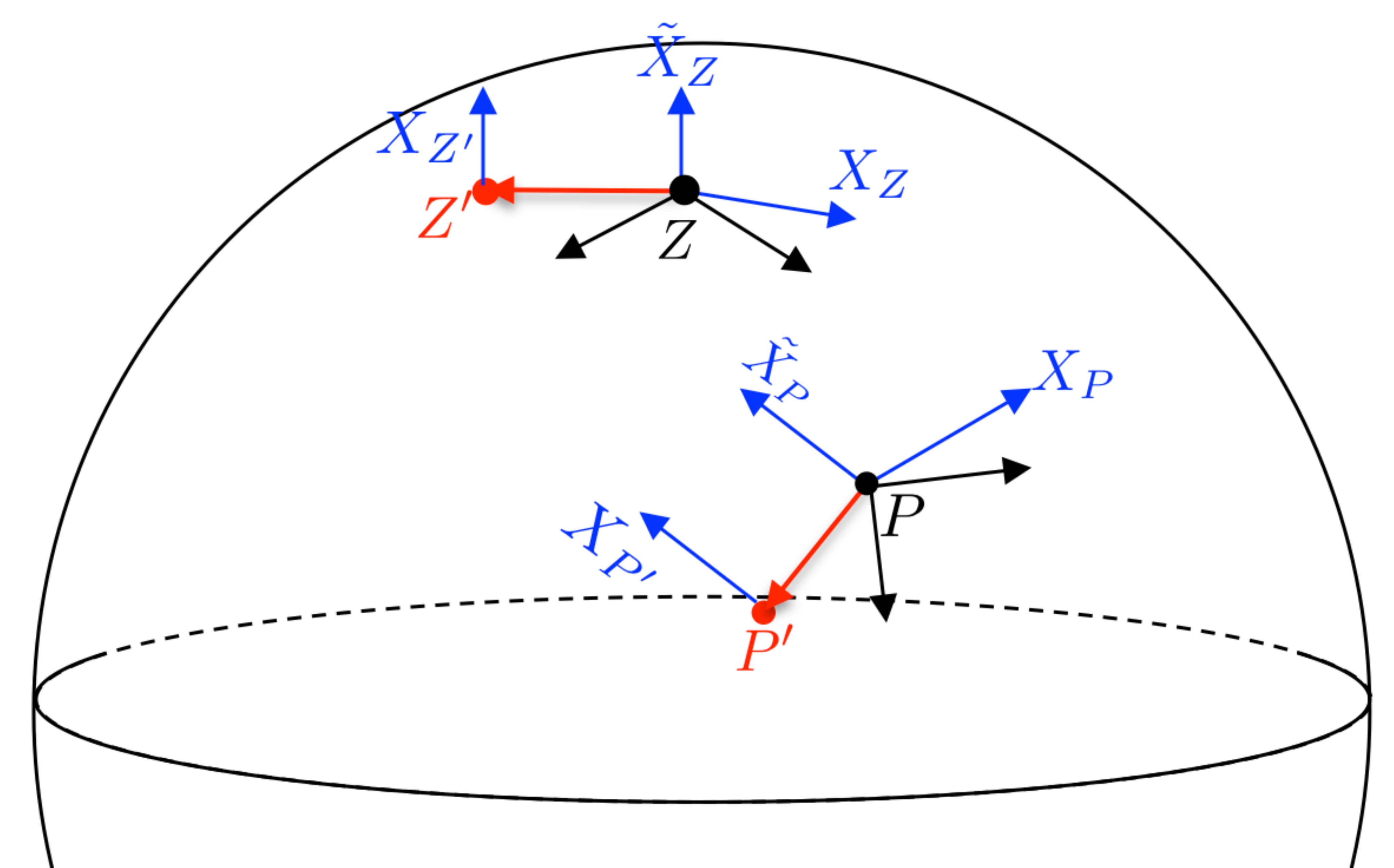}
\centering
\caption{The azimuthal point $Z$ is at the azimuthal direction $\gr{e}_z$, while the point $P$ is located in $R_{\bm{n}} \cdot \bm{e}_z$. Because of lensing the signal observed in $P$ is coming from $P'$ whose location is at $R_{\underline{\bm{n}\times {\bm \alpha}(\bm{n})}}\cdot R_{\bm{n}} \cdot \bm{e}_z$, which corresponds to a covariant transport along ${\bm \alpha}$ from the point $P$. As for the point $Z'$, it is obtained by applying $R_{\bm{n}}^{-1}$ on $P'$ meaning that it is located at $R_{\bm{n}}^{-1}\cdot R_{\underline{\bm{n}\times {\bm \alpha}(\bm{n})}} \cdot R_{\bm{n}} \cdot \bm{e}_z$. This is equivalent to $R_{\underline{\bm{e}_z\times [R^{-1}_{\bm{n}}{\bm \alpha}(\bm{n})]}}\cdot \bm{e}_z$ which corresponds to a covariant transport from the point $Z$, along the vector $R^{-1}_{\bm{n}}{\bm \alpha}(\bm{n})$. The lensed field at $P$ ($\tilde X_P$) is obtained by covariantly transporting back along ${\rm \alpha}$ the unlensed field at $P'$ ($X_{P'}$) by application of $R^{-1}_{\underline{\bm{n}\times {\bm \alpha}(\bm{n})}}$, and we get $\tilde X_P = \left(R^{-1}_{\underline{\bm{n}\times {\bm \alpha}(\bm{n})}} \cdot X\right)_P$, and it is in general different from the unlensed field at that point $X_P$. In order to read the components, everything is transported back into the azimuthal region by application of $R^{-1}_{\bm{n}}$. We get $X_{Z'} \equiv \left(R^{-1}_{\bm{n}} \cdot X\right)_{Z'}$, $X_Z \equiv \left(R^{-1}_{\bm{n}} \cdot X\right)_{Z}$, and $\tilde X_Z = \left(R^{-1}_{\bm{n}} \cdot \tilde X\right)_{Z}=\left(R^{-1}_{\bm{n}} \cdot R^{-1}_{\underline{\bm{n}\times {\bm \alpha}(\bm{n})}} \cdot X \right)_{Z}$. For the latter, there is an alternative expressions, which corresponds to covariantly transporting back the field $X_{Z'}$ along $R^{-1}_{\bm{n}}{\bm \alpha}$, and this leads to  $\tilde X_Z =  \left(R^{-1}_{\underline{\bm{e}_z\times [R^{-1}_{\bm{n}}{\bm \alpha}(\bm{n})]}}\cdot R^{-1}_{\bm{n}} \cdot X\right)_Z$, from which the components can be read by projection on the local helicity basis associated with ${\bm e}_x$ and ${\bm e}_y$. This is precisely the meaning of Eq.~\eqref{MagicLensing} which is used to compute the components of the lensed field.
}
\label{Fig3}
\end{figure}

Let us consider that, due to the lensing vector ${\bm \alpha}$, the tensor field we observe in the direction $\bm{n}$, is now the result of a parallel transport of the underlying tensor by this vector field ${\bm \alpha}$. Such a parallel transport is equivalent to a rotation around the axis $\bm{n} \times {\bm \alpha}$, so that the lensed tensor field is related to the unlensed one by 
\be
\tilde{\bm{X}}(\bm{n}) \equiv [R^{-1}_{\underline{\bm{n}\times {\bm
    \alpha}(\bm{n})}} \bm{X}](\bm{n})\,.
\ee
We use the notation $R_{\underline{\bm V}}$ to  indicate the rotation defined by the rotation vector ${\bm V}$. This is the rotation around the axis defined by the vector $\bm{V}$ with an angle obtained from the norm of ${\bm V}$. It must not be confused with the previous notation $R_{\bm{n}}$, which is the rotation that brings the azimuthal direction toward the direction $\bm{n}$. 
As emphasized previously, the components of the lensed tensor field, as any tensor field, can be obtained by transportation to the azimuthal direction, that is
\be\label{EqB5}
\tilde X^s(\bm{n}) = \bm{m}^{-s}(\bm{n}). \tilde{\bm X}(\bm{n}) = \bm{m}^{-s}(\bm{e}_z)\cdot [R^{-1}_{\bm{n}}R^{-1}_{\underline{\bm{n}\times {\bm \alpha}(\bm{n})}} {\bm X}](\bm{e}_z) \,.
\ee
Using the general property of rotations $R_{\bm{n}}R_{\underline{\bm{V}}} R^{-1}_{\bm{n}} = R_{\underline{R_{\bm{n}}\cdot \bm{V}}}$ leads to 
\be
R^{-1}_{\bm{n}}R_{\underline{\bm{n}\times {\bm \alpha}(\bm{n})}} R_{\bm{n}}=
R_{\underline{\bm{e}_z\times [R^{-1}_{\bm{n}}{\bm \alpha}(\bm{n})]}}\,,
\ee
that is to
\be
R^{-1}_{\bm{n}}R^{-1}_{\underline{\bm{n}\times {\bm \alpha}(\bm{n})}} =
R^{-1}_{\underline{\bm{e}_z\times [R^{-1}_{\bm{n}}{\bm \alpha}(\bm{n})]}} R^{-1}_{\bm{n}}\,,
\ee
which can be used to recast Eq.~\eqref{EqB5} as 
\be\label{MagicLensing}
\tilde X^s(\bm{n}) = \bm{m}^{-s}(\bm{e}_z)\cdot
[R^{-1}_{\underline{\bm{e}_z\times [R^{-1}_{\bm{n}}{\bm
      \alpha}(\bm{n})]}} R^{-1}_{\bm{n}}{\bm X}](\bm{e}_z).
\ee
This can be understood easily once we extract the helicity components of the lensing vector. Indeed, the helicity basis components of the lensing vector field are obtained just like for any vector field as $\alpha^\pm(\bm{n}) = \bm{n}^\mp(\bm{n}).{\bm \alpha}(\bm{n}) =
\frac{1}{\sqrt{2}}(\bm{e}_x\mp \ii \bm{e}_y). [R^{-1}_{\bm{n}}{\bm\alpha}(\bm{n})]$. If we define
\bea\label{DefComponentsAlpha}
&&\alpha_x =\frac{1}{\sqrt{2}}(\alpha_+ + \alpha_-)\,,
\qquad  \alpha_y =\frac{\ii}{\sqrt{2}}(\alpha_+ - \alpha_-)\,,\\
&&
\alpha_x = \alpha_\theta \cos \alpha_\varphi\,,\qquad \alpha_y = \alpha_\theta \sin \alpha_\varphi\,,
\eea
where $(\alpha_x,\alpha_y)$ are the components of the lensing field once transported to the azimuthal direction, and $(\alpha_\theta,\alpha_\varphi)$ their associated polar components, we obtain that
\be
R_{\underline{\bm{e}_z\times [R^{-1}_{\bm{n}}{\bm \alpha}(\bm{n})]}} =
R(\alpha_\varphi,\alpha_\theta,-\alpha_\varphi) \,.
\ee
This means that instead of lensing the tensor field at the point $\bm{n}$ and subsequently extracting the component, it is equivalent to transport both the field and the lensing vector at the azimuthal direction with $R_{\bm{n}}^{-1}$ and then let the transported lensing vector act on the transported tensor field. This procedure is explained graphically in Fig.~\ref{Fig3}.

With this crucial result at hand, we just need to compose the rotations of Eq.~\eqref{MagicLensing} in order to obtain the components of the lensed field in terms of the multipole components of the unlensed tensor field. Using the general transformation law~(\ref{RotationSpinnedField}), we get
\bea\label{MainLensingresult}
\tilde{X}^s(\bm{n})&=& \sum_{\ell m m' m''}Y_{\ell
  m''}^{s}(\bm{e}_z)D^\ell_{m''
  m'}[R^{-1}(\alpha_\varphi,\alpha_\theta,-\alpha_\varphi)]D^{\ell}_{m'
  m}[R^{-1}_{\bm{n}}]  X^s_{\ell m}  \\
&=&\sum_{\ell m m'} (-1)^s \sqrt{\frac{2 \ell+1}{4 \pi}} D^\ell_{-s\,
  m'}[R^{-1}(\alpha_\varphi,\alpha_\theta,-\alpha_\varphi)]D^{\ell}_{m'
  m}[R^{-1}_{\bm{n}}] X^s_{\ell\, m}\\
&=&\sum_{\ell m m'} e^{\ii s \alpha_\varphi }Y_{\ell m'}^s(\alpha_\theta,\alpha_\varphi)]D^{\ell}_{m'
  m}[R^{-1}_{\bm{n}}] X^s_{\ell\, m}\,,
\eea
where it is understood that the components $(\alpha_\theta,\alpha_\varphi)$ correspond to the lensing vector field at the position $\bm{n}$ considered, and these should be obtained from the definitions~\eqref{DefComponentsAlpha}.

From this relation between a tensor acting as a source for an observable and its lensed version due to the geodesic structure between the source and the observer, it is possible in principle, to obtain the correlations functions (see e.g. Ref.~\cite{Lewis2006} for the case of CMB lensing). 
A simplification can be obtained by expanding the spherical harmonics in a small angle approximation. Indeed, very close to the azimuthal direction, the spin-weighted spherical harmonics are approximated by
\be
Y^s_{\ell m}(\theta,\varphi) \simeq (-1)^m\sqrt{\frac{2\ell+1}{4 \pi}} e^{\ii m \varphi}J_{m+s}[(\ell+1/2)\theta]
\ee
and given that $J_{m+s}(x)$ behaves like $x^{m+s}$ when $x \to 0$ there is a natural way to expand Eq.~\eqref{MainLensingresult} in powers of the lensing angle. More details can be found in Ref.~\cite{Lewis2006}.

\section{Mathematical tool-box}\label{secD}

\subsection{From Cartesian to spherical derivatives}\label{App1plus2Cartesian}

In this section $D_i$ refers to the covariant derivative on the unit sphere in Cartesian coordinates, that is $D_i^\Rthreesymbol$ defined in Eq.~(\ref{DefRthreesymbol}). The key relation to derive all the decompositions from Cartesian derivatives $\partial_i$ to radial and covariant spherical derivatives is the simple relation
\be
r \partial_i \hat x_j =  S_{ij}, 
\ee
which is just the statement that the extrinsic curvature on a unit
sphere is equal to the metric on this sphere.
For a scalar, a projected vector and a projected tensor, we have

\bea
\partial_i \varphi &=& \frac{D_i \varphi}{r} + \varphi_{,r} \hat x_i\\
\partial_i \tilde B_j&=& \frac{D_i \tilde B_j}{r} + (\tilde B_j)_{,r} \hat
x_i - \frac{\hat x_j}{r} \tilde B_i\\
\partial_i \tilde E_{jk}&=& \frac{D_i \tilde E_{jk}}{r} + (\tilde E_{jk})_{,r} \hat
x_i - \frac{\hat x_j}{r} \tilde E_{ik}- \frac{\hat x_k}{r} \tilde E_{ji}\,,
\eea
where we use the notation $X_{,r} \equiv \hat x^i\partial_i X$ for the radial derivative. Note that this radial derivative and the covariant derivative on the unit sphere $D_i$ commute, as they are just the geometric versions of derivatives in spherical coordinates.

Iterating these relations we obtain for scalar perturbations
\bea
\partial_i \partial_j \varphi &=& 2 \hat x_{(i}
D_{j)}\left(\frac{\varphi}{r}\right)_{,r}+\frac{D_i D_j
  \varphi}{r^2}+\varphi_{,rr} \hat x_i \hat x_j+
S_{ij}\frac{\varphi_{,r}}{r}\\
S^{ij} \partial_i \partial_j \varphi&=& \frac{D_i D^i \varphi}{r^2} +
2 \frac{\varphi_{,r}}{r}\\
P[\hat x^j \partial_i \partial_j \varphi] &=& D_i
\left(\frac{\varphi}{r}\right)_{,r}\\
D_i \varphi_{,r} &=& (D_i \varphi)_{,r}\\
S_i^p S_j^q \partial_k \partial_p \partial_q \varphi&=& \frac{2}{r} S_{k(j}
D_{i)} \left(\frac{\varphi}{r}\right)_{,r} +S_{ij} \partial_k \left(
  \frac{\varphi_{,r}}{r}\right)+\partial_k \left(\frac{D_i D_j
    \varphi}{r^2}\right)\\
\hat x^k S_i^p S_j^q \partial_k \partial_p \partial_q \varphi&=&
\left(\frac{D_i D_j \varphi}{r^2}\right)_{,r} +S_{ij} \left(
  \frac{\varphi_{,r}}{r}\right)_{,r}\\
S_k^r S_i^p S_j^q \partial_r \partial_p \partial_q \varphi&=& \frac{2}{r} S_{k(j}
D_{i)} \left(\frac{\varphi}{r}\right)_{,r} +S_{ij} D_k \left(
  \frac{\varphi_{,r}}{r^2}\right)+\left(\frac{D_k D_i D_j
    \varphi}{r^3}\right)\,.
\eea
As for vectors and tensor, the useful relations are
\bea
\hat x^i \hat x^j \partial_k \partial_l
E_{ij}&=& \partial_k \partial_l E_r -\frac{4 \partial_{(k} \tilde
E_{l)}}{r}+\frac{2 \tilde E_{kl}}{r^2}-\frac{2 S_{kl} E_r}{r^2}-2
\tilde E_{(k} \hat x_{l)}\\
\hat x^l \partial_i \partial_j B_l&=&\frac{D_i D^i B_r}{r^2} - \frac{2
D^i \tilde B_i}{r^2}+(B_r)_{,rr} +2\frac{(B_r)_{,r}}{r}-\frac{2 B_r}{r^2}\,.
\eea
Finally, using the fact that the vector modes are transverse and that the tensor modes are transverse and traceless, we get
\bea
B_i &\equiv& \tilde B_i + \hat x_i B_r \label{FirstSplittingoftensor}\\
D^i \tilde B_i & = & -2 B_r - r (B_r)_{,r}\\
E_{ij} &\equiv& \tilde E_{ij} + 2 \tilde E_{(i} \hat x_{j)} + E_r \hat x_i
\hat x_j \\
S^{ij} \tilde E_{ij}&=& - E_r=0\\
D^i \tilde E_i &=& -3E_r-r ({E_r})_{,r}\\
D^i \tilde E_{ij} &=& -3\tilde E_j- r (\tilde E_j)_{,r}\label{LastSplittingoftensor}\,.
\eea

\subsection{Spin-weighted spherical harmonics}

Spin-weighted spherical harmonics are defined in terms of Wigner $D$-matrices as~\cite{Goldberg1967}
\bea
Y^s_{\ell m}(\alpha,\beta) & = &\sqrt{\frac{2 \ell+1}{4 \pi}} (-1)^m
e^{\ii s \gamma} D^{\ell}_{-m s}(\alpha,\beta,\gamma)\,, \\
& = & \sqrt{\frac{2 \ell+1}{4 \pi}} (-1)^m D^{\ell}_{-m s}(\alpha,\beta,0)\,,\\
& = & (-1)^{m+s}Y^{-s\,\star}_{\ell,- m}(\alpha,\beta)
\eea
where $\alpha$, $\beta$ and $\gamma$ are the Euler angles. Wigner $D$-matrices are in turn defined in terms of infinitesimal generators of three dimensional rotations as~\cite{edmonds}
\be
D^{\ell}_{m_1m_2}=\langle\ell m_1|U(R)|\ell m_2\rangle\,,\quad {\rm{where}}\;\quad 
U(\alpha,\beta,\gamma) = e^{-i\alpha J_z}e^{-i\beta J_y}e^{-i\gamma J_z}\,.
\ee
In the special case in which the direction is aligned with the $z$-axis we have
\be
Y^{-s}_{\ell m}(\bm{e}_z)  = \delta_{m s}(-1)^m \sqrt{\frac{2 \ell + 1}{4 \pi}}\,.
\ee
Under the parity transformation
\[
\alpha \rightarrow \alpha+\pi\,,\quad
\beta \rightarrow \pi - \beta\,,\quad
\gamma \rightarrow \gamma + \pi
\]
the spherical harmonics and Wigner $D$-matrices transform as
\begin{align}
\label{parity1}
Y_{\ell m}(\alpha,\beta) & \rightarrow (-1)^\ell Y_{\ell m}(\alpha,\beta)\,, \\
D^{\ell}_{m m'}(\alpha,\beta,\gamma) & \rightarrow (-1)^{\ell+m}D^{\ell}_{-mm'}(\alpha,\beta,\gamma) 
 = (-1)^{\ell+m'}D^{\ell}_{m,-m'}(\alpha,\beta,\gamma)\,.
\end{align}
In particular, it follows that under parity transformation the Spin-weighted spherical harmonics behaves as
\be
\label{parity2}
Y^{s}_{\ell m}(\alpha,\beta) \rightarrow (-1)^\ell Y^{-s}_{\ell m}(\alpha,\beta)\,.
\ee

\subsection{Rotation of fields on the sphere}

The transformation of the spherical harmonics is given by
\be
[R Y_{\ell m}](\bm{n}) \equiv Y_{\ell m}(R^{-1}.\bm{n})  = \sum_{m'}
\langle \bm{n}| \ell\, m' \rangle \langle \ell m' | R | \ell\, m\rangle  =\sum_{m'} Y_{\ell m'}(\bm{n}) D^\ell_{m' m}(R)
\ee
where the first equality is the definition of the transformation of a function on a sphere under a rotation. For a scalar field on the sphere, we can deduce the transformation of its multipolar components of its expansion in spherical harmonics,
$$
X(\bm{n}) = \sum_{\ell m} X_{\ell m}Y_{\ell m}(\bm{n}),
$$
to be
\be
[RX](\bm{n}) = X(R^{-1}.\bm{n}) =\sum_{\ell\,m\,m'}
Y_{\ell m'}(\bm{n}) D^\ell_{m' m}(R) X_{\ell m}
\ee
so that
\be
[RX]_{\ell m'} = \sum_m D^\ell_{m' m}(R) X_{\ell m}\,. 
\ee

The rotation of a tensor field on the sphere is very similar. Once it is broken down into symmetric traceless tensors then, by using that such tensors are decomposed as ${\bf X}(\bm{n}) \equiv X^{\pm s}(\bm{n}) \bm{m}^{\pm s}(\bm{n}) = [X^{\pm s} \bm{m}^{\pm s}](\bm{n})$, it can be expanded in Spin-weighted spherical harmonics as
\be
{\bf X}(\bm{n}) = \sum_{\ell m} \left[X^{+s}_{\ell m}Y^{+s}_{\ell m}(\bm{n}) \bm{m}^{+s}(\bm{n}) + X^{-s}_{\ell m}Y^{-s}_{\ell m}(\bm{n}) \bm{m}^{-s}(\bm{n})\right].
\ee
Under a rotation, it transforms as (see Appendix A of Ref.~\cite{Challinor:2005jy})
\be\label{RotationSpinnedField}
[R{\bf X}](\bm{n}) = R.{\bf X}(R^{-1}.\bm{n}) =\sum_{\ell\,m\,m'}
Y^{\pm s}_{\ell m'}(\bm{n}) D^\ell_{m' m}(R) X^{\pm s}_{\ell m} \bm{m}^{\pm s}(\bm{n})\,,
\ee
that is
\be
[RX]^{\pm s}_{\ell m'} = D^\ell_{m' m}(R) X^{\pm s}_{\ell m}\,.
\ee
We remark that it is exactly the same transformation law as for scalar fields because we have transformed the full tensor field $X^{\pm s}(\bm{n}) \bm{m}^{\pm s}(\bm{n}) $ and not just its component $X^{\pm s}(\bm{n})$ considered as a scalar function, for which the transformation law is more complicated~\cite{Challinor:2005jy}.

\subsection{Wigner 3j Symbols}\label{3j_identities}

The 3j symbols satisfy the following properties
\begin{align*}
\left(\begin{array}{ccc}
\ell_{1} & \ell_{2} & \ell_{3}\\
m_{1} & m_{2} & m_{3}
\end{array}\right) & =\left(\begin{array}{ccc}
\ell_{2} & \ell_{3} & \ell_{1}\\
m_{2} & m_{3} & m_{1}
\end{array}\right)=\left(\begin{array}{ccc}
\ell_{3} & \ell_{1} & \ell_{2}\\
m_{3} & m_{1} & m_{2}
\end{array}\right)\\
 & =(-1)^{\ell_{1}+\ell_{2}+\ell_{3}}\left(\begin{array}{ccc}
\ell_{1} & \ell_{3} & \ell_{2}\\
m_{1} & m_{3} & m_{2}
\end{array}\right)\\
 & =(-1)^{\ell_{1}+\ell_{2}+\ell_{2}}\left(\begin{array}{ccc}
\ell_{1} & \ell_{2} & \ell_{3}\\
-m_{1} & -m_{2} & -m_{2}
\end{array}\right)
\end{align*}
Moreover, they are identically zero whenever any of the following conditions are violated,
\[
m_1+m_2+m_3=0\,,\quad |\ell_i-\ell_j|\leq\ell_k\leq\ell_i+\ell_j\,,\quad\{i,j,k\}=\{1,2,3\}\,.
\]
They are also orthogonal in the sense that
\be\label{e:OrthoWigner1}
\sum_{m_1,m_2} \troisj{\ell_1}{\ell_2}{\ell}{m_1}{m_2}{m}
\troisj{\ell_1}{\ell_2}{\ell'}{m_1}{m_2}{m'}= \frac{1}{2 \ell+1}
\delta_{\ell \ell'} \delta_{m m'}
\ee
and that
\be \label{e:OrthoWigner2}
\sum_{\ell,m} (2\ell+1) \troisj{\ell_1}{\ell_2}{\ell}{m_1}{m_2}{m}
\troisj{\ell_1}{\ell_2}{\ell}{m'_1}{m'_2}{m}= \delta_{m_1 m'_1} \delta_{m_2 m_2'}\,.
\ee

Since Eq.~(\ref{e:OrthoWigner2}) holds for any set $\{m_1,m'_1,m_2,m'_2\}$, two important cases 
follow from this expression. First, consider the case where $m_2=-m_1$ and $m'_2=-m'_1$. Then, using the selection rule 
$m_1+m_2+m=0$, it follows that
\be
\label{e:OrthoWigner2-2}
\sum_{\ell} (2\ell+1) \troisj{\ell_1}{\ell_2}{\ell}{m_1}{-m_1}{0}
\troisj{\ell_1}{\ell_2}{\ell}{m'_1}{-m'_1}{0}= \delta_{m_1 m'_1}\,.
\ee
Second, note that if we further impose that $m_1=m'_1=0$, then
\be \label{e:OrthoWigner2-1}
\sum_{\ell} (2\ell+1) \troisj{\ell_1}{\ell_2}{\ell}{0}{0}{0}^2=1\,.
\ee
Another useful expression is
\be
\label{e:Wigner3}
\troisj{\ell_1}{\ell_2}{0}{0}{0}{0} = \delta_{\ell_1\ell_2}\frac{1}{\sqrt{2 \ell_1 +1}}\,.
\ee

A recurrent expression when dealing with deviations of isotropy is the integral of 
three spherical harmonics, also known as the Gaunt integral, and defined as
\begin{eqnarray}
\label{gaunt}
\int \dd^2\Omega\,
Y^{s_1}_{\ell_{1}m_{1}}(\hat{\mathbf{n}})
Y^{s_2}_{\ell_{2}m_{2}}(\hat{\mathbf{n}})
Y^{s_3}_{\ell_{3}m_{3}}(\hat{\mathbf{n}}) 
& = & \sqrt{\frac{(2\ell_{1}+1)(2\ell_{2}+1)(2\ell_{3}+1)}{4\pi}}\nonumber \\
&& \times\left(\begin{array}{ccc}
\ell_{1} & \ell_{2} & \ell_{3}\\
-s_1 & -s_2 & -s_3
\end{array}\right)\left(\begin{array}{ccc}
\ell_{1} & \ell_{2} & \ell_{3}\\
m_{1} & m_{2} & m_{3}
\end{array}\right)\,.
\end{eqnarray}
Note that, due to the symmetries of 3j symbols and the properties of Spin-weighted spherical harmonics under complex conjugation, the coefficients ${}^sC^{m_1m_2m_2}_{\ell_1\ell_2\ell_3}$ defined in Eq.~(\ref{C_Gaunt}) satisfy the following properties
\be\label{PropertiesGaunt}
{}^{-s} C^{m_1 m_2 m_3}_{\ell_1  \ell_2 \ell_3} = (-1)^{\ell_1+\ell_2+\ell_3}{}^s C^{m_1 m_2
  m_3}_{\ell_1  \ell_2 \ell_3} = 
{}^{s} C^{-m_1\, -m_2\, -m_3}_{\ell_1  \ell_2 \ell_3}\,.
\ee
From the definitions (\ref{gaunt}), (\ref{C_Gaunt}) and the closure relation of spherical harmonics one can also verify that
\be
\label{gaunt_ylm}
Y_{\ell_2m_2}(\bm{n})Y^s_{\ell_3m_3}(\bm{n})
=\sum_{\ell_1,m_1} {}^sC^{m_1m_2m_3}_{\ell_1\ell_2\ell_3}
Y^{s\,}_{\ell_1 m_1}(\bm{n})\,,
\ee
an identity which is needed in order to derive Eq.~(\ref{cov_matrix}).

Let us also define a useful integral for the gradient expansion approach of lensing by 
\be\label{IHU}
{}_{\pm s} I^{m_1 m_2 m_3}_{\ell_1  \ell_2 \ell_3}  \equiv  \int \dd^2\Omega\,
[D^a Y^{\pm s\star}_{\ell_{1}m_{1}}(\hat{\mathbf{n}})]\,
Y_{\ell_{2}m_{2}}(\hat{\mathbf{n}})\,
[D_a Y^{\pm s}_{\ell_{3}m_{3}}(\hat{\mathbf{n}})]
\ee
where the polarization basis is voluntarily omitted for a simpler notation. 
It has the useful property inherited from Eq.~\eqref{PropertiesGaunt}
\be
{}_{\pm s} I^{m_1 m_2 m_3}_{\ell_1  \ell_2 \ell_3}=(-1)^{\ell_1+\ell_2 +\ell_3}{}_{\mp s} I^{m_1 m_2 m_3}_{\ell_1  \ell_2 \ell_3}\,.
\ee
Its expression can be found using
\begin{align}
\label{PropertyIHU}
{}_{\pm s} I^{m_1 m_2 m_3}_{\ell_1  \ell_2 \ell_3} & =\frac{1}{2}\left[\ell_2(\ell_2+1)+\ell_3(\ell_3+1)-\ell_1(\ell_1+1) \right]\int \dd^2\Omega\,
Y^{\pm s \star}_{\ell_{1}m_{1}}(\hat{\mathbf{n}})
Y_{\ell_{2}m_{2}}(\hat{\mathbf{n}})
Y^{\pm s}_{\ell_{3}m_{3}}(\hat{\mathbf{n}})\,,\\
& = \frac{1}{2}\left[\ell_2(\ell_2+1)+\ell_3(\ell_3+1)-\ell_1(\ell_1+1) \right]{}^{\pm s} C^{m_1 m_2 m_3}_{\ell_1  \ell_2 \ell_3}\,,\\
& = {}_{\pm s}F_{\ell_1 \ell_2 \ell_3} \troisj{\ell_1}{\ell_2}{\ell_3}{-m_1}{m_2}{m_3}(-1)^{m_1+s}\,.
\end{align}
where, following~Ref.~\cite{Hu2000}, we defined the symbols
\be
{}_s F_{\ell \ell_1 \ell_2}\equiv \frac{1}{2}\left[ \ell_1(\ell_1+1)+\ell_2(\ell_2+1)-\ell(\ell+1)\right]\sqrt{\frac{(2 \ell+1)(2\ell_1+1)(2\ell_2 +1)}{4 \pi}}\troisj{\ell}{\ell_1}{\ell_2}{s}{0}{-s}
\ee
In particular we have
\bea
{}_2 F_{\ell \,2\, \ell +1}
&=&(\ell+4)\sqrt{\frac{5(2\ell+1)(2\ell+3)}{4 \pi}}
\troisj{\ell}{2}{\ell+1}{2}{0}{-2}=(-1)^\ell(\ell+4)\sqrt{\frac{15}{\pi}}\sqrt{\frac{(\ell+3)(\ell-1)}{\ell(\ell+1)(\ell+2)}}\nonumber\,,\\
{}_2 F_{\ell \,2\, \ell -1} &=&(3-\ell)\sqrt{\frac{5(2\ell+1)(2\ell-1)}{4 \pi}} \troisj{\ell}{2}{\ell-1}{2}{0}{-2}=(-1)^\ell(\ell-3)\sqrt{\frac{15}{\pi}}\sqrt{\frac{(\ell+2)(\ell-2)}{\ell(\ell+1)(\ell-1)}}\nonumber\,.
\eea

\end{document}